\documentclass[12pt]{article}
\usepackage{arxiv_jasa}

\begin{document}


\if1\anon
{
  \title{\bf Bayesian Nonparametric Causal Inference for Quantile Residual Life: An Application to Alzheimer's Disease}
  \author{
    Woojung Bae$^{1}$, Taekwon Hong$^{2}$, Sang Kyu Lee$^{3,*}$,\\
    Dongrak Choi$^{4}$, and Jong-Hyeon Jeong$^{5}$\\[0.5em]
    {\small $^{1}$Division of Biostatistics, Office of Biostatistics and Pharmacovigilance, CBER, U.S. FDA}\\
    {\small $^{2}$Division of Biometrics VII, Office of Biostatistics, CDER, U.S. FDA}\\
    {\small $^{3}$Department of Applied Statistics, Konkuk University}\\
    {\small $^{4}$Department of Biostatistics and Bioinformatics, Duke University}\\
    {\small $^{5}$Biometric Research Program, National Cancer Institute}\\[0.3em]
    {\small $^{*}$Corresponding author: \texttt{sangkyulee@konkuk.ac.kr}}
  }
  \date{}
  \maketitle
}
\fi

\if0\anon
{
  \bigskip
  \bigskip
  \bigskip
  \begin{center}
    {\LARGE\bf Bayesian Nonparametric Causal Inference for Quantile Residual Life: An Application to Alzheimer's Disease}
  \end{center}
  \medskip
}
\fi

\bigskip
\begin{abstract}
    In Alzheimer’s disease research, a clinically important question is how much longer individuals would remain dementia-free beyond a given time under different baseline amyloid statuses. We address this question using observational data from the Alzheimer’s Disease Neuroimaging Initiative (ADNI) treating baseline amyloid status as the exposure. Estimation is challenging because amyloid status is confounded, time to dementia onset is heterogeneous and heavily right censored, and the target population depends on joint potential event times. At each time point, we consider the always-survivor principal stratum comprising individuals who would remain dementia-free under both amyloid status and estimate quantile contrasts in residual time to dementia onset. We model the joint distribution of event time, exposure, and baseline covariates using an enriched Dirichlet process mixture and conduct posterior inference via Bayesian g-computation. The framework accommodates partially observed covariates under a within-subcluster missing-at-random assumption, estimates contrasts across multiple time points and quantiles from one posterior fit and supports sensitivity analyses for unmeasured confounding, cross-world dependence, and informative censoring. Simulations show favorable finite-sample performance under heterogeneity and heavy censoring. In ADNI, residual time to dementia onset was shorter under elevated than non-elevated baseline amyloid status, both overall and within baseline subgroups.
\end{abstract}

\noindent%

\vfill

\spacingset{1.8} 
\compactdisplays

\section{Introduction} \label{EDPMqrl.introduction}
    Alzheimer’s disease (AD) research increasingly requires clinically interpretable summaries of prognosis in real-world populations. Recent U.S. Food and Drug Administration approvals of amyloid beta--directed monoclonal antibodies for early AD have further underscored the clinical importance of amyloid pathology in disease progression \citep{fda2023leqembi, fda2024kisunla}. Beyond evaluating treatment efficacy, understanding how baseline amyloid status is associated with subsequent progression to dementia is also important for characterizing prognosis across heterogeneous patient populations. Because baseline amyloid status is a naturally occurring biomarker rather than a randomized exposure, causal inference regarding dementia progression relies on observational longitudinal data and is therefore susceptible to confounding. Such analyses are complicated by systematic differences between amyloid-status groups, heterogeneous disease trajectories, and substantial right censoring.

\subsection{ADNI Study} \label{EDPMqrl.introduction.adni}
    ADNI is a longitudinal, multicenter observational study that collects clinical, cognitive, imaging, genetic, and fluid-biomarker data to characterize the progression of mild cognitive impairment (MCI) and early AD (\href{http://adni.loni.usc.edu/}{adni.loni.usc.edu}). Its rich baseline information and long follow-up provide an important setting for studying the relationship between baseline amyloid status and subsequent dementia onset. Because baseline disease stage may modify the relationship between amyloid status and dementia prognosis, we examine amyloid-related differences both in the overall cohort and within baseline diagnostic subgroups defined by MCI and cognitively normal (CN) status. MCI represents an intermediate clinical stage in which cognitive impairment is present but the criteria for dementia have not yet been met, making the remaining time to dementia onset a particularly relevant prognostic quantity. CN represents an earlier clinical stage without substantial cognitive impairment at baseline. Examining these subgroups separately allows us to assess whether amyloid-related differences in dementia prognosis vary across clinically distinct baseline populations.
    
    \begin{figure}[!tbp]
    \centering
    \begin{subfigure}[b]{0.48\textwidth}
        \centering
        \includegraphics[width=\textwidth]{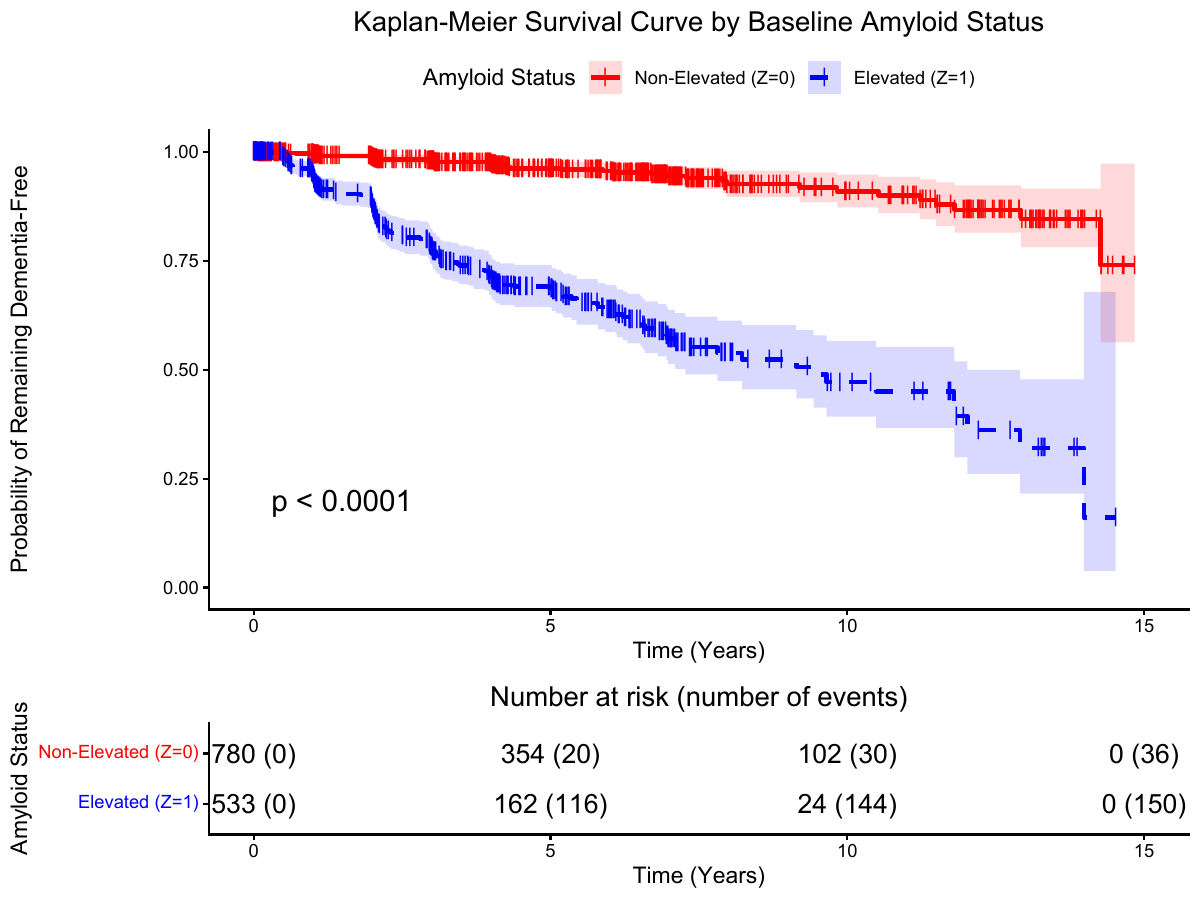}
        \caption{\label{figureEDPqrl_ADNIexploratory_km} Unadjusted Kaplan--Meier estimates of dementia-free survival by baseline amyloid status, with numbers at risk and cumulative dementia events.}
        \label{fig:adni_km}
    \end{subfigure}
    \hfill
    \begin{subfigure}[b]{0.48\textwidth}
        \centering
        \includegraphics[width=\textwidth]{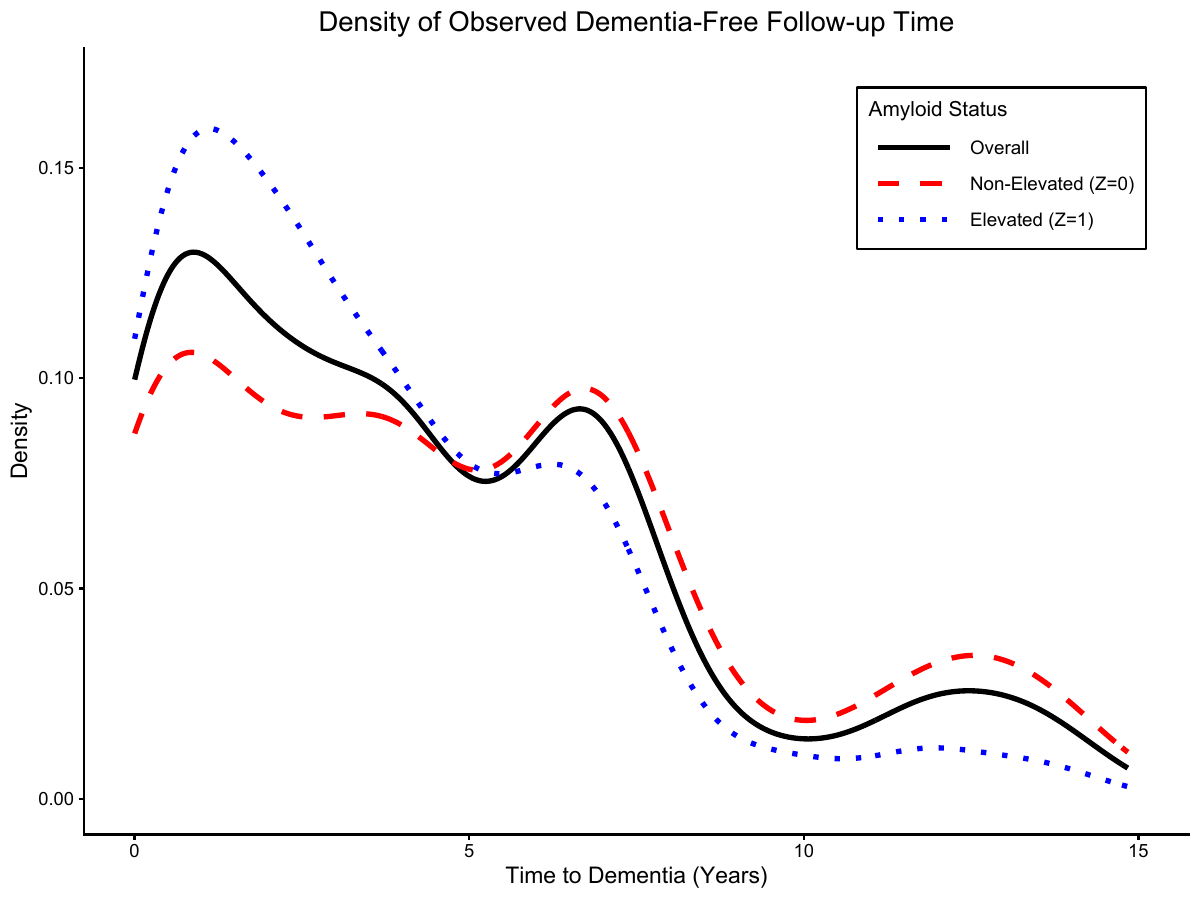}
        \caption{\label{figureEDPqrl_ADNIexploratory_density} Kernel density estimates of observed dementia-free follow-up time, overall and by baseline amyloid status.}
        \label{fig:adni_density}
    \end{subfigure}
    \caption{\label{figureEDPqrl_ADNIexploratory} Observed follow-up in the ADNI cohort.}
\end{figure}
    Figure~\ref{figureEDPqrl_ADNIexploratory} summarizes the observed follow-up data. The Kaplan--Meier curves in Figure~\ref{figureEDPqrl_ADNIexploratory_km} show substantial unadjusted separation in dementia-free survival by baseline amyloid status. Participants with elevated amyloid experience a more rapid decline in dementia-free survival than those with non-elevated amyloid, with the separation emerging early and persisting throughout follow-up. However, the observed separation represents a crude association rather than a causal contrast. The amyloid groups may differ systematically in demographic, genetic, and clinical characteristics associated with disease progression. These potential differences motivate adjustment for measured confounding in the ADNI analysis.

    By 5 years, 116 dementia events had occurred among participants with elevated amyloid, compared with 20 among those with non-elevated amyloid. Over the same period, the numbers at risk decreased from 533 to 162 and from 780 to 354, respectively. This substantial reduction reflects the accumulation of both dementia events and censoring and limits the information available about the tails of the dementia-free survival distributions. This loss of information is particularly consequential when residual life is evaluated later in follow-up, because fewer participants remain available to provide information about subsequent event times. These features motivate methods that account for right censoring and sensitivity analyses for potential departures from conditional non-informative censoring.
    
    Figure~\ref{figureEDPqrl_ADNIexploratory_density} displays kernel density estimates of observed dementia-free follow-up times for the overall cohort and by baseline amyloid status. The distributions are asymmetric and differ appreciably across amyloid groups, with irregular features suggestive of complex follow-up patterns. Because observed follow-up time is the minimum of the dementia-onset and censoring times, these densities should not be interpreted as distributions of latent dementia-onset times. Rather, they reflect the combined contributions of disease progression, censoring, and differences in the composition of the amyloid groups. Nevertheless, the observed complexity makes a single homogeneous parametric specification difficult to justify and motivates flexible covariate-dependent modeling.
    
\subsection{Existing Approaches} \label{EDPMqrl.introduction.existing}
    For individuals who have remained dementia-free for a given period, prognosis concerns not only whether dementia will eventually occur but also how much dementia-free time remains. Residual-life summaries address this question by characterizing the remaining time to dementia onset among individuals who remain dementia-free beyond a specified time point $\nu$, often referred to as a landmark time \citep{dafni2011landmark, yu2026learning}. In particular, the quantile residual life (QRL) function is the $\rho$-quantile of the remaining time to the event conditional on remaining event-free beyond $\nu$ \citep{jeong2008nonparametric, jung2009regression, kim2012censored, jeong2014statistical, lin2019quantile}. Compared with mean residual-life summaries, the QRL is less sensitive to extreme event times and directly describes the additional time by which a specified proportion of individuals who remain event-free experience the event.

    Although QRL provides a clinically interpretable prognostic summary, extending it to causal comparisons in observational studies introduces additional challenges. Baseline amyloid status is not randomized, requiring adjustment for confounding to compare potential residual-life distributions. Right censoring further limits information about the event-time distribution, particularly among individuals who remain dementia-free beyond later values of $\nu$. Beyond confounding and censoring, conditioning on remaining dementia-free beyond $\nu$ may induce selection because the individuals who would remain dementia-free through that time can differ across amyloid statuses. Consequently, a direct comparison of exposure-specific survivor populations may contrast groups with different underlying compositions rather than the same target population.
    
    To formalize causal comparisons in the presence of exposure-specific survivor populations, \citet{hong2026causal} introduced two causal QRL estimands: the observable survivor quantile contrast (OSQC) and the principal-stratum quantile contrast (PSQC). The OSQC compares residual-life distributions among those who would remain dementia-free beyond $\nu$ under each amyloid status separately and is therefore connected to the survivor population under each amyloid status. However, these populations need not contain the same individuals. The PSQC instead targets the time-specific always-survivor (TAS) principal stratum, comprising individuals who would remain dementia-free beyond $\nu$ under both elevated and non-elevated baseline amyloid statuses. It compares quantiles of the two potential residual-life distributions within this common population and is therefore more closely aligned with our scientific objective. As the TAS principal stratum depends jointly on the two potential event times, identification of the PSQC requires assumptions about their unobserved joint distribution \citep{frangakis2002principal, vanderweele2011principal}. 

    Bayesian nonparametric (BNP) models offer a natural strategy for estimating the complex covariate-conditional distributions required for PSQC inference. These models can accommodate nonlinear associations, distributional heterogeneity, and complex distributional shapes. Although Dirichlet process mixture models (DPMMs) have previously been used for QRL comparisons in randomized settings \citep{park2012bayesian}, that work did not address causal inference with observational exposures or target a TAS principal stratum. In addition, conventional DPMMs may perform poorly in conditional estimation when high-dimensional covariates dominate the induced clustering and obscure outcome-relevant structure \citep{roy2018bayesian, bae2024bayesian}. This limitation highlights the need for a joint model that retains flexible distributional estimation while reducing the influence of high-dimensional covariates on outcome-relevant clustering.
    
\subsection{Our Approach} \label{EDPMqrl.introduction.ourapproach}
    To address these inferential and modeling challenges, we propose a Bayesian causal framework for estimating the PSQC from observational right-censored data. We model the joint distribution of the event time, exposure, and baseline covariates using an enriched Dirichlet process mixture model (EDPMM), which flexibly accommodates heterogeneous event-time distributions while mitigating the covariate-domination problem of a conventional DPMM. The joint formulation supports posterior inference through Bayesian g-computation without requiring separate estimation of multiple nuisance models. Censored event times and partially observed baseline covariates are handled through data augmentation, with missing covariates modeled under a within-subcluster missing-at-random (MAR) assumption. A single posterior fit yields PSQC estimates across multiple follow-up times and quantile levels. Because the identifying assumptions cannot be verified from the observed data, we propose sensitivity analyses for conditional exchangeability, conditional cross-world independence, and conditional non-informative censoring. We evaluate the finite-sample performance of the proposed framework through simulations and apply it to the ADNI data.

\subsection{Article Outline} \label{EDPMqrl.introduction.outline}
    The remainder of the article is organized as follows. Section~\ref{EDPMqrl.causal} introduces the causal estimands and identification assumptions. Sections~\ref{EDPMqrl.edpm} and~\ref{EDPMqrl.posterior} present the EDPMM and posterior inference via Bayesian g-computation. Section~\ref{EDPMqrl.simulation} evaluates the finite-sample performance of the proposed approach. Section~\ref{EDPMqrl.sensitivity} develops sensitivity analyses for departures from the identifying assumptions. Section~\ref{EDPMqrl.application} presents the ADNI application, and Section~\ref{EDPMqrl.discussion} concludes with limitations and directions for future research.

\section{Causal Estimand and Identification Assumptions} \label{EDPMqrl.causal}
    In this section, we define the causal estimand of interest, introduce the potential-outcomes framework, and state the assumptions required for identification from the observed data. For $N$ independent individuals, let $Z_{i}$ denote baseline amyloid status for subject $i=1,\ldots,N$, with $Z_{i}=1$ for elevated amyloid and $Z_{i}=0$ otherwise. Let $\bbX_{i}$ denote the baseline covariate vector, and let $Y_{i}$ and $C_{i}$ denote the event and censoring times. We observe $T_{i}=\min\{Y_{i},C_{i}\}$ and $D_{i}=I(Y_{i}\le C_{i})$, together with $Z_{i}$ and the observed components of $\bbX_{i}$. Missing components of $\bbX_{i}$ are handled as described in Section~\ref{EDPMqrl.supp.mcmc.missing}.

\subsection{Potential Outcomes and Causal Estimands} \label{EDPMqrl.causal.potential}
    Let $Y^{z}$ and $C^{z}$ denote the potential event and censoring times under baseline amyloid status $z\in\{0,1\}$, respectively. Although amyloid status is an exposure rather than a randomized treatment, potential-outcomes notation provides a formal basis for adjusted contrasts under the assumptions below. For each time point $\nu$, the TAS principal stratum consists of individuals satisfying $Y^{0}>\nu$ and $Y^{1}>\nu$. The PSQC is defined as
    \begin{align} \label{EDPMqrl.eq.causal.potential.1}
        \Delta(\nu,\rho) = R_{\nu,\rho}^{1} - R_{\nu,\rho}^{0},
    \end{align}
    where $R_{\nu,\rho}^{z}$ is the $\rho$-quantile of $Y^{z}-\nu$ within the TAS principal stratum. In the ADNI application, it represents the contrast in quantiles of remaining time to dementia onset under elevated and non-elevated amyloid among individuals who would remain dementia-free beyond $\nu$ under both amyloid statuses.

    For example, $R_{2,0.25}^{1}=2.5$ means that, within the TAS principal stratum defined at year two, the 25th percentile of residual lifetime under elevated amyloid is 2.5 years. Equivalently, within this stratum, 25\% would experience dementia onset within 2.5 years after year two under elevated amyloid, whereas 75\% would remain dementia-free beyond that point. If $R_{2,0.25}^{0}=1.5$, then $\Delta(2,0.25)=1$ year, indicating that the 25th percentile of residual lifetime is one year longer under elevated than under non-elevated amyloid. Under heavy censoring, the mean residual lifetime may be difficult to estimate reliably, whereas a hazard ratio does not directly quantify differences in time to dementia onset. In contrast, $\Delta(\nu,\rho)$ is expressed directly in years and therefore has a clear interpretation on the prognostic time scale.

    Formally, for time point $\nu \geq 0$, define the residual-life survival function under amyloid status $z$ within the TAS principal stratum by
    \begin{align} \label{EDPMqrl.eq.causal.potential.2}
        \mS[y][\nu][z]
        &= \mP[Y^{z} > y+\nu \mid Y^{0}>\nu, Y^{1}>\nu] \notag \\
        &= \frac{\mP[Y^{z}>y+\nu,\;Y^{1-z}>\nu]}{\mP[Y^{0}>\nu,\;Y^{1}>\nu]},
        \qquad y \ge 0.
    \end{align}
    The corresponding $\rho$-quantile of residual lifetime is
    \begin{align} \label{EDPMqrl.eq.causal.potential.3}
        R_{\nu,\rho}^{z}
        =
        \inf\{ r \ge 0 : \mS[r][\nu][z] \le 1-\rho \},
        \qquad \rho \in (0,1).
    \end{align}

    In addition to the marginal PSQC, we consider subgroup-specific causal contrasts defined by a prespecified subset of baseline covariates. Let $\bbX' \subseteq \bbX$ denote such a subset and $\bbx'$ be a fixed value in its support. For example, in the ADNI application, $\bbX'$ may represent baseline diagnostic status yielding subgroup-specific contrasts for CN and MCI participants. We define the subgroup-specific residual-life survival function under baseline amyloid status $z$ as
    \begin{align*}
        \mS[y \mid \bbx'][\nu][z]
        =
        \mP[Y^{z} > y+\nu \mid Y^{0}>\nu,\;Y^{1}>\nu,\;\bbX'=\bbx'],
        \qquad y \ge 0,
    \end{align*}
    with corresponding subgroup-specific QRL
    \begin{align*}
        R_{\nu,\rho}^{z}(\bbx')
        =
        \inf\{ r \ge 0 : \mS[r \mid \bbx'][\nu][z] \le 1-\rho \},
        \qquad \rho \in (0,1).
    \end{align*}
    The subgroup-specific PSQC is then defined as
    \begin{align*}
        \Delta(\nu,\rho;\bbx')
        =
        R_{\nu,\rho}^{1}(\bbx') - R_{\nu,\rho}^{0}(\bbx').
    \end{align*}
    This estimand compares the two residual-life quantiles within the TAS principal stratum at time point $\nu$, restricted to the subgroup $\bbX'=\bbx'$. It reduces to the marginal PSQC when no subgroup variables are specified. For notational simplicity, the identification and estimation arguments below are presented for the marginal PSQC; the same arguments apply directly to subgroup-specific PSQCs defined by any prespecified subset of baseline covariates. 
    
\subsection{Identification Assumptions} \label{EDPMqrl.causal.identification}
    We now state five assumptions under which the PSQC is identified from the observed right censored data.
    \begin{itemize}
        \item[\eqref{EDPMqrl.causal.identification.A1}] Consistency \\
        For each individual and $z \in \{0,1\}$, if $Z=z$, then $Y=Y^{z}$ and $C=C^{z}$; that is,
        \begin{align} \label{EDPMqrl.causal.identification.A1} \tag{A1}
            Y = ZY^{1} + (1-Z)Y^{0}
            \quad \text{and} \quad
            C = ZC^{1} + (1-Z)C^{0}.
        \end{align}
        This assumption links the observed event and censoring times to the corresponding potential outcomes under the baseline amyloid classification recorded in ADNI \citep{cole2009consistency}. We also assume no interference, so that one participant's baseline amyloid status does not affect another participant's dementia onset or censoring time.

        \item[\eqref{EDPMqrl.causal.identification.A2}] Conditional exchangeability \\
        For $z \in \{0,1\}$,
        \begin{align} \label{EDPMqrl.causal.identification.A2} \tag{A2}
            Y^{z} \indep Z \mid \bbX.
        \end{align}
        This assumption identifies each marginal potential event-time distribution after adjustment for $\bbX$ \citep{rosenbaum1983central}. In ADNI, the adjustment set includes age, sex, years of education, baseline diagnosis, and APOE $\varepsilon$4 carrier status, which capture several major demographic, clinical, and genetic factors associated with both amyloid status and subsequent disease progression \citep{ossenkoppele2015prevalence}. These variables were selected a priori as measured baseline predictors of amyloid status and dementia progression. Adjustment for these variables makes conditional exchangeability more plausible, but the assumption remains unverifiable and may be violated by unmeasured or inadequately measured disease characteristics. We therefore assess sensitivity to residual unmeasured confounding in Section~\ref{EDPMqrl.sensitivity.unmeasured}.

        \item[\eqref{EDPMqrl.causal.identification.A3}] Conditional cross-world independence \\
        For all $\bbx$ in the support of $\bbX$, the potential event times under the two baseline amyloid statuses are assumed to be conditionally independent given baseline covariates: 
        \begin{align} \label{EDPMqrl.causal.identification.A3} \tag{A3}
            Y^{0} \indep Y^{1} \mid \bbX=\bbx.
        \end{align}
        Equivalently, for arbitrary thresholds $y_{0}$ and $y_{1}$,
        \begin{align*}
            \mP[Y^{0}>y_{0},Y^{1}>y_{1} \mid \bbX=\bbx]
            =
            \mP[Y^{0}>y_{0} \mid \bbX=\bbx]
            \mP[Y^{1}>y_{1} \mid \bbX=\bbx].
        \end{align*}
        This factorization identifies the conditional joint survivor probabilities required to define the TAS residual-life distributions from the two marginal potential event-time distributions. Because the two potential event times are never observed jointly, their cross-world dependence is not identified from the observed data. We therefore use conditional cross-world independence as a benchmark and examine departures through a Gaussian copula sensitivity analysis \citep{nelsen2007introduction, daniels2012bayesian}; see Section~\ref{EDPMqrl.sensitivity.cross}.

        
        \item[\eqref{EDPMqrl.causal.identification.A4}] Conditional non-informative censoring \\
        The event and censoring times are conditionally independent given baseline amyloid status and baseline covariates:
        \begin{align} \label{EDPMqrl.causal.identification.A4} \tag{A4}
            Y \indep C \mid Z, \bbX.
        \end{align}
        Under this assumption, censoring may depend on amyloid status and measured baseline characteristics but is conditionally independent of the latent event time given these variables \citep{robins2000correcting}. In ADNI, adjustment for demographic, diagnostic, genetic, and clinical characteristics associated with disease severity and study participation helps account for observed predictors of attrition \citep{burke2019factors}. Nevertheless, informative dropout related to unmeasured disease progression cannot be ruled out, particularly given the long follow-up and heavy censoring. We therefore examine departures from this assumption in Section~\ref{EDPMqrl.sensitivity.informative}.
        
        \item[\eqref{EDPMqrl.causal.identification.A5}] Positivity \\
        For all $\bbx$ in the support of $\bbX$ and $z \in \{0,1\}$ and $\nu \ge 0$,
        \begin{gather} 
            0 < \mP[Z = z \mid \bbX = \bbx] < 1, \notag \\
            \mP[T > \nu \mid Z = z, \bbX = \bbx] > 0, \label{EDPMqrl.causal.identification.A5} \tag{A5} \\
            \mP[Y^{0} > \nu, Y^{1} > \nu \mid \bbX = \bbx] > 0. \notag
        \end{gather}
        The first condition requires overlap in baseline amyloid status across covariate strata. The second requires sufficient observed follow-up beyond time point $\nu$ within each amyloid group, and the third ensures that the TAS principal stratum is nonempty within the covariate strata over which the estimand is averaged \citep{westreich2010positivity}. In ADNI, both amyloid groups are represented across the principal baseline subgroups, and participants remain under observation beyond each time point considered in the primary analysis. However, the empirical risk sets become smaller as $\nu$ increases, especially within diagnostic subgroups, making positivity and finite-sample information more fragile at later time points. We therefore restrict the primary analysis to $\nu \in \{0, 2.5, 5\}$.
        
    \end{itemize}

    \begin{figure}[!tbp]
    \centering
    \begin{tikzpicture}[
        scale=1.2,
        transform shape,
        >=stealth,
        node distance=1.8cm and 1.8cm,
        thick,
        state/.style={circle, draw, minimum size=0.9cm, align=center, inner sep=0pt},
        every edge/.style={draw, ->, shorten >=1pt}
    ]
        \node[state] (X) {$\bbX$};
        \node[state, right=of X] (Z) {$Z$};
        \node[state, above right=0.8cm and 1.2cm of Z] (Y) {$Y$};
        \node[state, below right=0.8cm and 1.2cm of Z] (C) {$C$};
        
        \draw[->, dashed] (X) -- (Z);
        \draw[->, dashed] (X) edge[bend left=15] (Y);
        \draw[->, dashed] (X) edge[bend right=15] (C);
        
        \draw[->] (Z) -- (Y);
        \draw[->] (Z) -- (C);
    \end{tikzpicture}
    \caption{\label{figureEDPqrl_DAG} Causal directed acyclic graph relating baseline covariates $\bbX$, baseline exposure $Z$, event time $Y$, and censoring time $C$. Dashed arrows denote associations of measured baseline covariates with $Z$, $Y$, and $C$; solid arrows denote relationships of amyloid status with the event and censoring times. }
\end{figure}
    Figure~\ref{figureEDPqrl_DAG} summarizes the structural relationships relevant to Assumptions~\eqref{EDPMqrl.causal.identification.A1}--\eqref{EDPMqrl.causal.identification.A5}. A formal identification proof is provided in Section~\ref{EDPMqrl.supp.identification}. Under these assumptions, for any $\nu \ge 0$, $\rho \in (0,1)$, and $z \in \{0,1\}$, the residual-life survival function within the TAS principal stratum is identified by
    \begin{align} \label{EDPMqrl.eq.causal.potential.4}
        \mS[y][\nu][z]
        =
        \frac{\mE[\mP[Y > y + \nu \mid Z = z, \bbX]\mP[Y > \nu \mid Z = 1-z, \bbX]][\bbX]
        }{\mE[\mP[Y > \nu \mid Z = 0, \bbX]\mP[Y > \nu \mid Z = 1, \bbX]][\bbX]},
        \qquad y \ge 0.
    \end{align}
    The corresponding exposure-specific QRL, $R_{\nu,\rho}^{z}$, is obtained as the value of $y$ satisfying
    \begin{align} \label{EDPMqrl.eq.causal.potential.5}
        \mS[y][\nu][z] = 1 - \rho.
    \end{align}
    Because a closed-form inverse is generally unavailable, we solve \eqref{EDPMqrl.eq.causal.potential.5} numerically. When the residual-life survival function is continuous, this numerical solution is equivalent to the definition in \eqref{EDPMqrl.eq.causal.potential.3}. The PSQC is then obtained as the difference between the resulting exposure-specific QRLs.
    
\section{Enriched Dirichlet Process Mixture Model} \label{EDPMqrl.edpm}
    Estimation of the PSQC requires flexible characterization of the exposure-specific event-time distributions conditional on baseline covariates. In particular, computing the QRLs through~\eqref{EDPMqrl.eq.causal.potential.4} requires the conditional survival function $\mP[Y>y\mid Z=z,\bbX]$. We estimate this function by modeling the joint distribution of $(Y,Z,\bbX)$ within a BNP framework. The joint model accommodates complex associations among the event time, amyloid status, and baseline covariates while supporting posterior imputation of missing covariate values.
    
    Under right censoring, the event time $Y_{i}$ is not always observed. Instead, we observe $T_{i}=\min\{Y_{i},C_{i}\}$ together with the event indicator $D_{i}=I(Y_{i}\le C_{i})$. For censored individuals with $D_{i}=0$, the event time is known only to satisfy $Y_{i}>T_{i}$. We therefore treat these unobserved event times as latent variables and update them subject to the observed censoring constraints; see Section~\ref{EDPMqrl.supp.mcmc.data}. This formulation connects the complete-data model for $(Y,Z,\bbX)$ to the observed right censored data $(T_{i},D_{i},Z_{i},\bbX_{i})$ through Bayesian data augmentation. We specify an EDPMM for the complete-data distribution of $(Y,Z,\bbX)$ \citep{wade2011enriched, wade2014improving}: 
    \begin{align} \label{EDPMqrl.eq.edpm.1}
        \begin{split}
            Y_{i} \mid Z_{i}, \bbX_{i}; \bbtheta_{i} & \sim f \left( y_{i} \mid z_{i}, \bbx_{i}; \bbtheta_{i} \right) \\
            Z_{i} ; \omega_{i}^{z} & \sim f \left( z_{i}; \omega_{i}^{z} \right) \\
            X_{i,q}; \omega_{i,q}^{\bbx} & \sim f \left( x_{i,q}; \omega_{i,q}^{\bbx} \right), \;\; q=1, \ldots , p_{\bbX}  \\
             \left( \bbtheta_{i} , \bbomega_{i} \right) \mid \nG & \sim \nG \\
            \nG & \sim \nedp \left( \alpha^{\bbtheta}, \alpha^{\bbomega}, \nG_{0} \right)
        \end{split}
    \end{align}
    where $\bbomega_{i} = \left( \omega_{i}^{z}, \bbomega_{i}^{\bbx} \right)$. The notation $\nG \sim \nedp \left( \alpha^{\bbtheta}, \alpha^{\bbomega}, \nG_{0} \right)$ means that $\nG^{\bbtheta} \sim \ndp \left( \alpha^{\bbtheta}, \nG_{0}^{\bbtheta} \right)$ and $\nG^{\bbomega \mid \bbtheta} \sim \ndp \left( \alpha^{\bbomega}, \nG_{0}^{\bbomega \mid \bbtheta} \right)$ with base measure $\nG_{0} = \nG_{0}^{\bbtheta} \times \nG_{0}^{\bbomega \mid \bbtheta}$. 

    The EDPMM in Equation~\ref{EDPMqrl.eq.edpm.1} has a square-breaking representation \citep{wade2011enriched},
    \begin{gather*}
        \mP[ y, z, \bbx \mid \nG ] = \sum_{k=1}^{\infty} \gamma_{k} f \left( y \mid z, \bbx; \bbtheta_{k} \right) \sum_{h=1}^{\infty} \gamma_{h \mid k} f \left( z; \omega_{h \mid k}^{z} \right) f \left( \bbx; \bbomega_{h \mid k}^{x} \right), 
    \end{gather*}
    where $k$ indexes the main outcome clusters, $h \mid k$ indexes the nested exposure--covariate subclusters and $f(\cdot)$ denotes the corresponding component distributions. The weights have priors $\gamma_{k}' \sim \mbeta{1, \alpha^{\bbtheta}} $ and $\gamma_{h \mid k}' \sim \mbeta{1, \alpha^{\bbomega}} $ where $\gamma_{k} = \gamma_{k}' \prod_{ l < k } \left( 1 - \gamma_{l}' \right)$ and $\gamma_{h \mid k} = \gamma_{h \mid k}' \prod_{d < h } \left( 1 - \gamma_{d \mid k}' \right)$. 

    Under the joint specification, the conditional and marginal distributions used in the posterior g-computation algorithm in Section~\ref{EDPMqrl.posterior} are given by:
    \begin{gather} \label{EDPMqrl.eq.edpm.2}
        \begin{split}
            \mP[ y \mid z, \bbx ] & = \sum_{k=1}^{ \infty} \Lambda_{k} \left( z, \bbx \right) f \left( y \mid z, \bbx; \bbtheta_{k} \right), \\
            \mP[ z, \bbx ] = \sum_{k=1}^{ \infty} \gamma_{k} & \sum_{h=1}^{\infty} \gamma_{h \mid k} f \left( z; \omega_{h \mid k}^{z} \right) \prod_{q=1}^{p_{\bbX}} f \left( x_{q}; \omega_{h \mid k, q}^{x} \right), \\
            \mP[ \bbx ] = \sum_{k=1}^{ \infty} & \gamma_{k} \sum_{h=1}^{\infty} \gamma_{h \mid k} \prod_{q=1}^{p_{\bbX}} f \left( x_{q}; \omega_{h \mid k, q}^{x} \right),
        \end{split}
    \end{gather}
    where
    \begin{gather*}
        \Lambda_{k} \left( z, \bbx \right) 
        = \frac{ \gamma_{k} \sum_{h=1}^{\infty} \gamma_{h \mid k} f \left( z; \omega_{h \mid k}^{z} \right) f \left( \bbx; \bbomega_{h \mid k}^{x} \right)}{\sum_{d=1}^{\infty} \gamma_{d} \sum_{h=1}^{\infty} \gamma_{h \mid d} f \left( z; \omega_{h \mid d}^{z} \right) f \left( \bbx; \bbomega_{h \mid d}^{x} \right)} 
        = \frac{\lambda_{k} \left( z, \bbx \right)}{\sum_{d=1}^{\infty} \lambda_{d} \left( z, \bbx \right)},
        \\
        \Lambda_{h \mid k} \left( z, \bbx \right) 
        = \frac{\gamma_{h \mid k} f \left( z; \omega_{h \mid k}^{z} \right) f \left( \bbx; \bbomega_{h \mid k}^{x} \right)}{ \sum_{s=1}^{\infty} \gamma_{s \mid k} f \left( z; \omega_{s \mid k}^{z} \right)  f \left( \bbx; \bbomega_{s \mid k}^{x} \right)} 
        = \frac{\lambda_{h \mid k} \left( z, \bbx \right)}{\sum_{s=1}^{\infty} \lambda_{s \mid k} \left( z, \bbx \right)}.
    \end{gather*}

    The corresponding conditional survival function can then be written as
    \begin{gather} \label{EDPMqrl.eq.edpm.3}
        \mS[y \mid z, \bbx]
        = \int \mS[y \mid z, \bbx ; \bbtheta] \, \nG \left( d \bbtheta | z, \bbx \right)
        = \sum_{k=1}^{\infty} \Lambda_{k} \left( z, \bbx \right) \mS[y \mid z, \bbx ; \bbtheta_{k}],
    \end{gather}
    where $\nG \left( d \bbtheta | z, \bbx \right)$ denotes the conditional mixing distribution of $\bbtheta$ given $(z, \bbx)$ induced by the joint EDPMM.

    For any time point $\nu$ and quantile level $\rho$, the conditional survival function in~\eqref{EDPMqrl.eq.edpm.3} supports Bayesian g-computation of $R_{\nu,\rho}^{z}$, $\Delta(\nu,\rho)$, and subgroup-specific contrasts $\Delta(\nu,\rho;\bbx')$ for prespecified subsets $\bbX' \subset \bbX$. 

    The EDPMM is particularly advantageous when the covariate distribution is high dimensional or heterogeneous. In a standard DPMM, clustering may be driven primarily by the covariates, producing highly fragmented outcome clusters and degrading estimation of the conditional outcome distribution \citep{roy2018bayesian}. The EDPMM mitigates this problem by separating primary outcome clusters from nested exposure--covariate subclusters. This hierarchical construction induces covariate-dependent mixture weights $\Lambda_k(z,\bbx)$, allowing relatively simple local kernels to represent complex global dependence among $Y$, $Z$, and $\bbX$. The resulting conditional survival distributions can then be used directly in the identifying functional for the PSQC.

    This joint modeling strategy may also improve finite-sample stability when covariate overlap is limited; see Assumption~\eqref{EDPMqrl.causal.identification.A5}. Unlike inverse-probability weighting (IPW), which can be sensitive to extreme weights in sparse regions, the EDPMM regularizes conditional distribution estimation by borrowing information across related mixture components. This model-based regularization can produce smoother conditional survival estimates without requiring ad hoc weight trimming. It cannot, however, recover information in covariate regions with no exposure or follow-up support and therefore does not eliminate the need for the underlying positivity assumption.

\section{Posterior Computation} \label{EDPMqrl.posterior}
    We obtain posterior draws using a Markov chain Monte Carlo (MCMC) algorithm that extends \citet[Algorithm~8]{neal2000markov} to the EDPMM \citep{wade2014improving}. For a censored individual, $Y_{i}$ is treated as latent and sampled subject to $Y_{i}>T_{i}$. This data-augmentation step restores conjugate updating of the local outcome-kernel parameters and avoids cluster-specific Metropolis--Hastings updates for those parameters. Missing baseline covariates are updated from their full conditional distributions under a within-subcluster MAR assumption. Conditional on the current latent subcluster, its parameters, and the observed quantities, the missingness indicator is assumed not to depend additionally on the missing covariate values. This local assumption need not imply marginal MAR after integrating over the latent partition. Posterior uncertainty in the missing covariates and latent subcluster assignments is propagated through the MCMC algorithm. Full details of the MCMC algorithm are provided in Section~\ref{EDPMqrl.supp.mcmc} of the supplementary materials, including data augmentation for censoring in Section~\ref{EDPMqrl.supp.mcmc.data}, imputation of missing covariates in Section~\ref{EDPMqrl.supp.mcmc.missing}, and updates of cluster assignments and model parameters in Section~\ref{EDPMqrl.supp.mcmc.params}.

    Posterior inference for the PSQC is conducted through Bayesian g-computation. Each retained MCMC draw induces exposure-specific conditional survival functions, which are integrated over the posterior-predictive covariate distribution using a synthetic cohort of size $M$. The resulting survival probabilities are combined according to the identification formula in Section~\ref{EDPMqrl.causal.identification}, and numerical root finding yields the exposure-specific residual-life quantiles within the TAS principal stratum. The predictive-weight derivations and complete post-processing algorithm are provided in Section~\ref{EDPMqrl.supp.poststep}. This post-processing strategy offers an important computational advantage: a single posterior fit yields posterior summaries and $(1-\alpha)$ credible intervals (CrIs) across multiple combinations of time points and quantile levels, $(\nu,\rho)$, as well as subgroup-specific PSQCs and other functionals of the survival distribution, without refitting the primary model.

\section{Simulation Study} \label{EDPMqrl.simulation}
    We compare the finite-sample performance of the EDPMM with that of a standard joint DPMM under dependent covariates, confounded exposure assignment, heterogeneous event-time distributions, and covariate-dependent censoring. Full data-generating and computational details are provided in Section~\ref{EDPMqrl.supp.simulation}.

\subsection{Simulation Setup} \label{EDPMqrl.simulation.setup}
    We generate datasets with primary sample sizes $N \in \{500, 1500\}$. Each dataset contains a five-dimensional baseline covariate vector $\bbX_{i}$ comprising two binary and three continuous variables generated under a dependent structure. Exposure $Z_{i} \in \{0, 1\}$ is assigned conditionally on the covariates, thereby inducing confounding. Conditional on $(Z_{i}, \bbX_{i})$, the log-event time follows a two-component mixture consisting of Student's $t$ and normal regression models with component-specific mean functions. This construction produces nonlinear marginal features, heavy tails, multimodality, and heterogeneous exposure effects, reflecting the distributional complexity that motivates flexible modeling in the ADNI analysis. Censoring is covariate dependent but conditionally independent of the event time with parameters selected to yield approximately 20\%, 40\%, 60\%, or 80\% censoring. Complete data-generating mechanisms are provided in Section~\ref{EDPMqrl.supp.simulation.scenarios}.

    We fit the EDPMM using a normal regression kernel for the log-event time, a Bernoulli kernel for exposure, and Bernoulli and normal kernels for the binary and continuous covariates, respectively. The same EDPMM specification is used in the simulation study and the ADNI application. The DPMM comparator uses the same local kernels and base measure but places a single DP prior on the complete-data joint distribution. This comparison therefore isolates the contribution of the enriched clustering structure. Model, prior, and computational details are given in Sections~\ref{EDPMqrl.supp.simulation.specifications}, \ref{EDPMqrl.supp.simulation.prior}, and~\ref{EDPMqrl.supp.simulation.implementation}.

    For each primary sample size and censoring setting, we generate 1000 datasets. Performance is evaluated using empirical bias, root mean squared error (RMSE), and coverage of nominal 95\% CrIs at time points $\nu \in \{0,1,2,3\}$. Under 20\% and 40\% censoring, we consider quantile levels $\rho \in \{0.3,0.6\}$. Under 60\% and 80\% censoring, we instead consider $\rho \in \{0.1,0.2\}$ focusing on portions of the event-time distribution with greater observed support. Additional analyses reported in Section~\ref{EDPMqrl.supp.simulation.additional} examine performance under 60\% and 80\% censoring at the larger sample size $N=5000$ using nominal 95\% CrIs and evaluate nominal 99\% CrIs for $N=500$ and $N=1500$.

\subsection{Simulation Results} \label{EDPMqrl.simulation.results}
    \begin{table}[!tbp]
\centering
\caption{\label{tableEDPqrl_sim_main}
Simulation performance of the EDPMM and DPMM for estimating the marginal PSQC at landmark time $\nu$ and quantile level $\rho$. Results are reported for $N \in \{500,1500\}$ and approximately 20\%, 40\%, 60\%, and 80\% right censoring. For each setting, the table gives the true PSQC, empirical bias, RMSE, and empirical CP of the nominal 95\% credible interval across 1000 simulated datasets.}
\centering
\resizebox{\ifdim\width>\linewidth\linewidth\else\width\fi}{!}{
\fontsize{7}{9}\selectfont
\begin{tabular}[t]{lcc>{}c>{}c>{}c>{}c>{}c>{}c>{}c>{}c>{}c>{}c>{}c>{}c>{}c}
\toprule
\multicolumn{4}{c}{ } & \multicolumn{6}{c}{20\%} & \multicolumn{6}{c}{40\%} \\
\cmidrule(l{3pt}r{3pt}){5-10} \cmidrule(l{3pt}r{3pt}){11-16}
\multicolumn{4}{c}{ } & \multicolumn{3}{c}{N = 500} & \multicolumn{3}{c}{N = 1500} & \multicolumn{3}{c}{N = 500} & \multicolumn{3}{c}{N = 1500} \\
\cmidrule(l{3pt}r{3pt}){5-7} \cmidrule(l{3pt}r{3pt}){8-10} \cmidrule(l{3pt}r{3pt}){11-13} \cmidrule(l{3pt}r{3pt}){14-16}
Model & $\nu$ & $\rho$ & True & Bias & RMSE & CP & Bias & RMSE & CP & Bias & RMSE & CP & Bias & RMSE & CP\\
\midrule
EDPMM & 0 & 0.3 & \cellcolor{white}{0.53} & \cellcolor[HTML]{f2f2f2}{0.01} & \cellcolor[HTML]{f2f2f2}{0.33} & \cellcolor[HTML]{f2f2f2}{93.4} & \cellcolor[HTML]{d9d9d9}{0.00} & \cellcolor[HTML]{d9d9d9}{0.19} & \cellcolor[HTML]{d9d9d9}{96.2} & \cellcolor[HTML]{f2f2f2}{0.00} & \cellcolor[HTML]{f2f2f2}{0.34} & \cellcolor[HTML]{f2f2f2}{93.8} & \cellcolor[HTML]{d9d9d9}{0.01} & \cellcolor[HTML]{d9d9d9}{0.21} & \cellcolor[HTML]{d9d9d9}{95.9}\\
 &  & 0.6 & \cellcolor{white}{3.22} & \cellcolor[HTML]{f2f2f2}{0.04} & \cellcolor[HTML]{f2f2f2}{0.58} & \cellcolor[HTML]{f2f2f2}{94.3} & \cellcolor[HTML]{d9d9d9}{-0.02} & \cellcolor[HTML]{d9d9d9}{0.35} & \cellcolor[HTML]{d9d9d9}{96.2} & \cellcolor[HTML]{f2f2f2}{0.04} & \cellcolor[HTML]{f2f2f2}{0.68} & \cellcolor[HTML]{f2f2f2}{92.7} & \cellcolor[HTML]{d9d9d9}{0.01} & \cellcolor[HTML]{d9d9d9}{0.41} & \cellcolor[HTML]{d9d9d9}{94.3}\\
 & 1 & 0.3 & \cellcolor{white}{1.26} & \cellcolor[HTML]{f2f2f2}{0.08} & \cellcolor[HTML]{f2f2f2}{0.40} & \cellcolor[HTML]{f2f2f2}{94.7} & \cellcolor[HTML]{d9d9d9}{0.03} & \cellcolor[HTML]{d9d9d9}{0.25} & \cellcolor[HTML]{d9d9d9}{97.2} & \cellcolor[HTML]{f2f2f2}{0.10} & \cellcolor[HTML]{f2f2f2}{0.47} & \cellcolor[HTML]{f2f2f2}{92.1} & \cellcolor[HTML]{d9d9d9}{0.06} & \cellcolor[HTML]{d9d9d9}{0.30} & \cellcolor[HTML]{d9d9d9}{94.7}\\
 &  & 0.6 & \cellcolor{white}{4.24} & \cellcolor[HTML]{f2f2f2}{0.01} & \cellcolor[HTML]{f2f2f2}{0.69} & \cellcolor[HTML]{f2f2f2}{93.2} & \cellcolor[HTML]{d9d9d9}{0.00} & \cellcolor[HTML]{d9d9d9}{0.44} & \cellcolor[HTML]{d9d9d9}{95.7} & \cellcolor[HTML]{f2f2f2}{0.04} & \cellcolor[HTML]{f2f2f2}{0.86} & \cellcolor[HTML]{f2f2f2}{92.3} & \cellcolor[HTML]{d9d9d9}{0.03} & \cellcolor[HTML]{d9d9d9}{0.54} & \cellcolor[HTML]{d9d9d9}{94.1}\\
 & 2 & 0.3 & \cellcolor{white}{2.09} & \cellcolor[HTML]{f2f2f2}{0.11} & \cellcolor[HTML]{f2f2f2}{0.39} & \cellcolor[HTML]{f2f2f2}{93.6} & \cellcolor[HTML]{d9d9d9}{0.03} & \cellcolor[HTML]{d9d9d9}{0.24} & \cellcolor[HTML]{d9d9d9}{95.9} & \cellcolor[HTML]{f2f2f2}{0.14} & \cellcolor[HTML]{f2f2f2}{0.51} & \cellcolor[HTML]{f2f2f2}{90.6} & \cellcolor[HTML]{d9d9d9}{0.06} & \cellcolor[HTML]{d9d9d9}{0.31} & \cellcolor[HTML]{d9d9d9}{93.6}\\
 &  & 0.6 & \cellcolor{white}{4.94} & \cellcolor[HTML]{f2f2f2}{0.04} & \cellcolor[HTML]{f2f2f2}{0.73} & \cellcolor[HTML]{f2f2f2}{92.9} & \cellcolor[HTML]{d9d9d9}{0.01} & \cellcolor[HTML]{d9d9d9}{0.46} & \cellcolor[HTML]{d9d9d9}{95.6} & \cellcolor[HTML]{f2f2f2}{0.09} & \cellcolor[HTML]{f2f2f2}{0.94} & \cellcolor[HTML]{f2f2f2}{91.5} & \cellcolor[HTML]{d9d9d9}{0.05} & \cellcolor[HTML]{d9d9d9}{0.57} & \cellcolor[HTML]{d9d9d9}{93.6}\\
 & 3 & 0.3 & \cellcolor{white}{2.65} & \cellcolor[HTML]{f2f2f2}{0.11} & \cellcolor[HTML]{f2f2f2}{0.42} & \cellcolor[HTML]{f2f2f2}{93.5} & \cellcolor[HTML]{d9d9d9}{0.03} & \cellcolor[HTML]{d9d9d9}{0.25} & \cellcolor[HTML]{d9d9d9}{96.3} & \cellcolor[HTML]{f2f2f2}{0.15} & \cellcolor[HTML]{f2f2f2}{0.55} & \cellcolor[HTML]{f2f2f2}{89.9} & \cellcolor[HTML]{d9d9d9}{0.05} & \cellcolor[HTML]{d9d9d9}{0.32} & \cellcolor[HTML]{d9d9d9}{93.8}\\
 &  & 0.6 & \cellcolor{white}{5.60} & \cellcolor[HTML]{f2f2f2}{0.04} & \cellcolor[HTML]{f2f2f2}{0.78} & \cellcolor[HTML]{f2f2f2}{92.9} & \cellcolor[HTML]{d9d9d9}{0.02} & \cellcolor[HTML]{d9d9d9}{0.47} & \cellcolor[HTML]{d9d9d9}{95.4} & \cellcolor[HTML]{f2f2f2}{0.11} & \cellcolor[HTML]{f2f2f2}{1.01} & \cellcolor[HTML]{f2f2f2}{91.7} & \cellcolor[HTML]{d9d9d9}{0.06} & \cellcolor[HTML]{d9d9d9}{0.59} & \cellcolor[HTML]{d9d9d9}{94.7}\\

\cmidrule(l{3pt}r{3pt}){1-16}

DPMM & 0 & 0.3 & \cellcolor{white}{0.53} & \cellcolor[HTML]{f2f2f2}{0.20} & \cellcolor[HTML]{f2f2f2}{0.28} & \cellcolor[HTML]{f2f2f2}{81.3} & \cellcolor[HTML]{d9d9d9}{0.19} & \cellcolor[HTML]{d9d9d9}{0.22} & \cellcolor[HTML]{d9d9d9}{61.9} & \cellcolor[HTML]{f2f2f2}{0.14} & \cellcolor[HTML]{f2f2f2}{0.26} & \cellcolor[HTML]{f2f2f2}{90.2} & \cellcolor[HTML]{d9d9d9}{0.14} & \cellcolor[HTML]{d9d9d9}{0.19} & \cellcolor[HTML]{d9d9d9}{79.5}\\
 &  & 0.6 & \cellcolor{white}{3.22} & \cellcolor[HTML]{f2f2f2}{-1.19} & \cellcolor[HTML]{f2f2f2}{1.31} & \cellcolor[HTML]{f2f2f2}{42.7} & \cellcolor[HTML]{d9d9d9}{-1.22} & \cellcolor[HTML]{d9d9d9}{1.26} & \cellcolor[HTML]{d9d9d9}{4.0} & \cellcolor[HTML]{f2f2f2}{-1.28} & \cellcolor[HTML]{f2f2f2}{1.43} & \cellcolor[HTML]{f2f2f2}{48.3} & \cellcolor[HTML]{d9d9d9}{-1.27} & \cellcolor[HTML]{d9d9d9}{1.32} & \cellcolor[HTML]{d9d9d9}{6.2}\\
 & 1 & 0.3 & \cellcolor{white}{1.26} & \cellcolor[HTML]{f2f2f2}{-0.50} & \cellcolor[HTML]{f2f2f2}{0.54} & \cellcolor[HTML]{f2f2f2}{36.1} & \cellcolor[HTML]{d9d9d9}{-0.51} & \cellcolor[HTML]{d9d9d9}{0.53} & \cellcolor[HTML]{d9d9d9}{1.2} & \cellcolor[HTML]{f2f2f2}{-0.54} & \cellcolor[HTML]{f2f2f2}{0.59} & \cellcolor[HTML]{f2f2f2}{37.8} & \cellcolor[HTML]{d9d9d9}{-0.54} & \cellcolor[HTML]{d9d9d9}{0.56} & \cellcolor[HTML]{d9d9d9}{2.3}\\
 &  & 0.6 & \cellcolor{white}{4.24} & \cellcolor[HTML]{f2f2f2}{-1.97} & \cellcolor[HTML]{f2f2f2}{2.06} & \cellcolor[HTML]{f2f2f2}{14.6} & \cellcolor[HTML]{d9d9d9}{-2.02} & \cellcolor[HTML]{d9d9d9}{2.04} & \cellcolor[HTML]{d9d9d9}{0.0} & \cellcolor[HTML]{f2f2f2}{-2.03} & \cellcolor[HTML]{f2f2f2}{2.16} & \cellcolor[HTML]{f2f2f2}{23.1} & \cellcolor[HTML]{d9d9d9}{-2.05} & \cellcolor[HTML]{d9d9d9}{2.09} & \cellcolor[HTML]{d9d9d9}{0.1}\\
 & 2 & 0.3 & \cellcolor{white}{2.09} & \cellcolor[HTML]{f2f2f2}{-1.27} & \cellcolor[HTML]{f2f2f2}{1.29} & \cellcolor[HTML]{f2f2f2}{0.0} & \cellcolor[HTML]{d9d9d9}{-1.29} & \cellcolor[HTML]{d9d9d9}{1.30} & \cellcolor[HTML]{d9d9d9}{0.0} & \cellcolor[HTML]{f2f2f2}{-1.31} & \cellcolor[HTML]{f2f2f2}{1.33} & \cellcolor[HTML]{f2f2f2}{0.6} & \cellcolor[HTML]{d9d9d9}{-1.31} & \cellcolor[HTML]{d9d9d9}{1.32} & \cellcolor[HTML]{d9d9d9}{0.0}\\
 &  & 0.6 & \cellcolor{white}{4.94} & \cellcolor[HTML]{f2f2f2}{-2.43} & \cellcolor[HTML]{f2f2f2}{2.53} & \cellcolor[HTML]{f2f2f2}{8.8} & \cellcolor[HTML]{d9d9d9}{-2.49} & \cellcolor[HTML]{d9d9d9}{2.52} & \cellcolor[HTML]{d9d9d9}{0.0} & \cellcolor[HTML]{f2f2f2}{-2.48} & \cellcolor[HTML]{f2f2f2}{2.61} & \cellcolor[HTML]{f2f2f2}{18.3} & \cellcolor[HTML]{d9d9d9}{-2.51} & \cellcolor[HTML]{d9d9d9}{2.55} & \cellcolor[HTML]{d9d9d9}{0.0}\\
 & 3 & 0.3 & \cellcolor{white}{2.65} & \cellcolor[HTML]{f2f2f2}{-1.77} & \cellcolor[HTML]{f2f2f2}{1.79} & \cellcolor[HTML]{f2f2f2}{0.0} & \cellcolor[HTML]{d9d9d9}{-1.79} & \cellcolor[HTML]{d9d9d9}{1.80} & \cellcolor[HTML]{d9d9d9}{0.0} & \cellcolor[HTML]{f2f2f2}{-1.80} & \cellcolor[HTML]{f2f2f2}{1.82} & \cellcolor[HTML]{f2f2f2}{0.0} & \cellcolor[HTML]{d9d9d9}{-1.81} & \cellcolor[HTML]{d9d9d9}{1.82} & \cellcolor[HTML]{d9d9d9}{0.0}\\
 &  & 0.6 & \cellcolor{white}{5.60} & \cellcolor[HTML]{f2f2f2}{-2.88} & \cellcolor[HTML]{f2f2f2}{2.97} & \cellcolor[HTML]{f2f2f2}{5.3} & \cellcolor[HTML]{d9d9d9}{-2.95} & \cellcolor[HTML]{d9d9d9}{2.98} & \cellcolor[HTML]{d9d9d9}{0.0} & \cellcolor[HTML]{f2f2f2}{-2.92} & \cellcolor[HTML]{f2f2f2}{3.05} & \cellcolor[HTML]{f2f2f2}{14.2} & \cellcolor[HTML]{d9d9d9}{-2.96} & \cellcolor[HTML]{d9d9d9}{3.00} & \cellcolor[HTML]{d9d9d9}{0.0}\\

\midrule
\multicolumn{4}{c}{} & \multicolumn{6}{c}{60\%} & \multicolumn{6}{c}{80\%} \\

\midrule

EDPMM & 0 & 0.1 & \cellcolor{white}{0.10} & \cellcolor[HTML]{f2f2f2}{0.00} & \cellcolor[HTML]{f2f2f2}{0.07} & \cellcolor[HTML]{f2f2f2}{94.3} & \cellcolor[HTML]{d9d9d9}{0.00} & \cellcolor[HTML]{d9d9d9}{0.05} & \cellcolor[HTML]{d9d9d9}{94.2} & \cellcolor[HTML]{f2f2f2}{0.00} & \cellcolor[HTML]{f2f2f2}{0.09} & \cellcolor[HTML]{f2f2f2}{92.6} & \cellcolor[HTML]{d9d9d9}{0.00} & \cellcolor[HTML]{d9d9d9}{0.06} & \cellcolor[HTML]{d9d9d9}{91.5}\\
 &  & 0.2 & \cellcolor{white}{0.22} & \cellcolor[HTML]{f2f2f2}{0.00} & \cellcolor[HTML]{f2f2f2}{0.18} & \cellcolor[HTML]{f2f2f2}{93.8} & \cellcolor[HTML]{d9d9d9}{0.00} & \cellcolor[HTML]{d9d9d9}{0.11} & \cellcolor[HTML]{d9d9d9}{95.4} & \cellcolor[HTML]{f2f2f2}{0.02} & \cellcolor[HTML]{f2f2f2}{0.24} & \cellcolor[HTML]{f2f2f2}{91.1} & \cellcolor[HTML]{d9d9d9}{0.00} & \cellcolor[HTML]{d9d9d9}{0.14} & \cellcolor[HTML]{d9d9d9}{93.0}\\
 & 1 & 0.1 & \cellcolor{white}{0.17} & \cellcolor[HTML]{f2f2f2}{0.02} & \cellcolor[HTML]{f2f2f2}{0.13} & \cellcolor[HTML]{f2f2f2}{94.5} & \cellcolor[HTML]{d9d9d9}{0.02} & \cellcolor[HTML]{d9d9d9}{0.08} & \cellcolor[HTML]{d9d9d9}{94.7} & \cellcolor[HTML]{f2f2f2}{0.06} & \cellcolor[HTML]{f2f2f2}{0.21} & \cellcolor[HTML]{f2f2f2}{91.1} & \cellcolor[HTML]{d9d9d9}{0.02} & \cellcolor[HTML]{d9d9d9}{0.12} & \cellcolor[HTML]{d9d9d9}{90.8}\\
 &  & 0.2 & \cellcolor{white}{0.55} & \cellcolor[HTML]{f2f2f2}{0.10} & \cellcolor[HTML]{f2f2f2}{0.36} & \cellcolor[HTML]{f2f2f2}{93.4} & \cellcolor[HTML]{d9d9d9}{0.05} & \cellcolor[HTML]{d9d9d9}{0.21} & \cellcolor[HTML]{d9d9d9}{94.5} & \cellcolor[HTML]{f2f2f2}{0.16} & \cellcolor[HTML]{f2f2f2}{0.55} & \cellcolor[HTML]{f2f2f2}{88.5} & \cellcolor[HTML]{d9d9d9}{0.07} & \cellcolor[HTML]{d9d9d9}{0.32} & \cellcolor[HTML]{d9d9d9}{90.4}\\
 & 2 & 0.1 & \cellcolor{white}{0.54} & \cellcolor[HTML]{f2f2f2}{0.10} & \cellcolor[HTML]{f2f2f2}{0.25} & \cellcolor[HTML]{f2f2f2}{92.2} & \cellcolor[HTML]{d9d9d9}{0.04} & \cellcolor[HTML]{d9d9d9}{0.14} & \cellcolor[HTML]{d9d9d9}{93.5} & \cellcolor[HTML]{f2f2f2}{0.14} & \cellcolor[HTML]{f2f2f2}{0.42} & \cellcolor[HTML]{f2f2f2}{87.0} & \cellcolor[HTML]{d9d9d9}{0.06} & \cellcolor[HTML]{d9d9d9}{0.23} & \cellcolor[HTML]{d9d9d9}{90.3}\\
 &  & 0.2 & \cellcolor{white}{1.28} & \cellcolor[HTML]{f2f2f2}{0.15} & \cellcolor[HTML]{f2f2f2}{0.47} & \cellcolor[HTML]{f2f2f2}{92.3} & \cellcolor[HTML]{d9d9d9}{0.06} & \cellcolor[HTML]{d9d9d9}{0.28} & \cellcolor[HTML]{d9d9d9}{93.6} & \cellcolor[HTML]{f2f2f2}{0.17} & \cellcolor[HTML]{f2f2f2}{0.76} & \cellcolor[HTML]{f2f2f2}{87.5} & \cellcolor[HTML]{d9d9d9}{0.07} & \cellcolor[HTML]{d9d9d9}{0.44} & \cellcolor[HTML]{d9d9d9}{90.0}\\
 & 3 & 0.1 & \cellcolor{white}{0.87} & \cellcolor[HTML]{f2f2f2}{0.15} & \cellcolor[HTML]{f2f2f2}{0.34} & \cellcolor[HTML]{f2f2f2}{90.8} & \cellcolor[HTML]{d9d9d9}{0.05} & \cellcolor[HTML]{d9d9d9}{0.18} & \cellcolor[HTML]{d9d9d9}{93.4} & \cellcolor[HTML]{f2f2f2}{0.19} & \cellcolor[HTML]{f2f2f2}{0.57} & \cellcolor[HTML]{f2f2f2}{87.4} & \cellcolor[HTML]{d9d9d9}{0.08} & \cellcolor[HTML]{d9d9d9}{0.30} & \cellcolor[HTML]{d9d9d9}{90.1}\\
 &  & 0.2 & \cellcolor{white}{1.77} & \cellcolor[HTML]{f2f2f2}{0.18} & \cellcolor[HTML]{f2f2f2}{0.54} & \cellcolor[HTML]{f2f2f2}{90.9} & \cellcolor[HTML]{d9d9d9}{0.06} & \cellcolor[HTML]{d9d9d9}{0.31} & \cellcolor[HTML]{d9d9d9}{92.8} & \cellcolor[HTML]{f2f2f2}{0.21} & \cellcolor[HTML]{f2f2f2}{0.94} & \cellcolor[HTML]{f2f2f2}{88.0} & \cellcolor[HTML]{d9d9d9}{0.09} & \cellcolor[HTML]{d9d9d9}{0.51} & \cellcolor[HTML]{d9d9d9}{90.1}\\

\cmidrule(l{3pt}r{3pt}){1-16}

DPMM & 0 & 0.1 & \cellcolor{white}{0.10} & \cellcolor[HTML]{f2f2f2}{0.12} & \cellcolor[HTML]{f2f2f2}{0.15} & \cellcolor[HTML]{f2f2f2}{74.0} & \cellcolor[HTML]{d9d9d9}{0.11} & \cellcolor[HTML]{d9d9d9}{0.12} & \cellcolor[HTML]{d9d9d9}{40.1} & \cellcolor[HTML]{f2f2f2}{0.08} & \cellcolor[HTML]{f2f2f2}{0.15} & \cellcolor[HTML]{f2f2f2}{87.0} & \cellcolor[HTML]{d9d9d9}{0.08} & \cellcolor[HTML]{d9d9d9}{0.10} & \cellcolor[HTML]{d9d9d9}{75.4}\\
 &  & 0.2 & \cellcolor{white}{0.22} & \cellcolor[HTML]{f2f2f2}{0.19} & \cellcolor[HTML]{f2f2f2}{0.25} & \cellcolor[HTML]{f2f2f2}{79.2} & \cellcolor[HTML]{d9d9d9}{0.18} & \cellcolor[HTML]{d9d9d9}{0.21} & \cellcolor[HTML]{d9d9d9}{49.8} & \cellcolor[HTML]{f2f2f2}{0.13} & \cellcolor[HTML]{f2f2f2}{0.26} & \cellcolor[HTML]{f2f2f2}{89.2} & \cellcolor[HTML]{d9d9d9}{0.12} & \cellcolor[HTML]{d9d9d9}{0.18} & \cellcolor[HTML]{d9d9d9}{81.6}\\
 & 1 & 0.1 & \cellcolor{white}{0.17} & \cellcolor[HTML]{f2f2f2}{0.02} & \cellcolor[HTML]{f2f2f2}{0.09} & \cellcolor[HTML]{f2f2f2}{92.6} & \cellcolor[HTML]{d9d9d9}{0.02} & \cellcolor[HTML]{d9d9d9}{0.05} & \cellcolor[HTML]{d9d9d9}{92.6} & \cellcolor[HTML]{f2f2f2}{0.01} & \cellcolor[HTML]{f2f2f2}{0.12} & \cellcolor[HTML]{f2f2f2}{93.7} & \cellcolor[HTML]{d9d9d9}{0.00} & \cellcolor[HTML]{d9d9d9}{0.06} & \cellcolor[HTML]{d9d9d9}{95.0}\\
 &  & 0.2 & \cellcolor{white}{0.55} & \cellcolor[HTML]{f2f2f2}{-0.13} & \cellcolor[HTML]{f2f2f2}{0.22} & \cellcolor[HTML]{f2f2f2}{86.0} & \cellcolor[HTML]{d9d9d9}{-0.15} & \cellcolor[HTML]{d9d9d9}{0.17} & \cellcolor[HTML]{d9d9d9}{66.2} & \cellcolor[HTML]{f2f2f2}{-0.17} & \cellcolor[HTML]{f2f2f2}{0.31} & \cellcolor[HTML]{f2f2f2}{85.5} & \cellcolor[HTML]{d9d9d9}{-0.19} & \cellcolor[HTML]{d9d9d9}{0.23} & \cellcolor[HTML]{d9d9d9}{66.1}\\
 & 2 & 0.1 & \cellcolor{white}{0.54} & \cellcolor[HTML]{f2f2f2}{-0.33} & \cellcolor[HTML]{f2f2f2}{0.34} & \cellcolor[HTML]{f2f2f2}{6.3} & \cellcolor[HTML]{d9d9d9}{-0.34} & \cellcolor[HTML]{d9d9d9}{0.34} & \cellcolor[HTML]{d9d9d9}{0.0} & \cellcolor[HTML]{f2f2f2}{-0.35} & \cellcolor[HTML]{f2f2f2}{0.37} & \cellcolor[HTML]{f2f2f2}{24.1} & \cellcolor[HTML]{d9d9d9}{-0.36} & \cellcolor[HTML]{d9d9d9}{0.37} & \cellcolor[HTML]{d9d9d9}{0.1}\\
 &  & 0.2 & \cellcolor{white}{1.28} & \cellcolor[HTML]{f2f2f2}{-0.84} & \cellcolor[HTML]{f2f2f2}{0.86} & \cellcolor[HTML]{f2f2f2}{2.4} & \cellcolor[HTML]{d9d9d9}{-0.85} & \cellcolor[HTML]{d9d9d9}{0.86} & \cellcolor[HTML]{d9d9d9}{0.0} & \cellcolor[HTML]{f2f2f2}{-0.87} & \cellcolor[HTML]{f2f2f2}{0.91} & \cellcolor[HTML]{f2f2f2}{16.1} & \cellcolor[HTML]{d9d9d9}{-0.89} & \cellcolor[HTML]{d9d9d9}{0.90} & \cellcolor[HTML]{d9d9d9}{0.0}\\
 & 3 & 0.1 & \cellcolor{white}{0.87} & \cellcolor[HTML]{f2f2f2}{-0.65} & \cellcolor[HTML]{f2f2f2}{0.66} & \cellcolor[HTML]{f2f2f2}{0.0} & \cellcolor[HTML]{d9d9d9}{-0.66} & \cellcolor[HTML]{d9d9d9}{0.66} & \cellcolor[HTML]{d9d9d9}{0.0} & \cellcolor[HTML]{f2f2f2}{-0.66} & \cellcolor[HTML]{f2f2f2}{0.68} & \cellcolor[HTML]{f2f2f2}{2.4} & \cellcolor[HTML]{d9d9d9}{-0.68} & \cellcolor[HTML]{d9d9d9}{0.68} & \cellcolor[HTML]{d9d9d9}{0.0}\\
 &  & 0.2 & \cellcolor{white}{1.77} & \cellcolor[HTML]{f2f2f2}{-1.29} & \cellcolor[HTML]{f2f2f2}{1.31} & \cellcolor[HTML]{f2f2f2}{0.1} & \cellcolor[HTML]{d9d9d9}{-1.31} & \cellcolor[HTML]{d9d9d9}{1.31} & \cellcolor[HTML]{d9d9d9}{0.0} & \cellcolor[HTML]{f2f2f2}{-1.32} & \cellcolor[HTML]{f2f2f2}{1.36} & \cellcolor[HTML]{f2f2f2}{4.4} & \cellcolor[HTML]{d9d9d9}{-1.35} & \cellcolor[HTML]{d9d9d9}{1.36} & \cellcolor[HTML]{d9d9d9}{0.0}\\
\bottomrule
\end{tabular}}
\end{table}
    Table~\ref{tableEDPqrl_sim_main} summarizes performance across the four censoring settings and two primary sample sizes. Overall, the EDPMM estimates the target PSQCs more accurately than the DPMM, with smaller bias and RMSE and empirical coverage generally closer to the nominal 95\% level. Under 20\% and 40\% censoring, the EDPMM performs well at both $\rho=0.3$ and $\rho=0.6$. Some positive bias and undercoverage arise for $N=500$ at later time points, particularly under 40\% censoring, but both are substantially reduced when the sample size increases to $N=1500$. In contrast, the DPMM performs poorly at $\rho=0.6$ and later time points, with increasingly negative bias, larger RMSE, and severe undercoverage.
    
    Under 60\% and 80\% censoring, the finite-sample limitations of the EDPMM become more evident. Increasing the sample size from $N=500$ to $N=1500$ reduces both bias and RMSE, with absolute bias below $0.1$ across the reported settings. Some undercoverage nevertheless remains under 80\% censoring, particularly at later time points. This pattern is consistent with the limited observed failure-time information available within the TAS principal stratum. For example, under 80\% censoring at $\nu=3$, an observed failure time is available for only 6.9\% of individuals in the TAS principal stratum, corresponding to 2.5\% of the full simulated population; see Supplementary Table~\ref{tableEDPqrl_SIM_pop_supp}. Despite this limited information, the EDPMM remains more accurate and stable than the DPMM, especially at later time points and for $\rho=0.2$.
    
    The supplementary analyses further clarify these patterns. Increasing the sample size to $N=5000$ reduces EDPMM bias and RMSE and brings empirical coverage closer to the nominal 95\% level. For $N \in \{500,1500\}$, using 99\% CrIs also improves coverage under heavy censoring. The DPMM continues to exhibit larger bias, RMSE, and undercoverage at $N=5000$ and when wider CrIs are used, indicating that its inferior performance is not explained solely by limited sample size or interval width. Detailed results are reported in Tables~\ref{tableEDPqrl_sim34_suppN5000} and~\ref{tableEDPqrl_sim34_supp}.
    
    Because the 95\% CrIs exhibit undercoverage in the most heavily censored settings, we report 99\% CrIs in the ADNI application. The supplementary simulations indicate that these wider intervals improve frequentist coverage for the EDPMM under limited event-time information without altering the qualitative conclusions.

\section{Sensitivity Analysis} \label{EDPMqrl.sensitivity}
    We assess the robustness of the estimated PSQCs to departures from conditional exchangeability, conditional cross-world independence, and conditional non-informative censoring.
    
\subsection{Sensitivity Analysis for Unmeasured Confounding} \label{EDPMqrl.sensitivity.unmeasured}
    Although the baseline covariates available in ADNI help support Assumption~\eqref{EDPMqrl.causal.identification.A2} (conditional exchangeability), this assumption is fundamentally untestable from the observed data. Unmeasured confounding therefore remains an inherent concern in observational analyses of baseline amyloid status. To assess the robustness of the estimated contrasts in time to dementia onset, we conduct a sensitivity analysis for residual unmeasured confounding. A practical advantage of the proposed framework is that this analysis can be implemented entirely through post-processing of the posterior draws, without refitting the MCMC sampler. 

    We adopt a shift-bias model on the log-time scale implied by the local log-normal AFT kernel. Related shift-based sensitivity models have been used to assess departures from identifying assumptions in the missing-data literature \citep{wang2011note, linero2015flexible}. Throughout this subsection, $Y^{z}$ denotes 
    potential  
    log-event times. Let $\psi_{z} \in \mathbb{R}$ denote a sensitivity parameter for amyloid group $z$, interpreted as the magnitude of residual bias induced by unmeasured confounding in the conditional distribution of the potential outcome $Y^{z}$. Under Assumption~\eqref{EDPMqrl.causal.identification.A2}, the observed conditional distribution among individuals with $Z=z$ is identified with the corresponding counterfactual distribution. When conditional exchangeability fails, however, these two distributions may differ systematically because of omitted factors associated with both baseline amyloid status and dementia risk. We represent this discrepancy through a constant shift on the log-time scale:
    \begin{align*}
        \mP[ Y^{z} > y \mid \bbX ]
        &= \mP[ Y^{z} > y - \psi_{z} \mid Z=z, \bbX ] \quad \text{(sensitivity shift model)} \\
        &= \mP[ Y > y - \psi_{z} \mid Z=z, \bbX ] \quad \text{(by Assumption~\eqref{EDPMqrl.causal.identification.A1})},
    \end{align*}
    where the second equality follows from Assumption~\eqref{EDPMqrl.causal.identification.A1}. When $\psi_{z}=0$, this reduces to the standard identifying relation under conditional exchangeability. Values $\psi_{z}>0$ correspond to unmeasured factors that decelerate disease progression, thereby delaying dementia onset in group $z$, whereas $\psi_{z}<0$ corresponds to unmeasured factors that accelerate dementia onset. Thus, $\psi_{z}$ is not interpreted as a specific omitted covariate, but rather as the aggregate directional bias induced by residual confounding within amyloid group $z$. 

    Because the local AFT kernel is specified on the log-time scale, the shift $\psi_{z}$ induces a multiplicative acceleration on the original time scale. Equivalently, for any original time $t=\exp(y)$,
    \begin{align*}
        \mP[ \exp(Y^{z}) > t \mid \bbX ]
        =
        \mP[ \exp(Y) > t\,\exp(-\psi_{z}) \mid Z=z, \bbX ].
    \end{align*}
    Accordingly, the sensitivity analysis is implemented in the g-computation post-processing step by evaluating the conditional survival function at the rescaled original-time arguments $t\exp(-\psi_{z})$. Because the estimand is defined at time point $\nu$, the same scaling is applied to the two arguments of the residual-life survival function corresponding to $y+\nu$ and $\nu$. Varying $(\psi_{0},\psi_{1})$ over prespecified settings yields the corresponding PSQC estimates and allows us to assess whether residual unmeasured confounding could materially attenuate or strengthen the estimated contrast. In the ADNI application, we consider the primary specification $(\psi_{0},\psi_{1})=(0,0)$ and the two directional alternatives $(\psi_{0},\psi_{1})=(-0.25,0.25)$ and $(\psi_{0},\psi_{1})=(0.25,-0.25)$. These values correspond to multiplicative acceleration factors $\exp(\psi_{z}) \in \{0.78, 1, 1.28\}$ on the original time scale. Because the sensitivity parameter acts multiplicatively on the original time scale, its effect is not a fixed additive shift in years. Specifically, $\exp(\psi_{z}) = 0.78$ and $\exp(\psi_{z}) = 1.28$ correspond to approximately 22\% shorter and 28\% longer event times, respectively, relative to the distribution identified under conditional exchangeability. For illustration, at an event time of $5$ years, these factors correspond to approximately $3.9$ and $6.4$ years on the original time scale.

\subsection{Sensitivity Analysis for Cross-World Dependence} \label{EDPMqrl.sensitivity.cross}
    The primary analysis relies on the conditional cross-world independence assumption in Assumption~\eqref{EDPMqrl.causal.identification.A3}. Because the joint distribution of $(Y^{0}, Y^{1})$ is not identified from the observed data, we assess the sensitivity of the PSQC to departures from this benchmark using the Gaussian copula sensitivity model introduced in Section~\ref{EDPMqrl.supp.identification}. This analysis is easily implemented, again, as a posterior post-processing step and does not require refitting the EDPMM.

    The sensitivity parameter $\kappa$ indexes the residual cross-world dependence between $Y^{0}$ and $Y^{1}$ after conditioning on $\bbX$. The benchmark value $\kappa = 0$ corresponds to conditional cross-world independence. Positive values of $\kappa$ represent positive residual dependence between the two potential event times, whereas allowing $\kappa$ to vary over a prespecified range represents uncertainty about the unidentified cross-world dependence structure.
    
    For each posterior draw, we combine the posterior draws of the marginal exposure-specific event-time distributions with the Gaussian copula-induced joint survivor function $S^{0,1} (\cdot, \cdot \mid \bbX ; \kappa)$ defined in Section~\ref{EDPMqrl.supp.identification}. We then compute the sensitivity versions of the residual-life survival functions within the TAS principal stratum in \eqref{EDPMqrl.eq.supp.identification.2}, solve for $R_{\nu, \rho}^{z}(\kappa)$, and summarize the corresponding posterior distribution of $\Delta (\nu, \rho ; \kappa)$.
    
    The primary analysis uses the identifying benchmark $\kappa = 0$. As sensitivity analyses, we additionally consider $\kappa \sim \munif{0, 1}$ and $\kappa \sim \munif{-1, 1}$. For each posterior draw under a sensitivity distribution, we draw a value of $\kappa$ during posterior post-processing and propagate it through the computation of $\Delta (\nu, \rho ; \kappa)$.

\subsection{Sensitivity Analysis for Informative censoring} \label{EDPMqrl.sensitivity.informative}
    The primary analysis relies on Assumption~\eqref{EDPMqrl.causal.identification.A4}, which requires censoring to be conditionally independent of the event time given baseline amyloid status and baseline covariates. Although this assumption is standard in survival analysis, it may be violated if dropout or loss to follow-up depends on unobserved disease progression. Related distributional sensitivity models have been used to assess departures from identifying assumptions in the missing-data literature \citep{wang2011note, linero2015flexible}. We therefore assess the sensitivity of the estimated PSQCs to alternative distributions for the latent event times of censored participants.

    This sensitivity analysis modifies the data-augmentation step for censored outcomes described in Section~\ref{EDPMqrl.supp.mcmc.data}. Under the primary analysis,
    \begin{align*}
        Y_{i} \mid Z_{i}=z, \bbX_{i}, D_{i}=0; \bbbeta_{i}, \sigma_{i}^{2}
        \sim
        \mnormal{\mathbb{X}_{i}\bbbeta_{i}, \sigma_{i}^{2}}
        \mbox{ truncated from below at } T_{i}.
    \end{align*}
    To represent departures from conditional non-informative censoring, we replace this distribution with
    \begin{align*}
        Y_{i} \mid Z_{i}=z, \bbX_{i}, D_{i}=0; \bbbeta_{i}, \sigma_{i}^{2}
        \sim
        \mnormal{ \mathbb{X}_{i} \bbbeta_{i}, \eta_{z} \sigma_{i}^{2} }
        \mbox{ truncated from below at } T_{i},
    \end{align*}
    where $\eta_{z} \in \mathbb{R}^{+}$ is an exposure-specific variance-scaling sensitivity parameter. The primary specification $(\eta_{0},\eta_{1})=(1,1)$ recovers the imputation distribution used under conditional non-informative censoring. Values of $\eta_{z}$ above or below $1$ increase or decrease the conditional dispersion of the latent event times for censored participants in amyloid group $z$. Because the distribution is truncated from below at $T_{i}$, changing its variance also changes the implied distribution of event times beyond the observed censoring time.

    In the ADNI application, we examine the primary specification $(\eta_{0},\eta_{1})=(1,1)$ and the alternatives $(\eta_{0},\eta_{1})=(0.9,1.1)$ and $(\eta_{0},\eta_{1})=(1.1,0.9)$. These alternatives allow the latent event-time distributions of censored participants to differ between the two amyloid groups. Because the sensitivity parameters modify the event-time imputation step within the MCMC algorithm, the model is refitted under each specification.

\section{ADNI Application} \label{EDPMqrl.application}
    We apply the proposed framework to estimate PSQCs comparing elevated with non-elevated baseline amyloid status for the overall cohort and the baseline MCI and CN subgroups, with residual life evaluated at follow-up times $\nu \in \{0, 2.5, 5\}$ and quantile levels $\rho \in \{0.1, 0.2, 0.3\}$. Sensitivity figures additionally include $\rho \in \{ 0.05, 0.15, 0.25 \}$. The main sensitivity figure focuses on MCI, where more observed dementia events provide greater direct support.

\subsection{ADNI Data and Analysis Setup} \label{EDPMqrl.application.setup}
    The adjustment set $\bbX$ includes age, sex, years of education, baseline diagnosis, and APOE $\varepsilon$4 carrier status. Missing baseline covariates are imputed within the EDPMM as described in Section~\ref{EDPMqrl.supp.mcmc.missing}. Cohort construction, missingness, and baseline characteristics are reported in Section~\ref{EDPMqrl.supp.application.cohort}.
    
    The final analytic sample includes $N=1313$ participants who were CN or had MCI at baseline and had available amyloid status and valid follow-up information. Of these, $533$ (40.59\%) had elevated amyloid. During follow-up, $186$ (14.17\%) experienced dementia onset and $1127$ (85.83\%) were right censored. Dementia onset occurred in $20$ of $688$ CN participants (2.91\%) and $166$ of $625$ MCI participants (26.56\%). Thus, the available event-time information varies substantially across diagnostic subgroups and decreases further after conditioning on remaining dementia-free beyond later time points. The EDPMM provides flexible conditional survival estimation while borrowing information across related covariate profiles, although the limited number of observed events remains an important source of uncertainty.
    
    We fit the EDPMM in \eqref{EDPMqrl.supp.simulation.specifications.edpmm.1} using the prior and MCMC specifications from the simulation study. Bayesian g-computation yields posterior means and 99\% CrIs for marginal and subgroup-specific PSQCs. We report 99\% CrIs because, in the supplementary simulations, they provided more conservative uncertainty summaries and reduced the coverage shortfall under the heavy-censoring settings most comparable to ADNI. Supporting estimates of the dementia-free survival functions under elevated and non-elevated baseline amyloid status, together with the corresponding Kaplan--Meier curves, are reported in Section~\ref{EDPMqrl.supp.application.survival}.

\subsection{Results} \label{EDPMqrl.application.results}
    \begin{table}[!tbp]
\centering
\caption{\label{tableEDPqrl_ADNIpsqc}
Posterior summaries of the PSQC under the primary identifying assumptions, $(\psi_{0},\psi_{1})=(0,0)$, $\kappa=0$, and $(\eta_{0},\eta_{1})=(1,1)$, for the overall ADNI cohort and the baseline MCI and CN subgroups. For each population and quantile level $\rho\in\{0.1,0.2,0.3\}$, the posterior mean and 99\% credible interval, defined by the 0.5\% and 99.5\% posterior quantiles (PQ), are reported at landmark times $\nu\in\{0,2.5,5\}$. Negative PSQC values indicate shorter remaining dementia-free time under elevated baseline amyloid status relative to non-elevated baseline amyloid status.}
\resizebox{\ifdim\width>\linewidth\linewidth\else\width\fi}{!}{
\fontsize{9}{8}\selectfont
\begin{tabular}[t]{>{}l>{}l>{}c>{}c>{}c>{}c>{}c>{}c>{}c>{}c>{}c}
\toprule
\multicolumn{2}{c}{ } & \multicolumn{3}{c}{$\nu=0$} & \multicolumn{3}{c}{$\nu=2.5$} & \multicolumn{3}{c}{$\nu=5$} \\
\cmidrule(l{3pt}r{3pt}){3-5} \cmidrule(l{3pt}r{3pt}){6-8} \cmidrule(l{3pt}r{3pt}){9-11}
Group & $\rho$ & Mean & 0.5\% PQ & 99.5\% PQ & Mean & 0.5\% PQ & 99.5\% PQ & Mean & 0.5\% PQ & 99.5\% PQ\\
\midrule
Overall & 0.1 & \cellcolor{gray!5}{-5.4} & \cellcolor{gray!5}{-8.4} & \cellcolor{gray!5}{-3.4} & \cellcolor{gray!10}{-5.5} & \cellcolor{gray!10}{-9.0} & \cellcolor{gray!10}{-3.3} & \cellcolor{gray!15}{-6.2} & \cellcolor{gray!15}{-10.5} & \cellcolor{gray!15}{-3.6}\\
 & 0.2 & \cellcolor{gray!5}{-10.5} & \cellcolor{gray!5}{-17.1} & \cellcolor{gray!5}{-6.4} & \cellcolor{gray!10}{-11.6} & \cellcolor{gray!10}{-19.9} & \cellcolor{gray!10}{-6.8} & \cellcolor{gray!15}{-13.8} & \cellcolor{gray!15}{-24.7} & \cellcolor{gray!15}{-7.7}\\
 & 0.3 & \cellcolor{gray!5}{-17.3} & \cellcolor{gray!5}{-29.8} & \cellcolor{gray!5}{-10.2} & \cellcolor{gray!10}{-19.9} & \cellcolor{gray!10}{-35.7} & \cellcolor{gray!10}{-10.9} & \cellcolor{gray!15}{-23.8} & \cellcolor{gray!15}{-45.7} & \cellcolor{gray!15}{-12.6}\\
MCI & 0.1 & \cellcolor{gray!5}{-3.1} & \cellcolor{gray!5}{-4.9} & \cellcolor{gray!5}{-1.9} & \cellcolor{gray!10}{-2.4} & \cellcolor{gray!10}{-4.0} & \cellcolor{gray!10}{-1.4} & \cellcolor{gray!15}{-2.1} & \cellcolor{gray!15}{-3.6} & \cellcolor{gray!15}{-1.2}\\
 & 0.2 & \cellcolor{gray!5}{-5.4} & \cellcolor{gray!5}{-8.6} & \cellcolor{gray!5}{-3.3} & \cellcolor{gray!10}{-4.8} & \cellcolor{gray!10}{-7.9} & \cellcolor{gray!10}{-2.9} & \cellcolor{gray!15}{-4.5} & \cellcolor{gray!15}{-7.4} & \cellcolor{gray!15}{-2.6}\\
 & 0.3 & \cellcolor{gray!5}{-8.1} & \cellcolor{gray!5}{-13.0} & \cellcolor{gray!5}{-5.0} & \cellcolor{gray!10}{-7.5} & \cellcolor{gray!10}{-12.3} & \cellcolor{gray!10}{-4.4} & \cellcolor{gray!15}{-7.3} & \cellcolor{gray!15}{-12.2} & \cellcolor{gray!15}{-4.2}\\
CN & 0.1 & \cellcolor{gray!5}{-18.5} & \cellcolor{gray!5}{-34.5} & \cellcolor{gray!5}{-9.5} & \cellcolor{gray!10}{-18.1} & \cellcolor{gray!10}{-34.9} & \cellcolor{gray!10}{-9.0} & \cellcolor{gray!15}{-17.1} & \cellcolor{gray!15}{-33.8} & \cellcolor{gray!15}{-8.3}\\
 & 0.2 & \cellcolor{gray!5}{-32.3} & \cellcolor{gray!5}{-64.4} & \cellcolor{gray!5}{-15.4} & \cellcolor{gray!10}{-31.9} & \cellcolor{gray!10}{-64.1} & \cellcolor{gray!10}{-15.1} & \cellcolor{gray!15}{-31.2} & \cellcolor{gray!15}{-63.5} & \cellcolor{gray!15}{-14.3}\\
 & 0.3 & \cellcolor{gray!5}{-48.7} & \cellcolor{gray!5}{-102.4} & \cellcolor{gray!5}{-23.1} & \cellcolor{gray!10}{-48.5} & \cellcolor{gray!10}{-102.1} & \cellcolor{gray!10}{-22.7} & \cellcolor{gray!15}{-47.9} & \cellcolor{gray!15}{-101.9} & \cellcolor{gray!15}{-21.8}\\
\bottomrule
\end{tabular}}
\end{table}
    Table~\ref{tableEDPqrl_ADNIpsqc} reports negative posterior mean PSQCs for the overall cohort and both subgroups at every primary $(\nu,\rho)$ combination, with all 99\% CrIs below zero. Thus, within the TAS principal stratum, the reported residual-life quantiles are shorter under elevated amyloid. The magnitude generally increases with $\rho$. At $\nu=2.5$ and $\rho=0.2$, the posterior mean PSQC is $-11.6$ years in the overall cohort, with a 99\% CrI of $(-19.9,-6.8)$. The corresponding values are $-4.8$ years in MCI, with a 99\% CrI of $(-7.9,-2.9)$, and $-31.9$ years in CN, with a 99\% CrI of $(-64.1,-15.1)$. The MCI result is supported by substantially more observed events; the larger CN estimate is less precise and more dependent on extrapolation.

    \begin{figure}[!tbp]
    \centering
    \includegraphics[width=\textwidth]{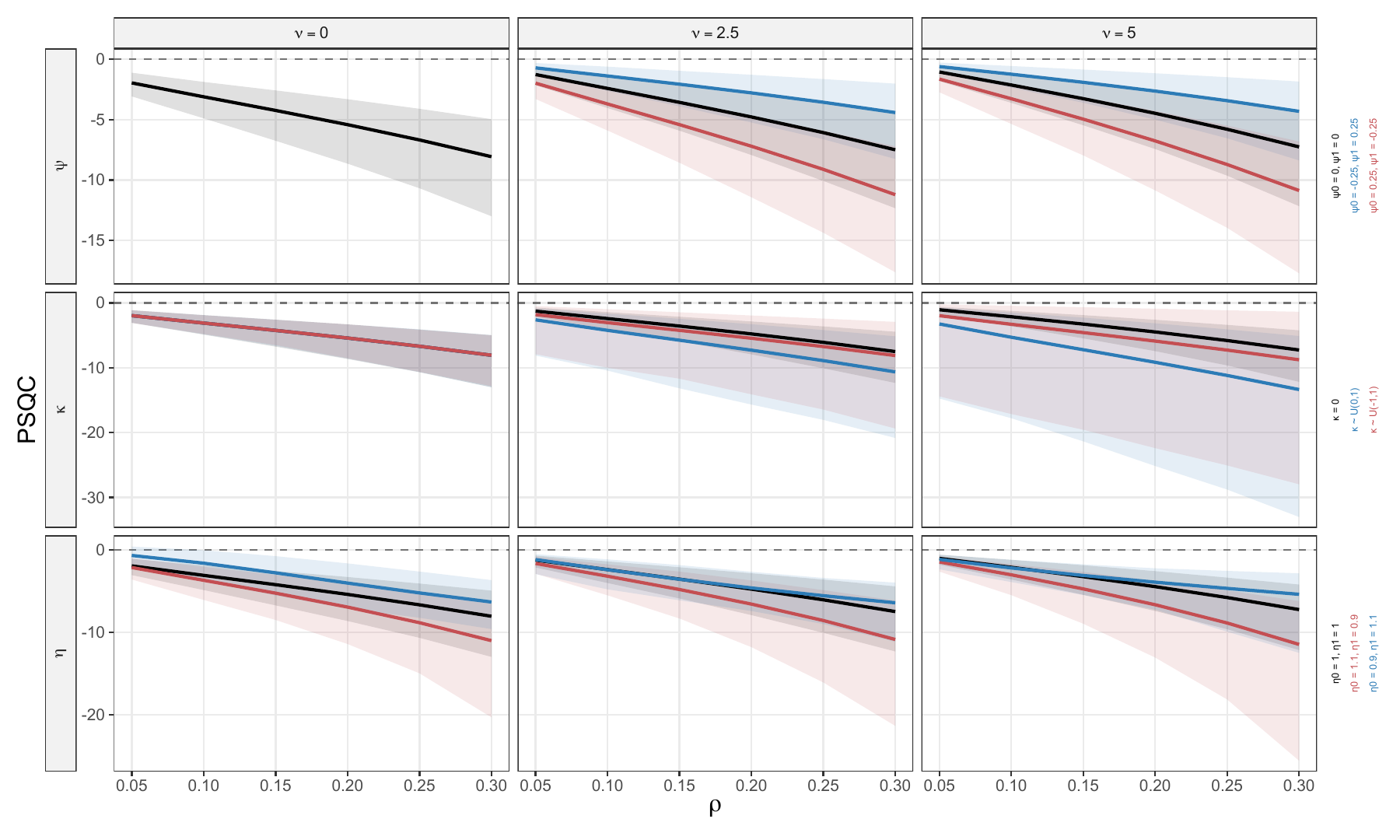}
    \caption{\label{figureEDPqrlPOST_PSQC_MCI}
    Primary and sensitivity analyses of the PSQC across quantile levels $\rho$ in the baseline MCI subgroup. Columns correspond to landmark times $\nu=0$, $2.5$, and $5$, and rows examine $(\psi_{0},\psi_{1})$, $\kappa$, and $(\eta_{0},\eta_{1})$. Curves and shaded bands denote posterior means and 99\% credible intervals. Black denotes the primary specification in each row, colored curves denote the alternatives shown in the legends, and the horizontal dashed line marks PSQC $=0$. More negative values indicate shorter remaining time to dementia onset under elevated amyloid.
    }
\end{figure}
    Figure~\ref{figureEDPqrlPOST_PSQC_MCI} presents the primary and sensitivity analyses for the MCI subgroup. The primary curves correspond to the estimates in Table~\ref{tableEDPqrl_ADNIpsqc}, while the alternative curves evaluate departures from the assumptions described in Section~\ref{EDPMqrl.sensitivity}. Across the examined specifications, the posterior mean PSQC remains predominantly negative, supporting shorter quantiles of remaining dementia-free time under elevated amyloid within the TAS principal stratum. Corresponding primary and sensitivity results for the overall cohort and CN subgroup are reported in Section~\ref{EDPMqrl.supp.application.results} and displayed in Supplementary Figures~\ref{figureEDPqrlPOST_PSQC_OVERALL} and~\ref{figureEDPqrlPOST_PSQC_CN}, respectively.

    The three rows of Figure~\ref{figureEDPqrlPOST_PSQC_MCI} assess sensitivity to residual unmeasured confounding, cross-world dependence, and informative censoring, respectively. Across these analyses, the alternative specifications affect the magnitude and uncertainty of the estimated PSQC more noticeably at later values of $\nu$ and higher quantile levels. For residual unmeasured confounding, $(\psi_{0}, \psi_{1}) = (-0.25, 0.25)$ generally attenuates the estimated contrast, whereas $(\psi_{0}, \psi_{1})=(0.25, -0.25)$ produces more negative values. For cross-world dependence, $\kappa \sim \munif{0,1}$ generally yields more negative estimates, while $\kappa \sim \munif{-1,1}$ remains closer to the primary analysis. For informative censoring, $(\eta_{0}, \eta_{1}) = (0.9, 1.1)$ generally attenuates the contrast, whereas $(\eta_{0}, \eta_{1}) = (1.1, 0.9)$ generally strengthens it, with uncertainty tending to increase at higher values of $\rho$. The posterior mean PSQC remains negative across all examined specifications. The only notable exception in interval inference occurs at $\nu = 0$ and $\rho = 0.05$ under $(\eta_{0}, \eta_{1})=(0.9, 1.1)$, where the 99\% CrI includes zero; apart from this setting, the sensitivity analyses consistently support shorter quantiles of remaining dementia-free time under elevated amyloid.

\section{Discussion} \label{EDPMqrl.discussion}
    Throughout the ADNI application, we estimated PSQCs comparing hypothetical elevated and non-elevated baseline amyloid statuses within the TAS principal stratum across all prespecified combinations of time point and quantile level. Under Assumptions~\eqref{EDPMqrl.causal.identification.A1}--\eqref{EDPMqrl.causal.identification.A5}, these contrasts have a causal interpretation. The estimated PSQC generally became more negative at higher quantile levels, indicating larger differences farther into the residual-lifetime distribution. The MCI estimates had stronger direct support from the observed event-time data, whereas the larger-magnitude CN estimates were less precise and relied more heavily on extrapolation. Overall, the results consistently indicated shorter remaining dementia-free time under hypothetical elevated than under hypothetical non-elevated baseline amyloid status. The use of 99\% CrIs provided more conservative uncertainty summaries, and the qualitative direction of the estimates remained stable across the examined sensitivity specifications.
    
    Our framework combines an EDPMM with posterior g-computation to estimate multiple PSQCs from one posterior fit. Joint modeling accommodates heterogeneous event-time distributions and partially observed covariates, while the enriched clustering structure reduces the covariate-domination problem of a conventional DPMM. In simulations, the EDPMM had smaller bias and RMSE and better-calibrated intervals than the DPMM. Remaining undercoverage under heavy censoring was concentrated at later time points, where observed failures were sparse within the TAS principal stratum; 99\% intervals improved calibration and motivated their use in ADNI.

    As in any causal analysis of observational data, identification of the PSQC depends on assumptions that cannot be fully verified from the observed data, including conditional exchangeability, conditional cross-world independence, and conditional non-informative censoring. Regarding this concern, we adjusted for important demographic, genetic, and clinical covariates and examined departures from these assumptions through sensitivity parameters $(\psi_{0},\psi_{1})$, $\kappa$, and $(\eta_{0},\eta_{1})$. These sensitivity analyses do not remove dependence on the identifying assumptions, but they provide a transparent assessment of how the estimated contrasts change under prespecified and interpretable departures from the primary analysis. The examined sensitivity specifications nevertheless represent only a limited range of possible violations, and the informative-censoring analysis requires refitting the posterior sampler. Future work could consider richer and more computationally efficient sensitivity procedures.

    Several additional extensions are also of interest. First, extending the proposed framework to settings with time-varying exposures or treatment regimes would further expand its usefulness for studying dynamic causal effects on residual lifetime in longitudinal observational studies. 
    Second, extending the framework to high-dimensional covariates would broaden its applicability to studies such as ADNI, which collect rich imaging, genetic, cognitive, and fluid-biomarker information \citep{tang2023ultra}.
    More broadly, incorporating multivariate longitudinal biomarkers could support dynamic prediction of remaining dementia-free time as patients' clinical and biomarker trajectories evolve during follow-up. This is particularly relevant in AD, where prognosis may change substantially as new cognitive, imaging, and fluid-biomarker measurements become available.

\section*{Supplementary Materials}
    Supplementary materials for this article are available online. They provide additional details on identification of the causal estimands (Section~\ref{EDPMqrl.supp.identification}), the MCMC algorithm (Section~\ref{EDPMqrl.supp.mcmc}), posterior inference via Bayesian g-computation (Section~\ref{EDPMqrl.supp.poststep}), simulation details (Section~\ref{EDPMqrl.supp.simulation}), and supplementary application results (Section~\ref{EDPMqrl.supp.application}). Reproducible code for the proposed methods, simulation studies, and ADNI application analyses is available at \href{https://github.com/WoojungBae/EDPqrl}{github.com/WoojungBae/EDPqrl}.
    
\section*{Disclaimer}
    The views expressed in this article should not be construed to represent those of the U.S. Food and Drug Administration.

\section*{Conflict of Interest}
    The authors declare no conflicts of interest.

\section*{Acknowledgements}
    Sang Kyu Lee was supported by the National Research Foundation of Korea (NRF) grant funded by the Korea government (RS-2026-25494847).
    
\section*{Data Availability}
    Data used in this study were obtained from the ADNI database. ADNI data are available to qualified researchers subject to ADNI data-use procedures (\href{http://adni.loni.usc.edu/}{adni.loni.usc.edu}). The analytic code used to preprocess the study variables, fit the models, and reproduce the reported analyses is available at \href{https://github.com/WoojungBae/EDPqrl}{github.com/WoojungBae/EDPqrl}.

    {
    \setlength{\bibsep}{0pt}
    \bibliographystyle{agsm}
    \bibliography{bibtex}
    }


\clearpage

\setcounter{section}{0}
\renewcommand{\thesection}{S\arabic{section}}
\renewcommand{\thesubsection}{\thesection.\arabic{subsection}}
\renewcommand{\thesubsubsection}{\thesubsection.\arabic{subsubsection}}
\renewcommand{\thefigure}{S\arabic{figure}}
\renewcommand{\thetable}{S\arabic{table}}

\setcounter{section}{0}
\setcounter{figure}{0}
\setcounter{table}{0}

\counterwithin*{figure}{section}
\counterwithin*{table}{section}

\section{Identification} \label{EDPMqrl.supp.identification}
    Under Assumptions~\eqref{EDPMqrl.causal.identification.A1}--\eqref{EDPMqrl.causal.identification.A5} stated in the main manuscript, the principal-stratum quantile contrast $\Delta(\nu,\rho)$ is identified from the observed data. For the PSQC, identifying only the marginal survival functions of $Y^{0}$ and $Y^{1}$ is insufficient because the TAS residual-life distributions depend on their joint distribution. Under conditional cross-world independence, the required conditional joint survivor probabilities factor into the two marginal conditional survivor probabilities.
    For $z \in \{0,1\}$, define
    \begin{align*}
        F^{z}(y \mid \bbX=\bbx)
        =
        \mP[Y^{z} \le y \mid \bbX=\bbx],
        \qquad
        S^{z}(y \mid \bbX=\bbx)
        =
        \mP[Y^{z} > y \mid \bbX=\bbx].
    \end{align*}
    The marginal potential-outcome survival functions are identified as
    \begin{align*}
        S^{z}(y \mid \bbX=\bbx)
        &=
        \mP[Y^{z} > y \mid \bbX=\bbx] \\
        &=
        \mP[Y^{z} > y \mid Z=z,\bbX=\bbx] \quad \text{by Assumption~\eqref{EDPMqrl.causal.identification.A2}} \\
        &=
        \mP[Y > y \mid Z=z,\bbX=\bbx] \quad \text{by Assumption~\eqref{EDPMqrl.causal.identification.A1}}.
    \end{align*}
    The final term is the exposure-specific conditional event-time survival function, which is identifiable from the observed right-censored data under Assumption~\eqref{EDPMqrl.causal.identification.A4} and the observed-follow-up support condition in Assumption~\eqref{EDPMqrl.causal.identification.A5}.
    
    Assumption~\eqref{EDPMqrl.causal.identification.A3} specifies conditional cross-world independence, $Y^{0}\indep Y^{1}\mid\bbX$. Therefore, the conditional joint survivor function is identified as
    \begin{align*}
        S^{0,1}(y_{0},y_{1}\mid\bbX=\bbx)
        &=
        \mP[Y^{0}>y_{0},Y^{1}>y_{1}\mid\bbX=\bbx] \\
        &=
        S^{0}(y_{0}\mid\bbX=\bbx) S^{1}(y_{1}\mid\bbX=\bbx).
    \end{align*}
    Under conditional cross-world independence, the marginal joint-survival probability appearing in the denominator of the TAS residual-life survival functions is identified as
    \begin{align*}
        \mP[Y^{0}>\nu,Y^{1}>\nu]
        =
        \mE[ S^{0,1}(\nu,\nu \mid \bbX) ][\bbX]
        =
        \mE[ S^{0}(\nu \mid \bbX) S^{1}(\nu \mid \bbX) ][\bbX].
    \end{align*}
    
    The residual-life survival functions within the TAS principal stratum are identified by
    \begin{align} \label{EDPMqrl.eq.supp.identification.1}
        S_{\nu}^{0}(y)
        &=
        \frac{\mE[ S^{0,1}(y+\nu,\nu \mid \bbX) ][\bbX]}
             {\mE[ S^{0,1}(\nu,\nu \mid \bbX) ][\bbX]}
        =
        \frac{\mE[ S^{0}(y+\nu \mid \bbX) S^{1}(\nu \mid \bbX) ][\bbX]}
             {\mE[ S^{0}(\nu \mid \bbX) S^{1}(\nu \mid \bbX) ][\bbX]}, \notag \\
        S_{\nu}^{1}(y)
        &=
        \frac{\mE[ S^{0,1}(\nu,y+\nu \mid \bbX) ][\bbX]}
             {\mE[ S^{0,1}(\nu,\nu \mid \bbX) ][\bbX]}
        =
        \frac{\mE[ S^{0}(\nu \mid \bbX) S^{1}(y+\nu \mid \bbX) ][\bbX]}
             {\mE[ S^{0}(\nu \mid \bbX) S^{1}(\nu \mid \bbX) ][\bbX]}.
    \end{align}
    Solving
    \begin{align*}
        S_{\nu}^{z}(r) = 1 - \rho
    \end{align*}
    for $r$ yields the causal quantile $R_{\nu,\rho}^{z}$. In general, a closed-form inverse is unavailable, so this equation is solved numerically using a standard root-finding algorithm. This approach avoids estimating the quantile from a finite simulated sample within the TAS principal stratum, which can be unstable at larger values of $\nu$ or when $\mP[Y^{0}>\nu,Y^{1}>\nu]$ is small. For a continuous survival function, the resulting solution is equivalent to the formal definition of the exposure-specific residual-life quantile within the TAS principal stratum. Finally, the PSQC is identified as
    \begin{align*}
        \Delta(\nu,\rho)
        =
        R_{\nu,\rho}^{1}
        -
        R_{\nu,\rho}^{0}.
    \end{align*}
    For subgroup-specific PSQCs, the same identification results apply after conditioning throughout on $\bbX'=\bbx'$, with the expectations taken over the remaining covariates under their conditional distribution given $\bbX'=\bbx'$.
    
    To assess departures from conditional cross-world independence, we also consider a Gaussian copula sensitivity model for the conditional joint distribution of $(Y^{0},Y^{1})$ given $\bbX$. For a fixed sensitivity parameter $\kappa$, define the copula-induced conditional joint survivor function, $\mP[Y^{0}>y_{0},Y^{1}>y_{1}\mid\bbX=\bbx;\kappa]$, as
    \begin{align*}
        \mS[y_{0},y_{1}\mid\bbX=\bbx;\kappa][][0,1]
        &=
        \mS[y_{0}\mid\bbX=\bbx][][0] + \mS[y_{1}\mid\bbX=\bbx][][1] -1 + \\
        & \qquad \Phi_{2} \left\{ \Phi_{1}^{-1}\left(F^{0}(y_{0}\mid\bbX=\bbx)\right),
        \Phi_{1}^{-1}\left(F^{1}(y_{1}\mid\bbX=\bbx)\right); \kappa \right\}.
    \end{align*}
    The value $\kappa=0$ recovers the conditional cross-world independence benchmark,
    \begin{align*}
        \mS[y_{0},y_{1}\mid\bbX=\bbx;\kappa=0][][0,1]
        =
        \mS[y_{0}\mid\bbX=\bbx][][0] \mS[y_{1}\mid\bbX=\bbx][][1].
    \end{align*}
    For a fixed value of $\kappa$, the corresponding sensitivity version of the residual-life survivor functions is identified conditional on $\kappa$ by
    \begin{align} 
        \begin{split} \label{EDPMqrl.eq.supp.identification.2}
            \mS[y;\kappa][\nu][0]
            &=
            \frac{\mE[ \mS[y+\nu,\nu \mid \bbX;\kappa][][0,1]][\bbX]}
                 {\mE[ \mS[\nu,\nu \mid \bbX;\kappa][][0,1]][\bbX]} \\
            \mS[y;\kappa][\nu][1] 
            &=
            \frac{\mE[ \mS[\nu,y+\nu \mid \bbX;\kappa][][0,1]][\bbX]}
                 {\mE[ \mS[\nu,\nu \mid \bbX;\kappa][][0,1]][\bbX]}.
        \end{split}
    \end{align}
    Solving
    \begin{align*}
        \mS[r;\kappa][\nu][z] =1-\rho
    \end{align*}
    for $r$ yields $R_{\nu,\rho}^{z}(\kappa)$, and the corresponding sensitivity estimand is
    \begin{align*}
        \Delta(\nu,\rho;\kappa)
        =
        R_{\nu,\rho}^{1}(\kappa)
        -
        R_{\nu,\rho}^{0}(\kappa).
    \end{align*}
    Details on the interpretation and choice of $\kappa$, including fixed values and prespecified sensitivity distributions, are provided in Section~\ref{EDPMqrl.sensitivity.cross}.

\section{Markov Chain Monte Carlo Algorithm} \label{EDPMqrl.supp.mcmc}
    Let $\bbtheta$ and $\bbomega$ denote the parameters for the outcome regressions and the exposure/confounders, respectively. To generate posterior samples, we adapt the nested clustering MCMC approach of \citet{roy2018bayesian} and \citet{bae2024bayesian}, which builds upon \cite[Algorithm 8]{neal2000markov}. Our algorithm sequentially proceeds through three main stages within each Gibbs iteration: augmenting censored survival times, imputing missing covariates, and updating cluster assignments and their associated parameters.

\subsection{Data Augmentation for the Censored Outcome} \label{EDPMqrl.supp.mcmc.data}
    To implement posterior inference for our model, we must address the presence of right censored event times. Let $\boldsymbol{O}_{i} = \left( T_{i}, D_{i}, Z_{i}, \bbX_{i} \right)$ denote the observed data for individual $i$. The observed data likelihood contribution is given by:
    \begin{align*}
        L \left( \boldsymbol{O}_{i} \mid \bbtheta_{i}, \bbomega_{i} \right) 
        &= \left\{ f \left( t_{i} \mid z_{i}, \bbx_{i} ; \bbtheta_{i} \right) \right\}^{D_{i}} 
        \left\{ S \left( t_{i} \mid z_{i}, \bbx_{i} ; \bbtheta_{i} \right) \right\}^{1 - D_{i}} 
        f \left( z_{i} \mid \bbx_{i} ; \omega_{i}^{z} \right)
        f \left( \bbx_{i} ; \bbomega_{i}^{x} \right),
    \end{align*}
    where the full parameter set for the $i$-th individual is denoted by $\left( \bbtheta_{i}, \bbomega_{i} \right)$. 
    
    The inclusion of the survival function $S \left( \cdot \right)$ for censored individuals ($D_{i} = 0$) presents a significant computational challenge. Because this term typically involves the cumulative distribution function (e.g., the Normal CDF, $\Phi$, in an accelerated failure time model), it fundamentally breaks the conjugacy of the model. Consequently, the full conditional posterior distributions for the outcome parameters $\bbtheta$ do not belong to standard families (such as the Normal or Inverse-Gamma distributions), meaning we cannot sample from them directly using a standard Gibbs sampler. 
    
    Without conjugacy, one standard alternative is to employ a Metropolis-within-Gibbs algorithm. In this approach, new candidate values for $\bbtheta$ must be drawn from a proposal distribution and accepted or rejected based on a calculated probability ratio. However, this method suffers from several practical drawbacks. It requires careful tuning of the proposal distributions to achieve an optimal acceptance rate; poor tuning can lead to highly inefficient sampling, slow exploration of the parameter space, and ultimately, poor mixing and failure to converge. In the context of flexible BNP models, tuning these steps across potentially infinite, dynamically changing clusters becomes computationally prohibitive.
    
    To circumvent these computational bottlenecks, we employ a Bayesian data augmentation strategy. By treating the true, unobserved event times for the censored individuals as latent variables, we can `augment' the data to restore the `complete-data' likelihood. This approach avoids cluster-specific Metropolis--Hastings updates for the local outcome-kernel parameters. In standard Metropolis-within-Gibbs algorithms, acceptance rates are typically much lower due to the lack of a universally efficient proposal distribution, necessitating continuous and often difficult tuning. In contrast, data augmentation yields direct Gibbs updates for the latent event times and restores conjugate updates for the local outcome parameters, avoiding the need to tune cluster-specific proposal distributions.
    
    To illustrate this data augmentation step, we consider a log-normal AFT local outcome model as in Sections~\ref{EDPMqrl.simulation} and \ref{EDPMqrl.application}. For notational simplicity, all event times are henceforth log-transformed:
    \begin{align*} 
        Y_{i} \mid Z_{i}, \bbX_{i}; \bbtheta_{i} & \sim \mnormal{ \mathbb{X}_{i} \bbbeta_{i}, \sigma_{i}^{2} },
    \end{align*}
    where $\bbtheta_{i} = \left( \bbbeta_{i}, \sigma_{i}^{2} \right)$ contains the cluster-specific parameters, and $\mathbb{X}_{i} = \left( 1, Z_{i}, \bbX_{i} \right)$ denotes the design vector.
    
    Within our Gibbs sampler, this data augmentation strategy is executed by alternating between the following two primary steps:
    \begin{enumerate}
        \item \textbf{Impute the Censored Data (Latent Variable Update):} \\
        Conditional on the current model parameters, we update the missing event times for all censored observations ($D_{i} = 0$). While the exact event time is unknown, it is constrained to be greater than the observed censoring time. Assuming $Y_{i}$ represents the log-event time and $T_{i}$ represents the log-observed time, we draw a new latent value $Y_{i}$ from its full conditional distribution, which is a truncated Normal distribution:
        \begin{align*}
            Y_{i} \mid Z_{i}, \bbX_{i}, D_{i} = 0; \bbbeta_{i}, \sigma_{i}^{2} \sim \mnormal{ \mathbb{X}_{i} \bbbeta_{i}, \sigma_{i}^{2} } \mbox{ truncated from below at } T_{i}.
        \end{align*}
        For uncensored observations ($D_{i} = 1$), $Y_{i}$ simply remains fixed at the observed $T_{i}$.
    
        \item \textbf{Update the Parameters via Conjugacy:} \\
        Let $\bby = \left( y_{1}, \ldots, y_{N} \right)$ denote the `complete' vector of log-survival times, comprising both the exactly observed times and the newly imputed latent times. Conditional on this completely augmented vector $\bby$, the complex survival terms vanish from the likelihood. The outcome model structurally reduces to standard Bayesian linear regression. Consequently, the full conditional posterior distributions for $\bbbeta$ and $\sigma^{2}$ revert to their standard, known conjugate forms. We can then draw updated parameters directly and efficiently from these standard distributions (e.g., Normal and Inverse-Gamma) without any need for Metropolis-Hastings rejection steps.
    \end{enumerate}

    Furthermore, we handle incomplete data under within-subcluster MAR assumption via a MCMC data augmentation algorithm conditional on the observed data and the EDPMM parameters; see Section~\ref{EDPMqrl.supp.mcmc.missing} of the supplementary materials for details.

\subsection{Handling Missingness in Covariates} \label{EDPMqrl.supp.mcmc.missing}
    In observational survival studies, missingness may arise in several data components. Individuals without observed exposure status or valid event-follow-up information are excluded from the analysis to avoid excessive reliance on model-based extrapolation, whereas right-censored event times are handled through the data-augmentation procedure in Section~\ref{EDPMqrl.supp.mcmc.data}. Missing baseline covariates are imputed under a within-subcluster MAR assumption: conditional on the current latent subcluster assignment, its parameters, and the observed quantities, the missingness indicator does not additionally depend on the missing covariate values. Because the subcluster assignment is latent and informed by the joint configuration of $(Y,Z,\bbX)$, this local assumption need not imply MAR after marginalizing over the latent partition. We therefore draw the missing covariates from their full conditional posterior distributions and propagate uncertainty in both the missing values and latent subcluster assignments through the MCMC algorithm. A related cluster-conditional data-augmentation strategy was used by \citet{bae2024bayesian}.

    Recall from Equation~\ref{EDPMqrl.supp.simulation.specifications.edpmm.1} that conditional on the cluster assignments (and their respective parameters $\bbtheta$ and $\bbomega$), the covariates are modeled using local parametric distributions: $X_{i,q} \sim \mbernoulli{ \pi_{i,q} }$ for binary confounders and $X_{i,q} \sim \mnormal{ \mu_{i,q}, \tau_{i,q}^{2} }$ for continuous confounders. Let $\bbX_{i, obs}$ and $\bbX_{i, mis}$ denote the observed and missing partitions of the covariate vector for subject $i$.
    
    During each iteration of the MCMC algorithm, we impute the missing values $\bbX_{i, mis}$ by drawing from their full conditional posterior distributions, given the current cluster parameters, the observed exposure $Z_{i}$, the observed covariates $\bbX_{i, obs}$, and the (potentially augmented) log-event time $Y_{i}$. 

    Although baseline covariates $\bbX$ are measured temporally prior to the event time $Y$, conditioning on $Y$ during the imputation of $\bbX_{i, mis}$ is practically essential. From a Bayesian joint modeling perspective, proper MCMC convergence to the true posterior requires sampling from the exact full conditional distribution, which by definition is proportional to the product of the local covariate prior and the outcome likelihood, $f (Y_{i} \mid Z_{i}, \bbX_{i}; \bbtheta_{i})$. Furthermore, from a missing data perspective, while the EDPMM's latent cluster assignments do capture global, non-linear dependence between the covariates and the outcome, the local within-cluster models explicitly link $\bbX$ and $Y$ via the regression coefficients $\bbbeta_{i}$. Imputing $\bbX_{i, mis}$ using only its local prior while ignoring the outcome would falsely force local conditional independence between the covariate and survival time. This omission would systematically attenuate the cluster-specific regression coefficients toward zero, distorting the conditional survival functions required for the g-computation algorithm. Therefore, incorporating $Y$ into the local imputation updates is necessary to preserve the true structural associations required for valid causal inference.
    
    Because the local covariate distributions are conditionally independent given the cluster assignment, the full conditional distribution for a single missing covariate $X_{i,q}$ depends only on its local base distribution and its contribution to the outcome model likelihood. We derive the full conditional updates as follows:
    \begin{itemize}[leftmargin=1.5em]
        \item \textbf{For a missing binary covariate ($q \in \{1, \ldots, p_{\bbX, 1}\}$):} \\
        The prior within the cluster is $\mbernoulli{\pi_{i,q}}$. The full conditional is also a Bernoulli distribution with updated success probability $\widetilde{\pi}_{i,q}^{*}$:
        \begin{align*}
            X_{i,q}^{(mis)} \mid \dots & \sim \mbernoulli{ \widetilde{\pi}_{i,q}^{*} },
        \end{align*}
        where the posterior probability is computed via Bayes' rule, weighing the outcome likelihood under both potential covariate states:
        \begin{align*}
            \widetilde{\pi}_{i,q}^{*} = \frac{ \pi_{i,q} f \left( Y_{i} \mid Z_{i}, X_{i,q}=1, \bbX_{i, obs} ; \bbtheta_{i} \right) }{ \pi_{i,q} f \left( Y_{i} \mid Z_{i}, X_{i,q}=1, \bbX_{i, obs} ; \bbtheta_{i} \right) + \left(1 - \pi_{i,q}\right) f \left( Y_{i} \mid Z_{i}, X_{i,q}=0, \bbX_{i, obs} ; \bbtheta_{i} \right) }.
        \end{align*}
        
        \item \textbf{For a missing continuous covariate ($q \in \{1 + p_{\bbX, 1}, \ldots, p_{\bbX}\}$):} \\
        The prior within the cluster is $\mnormal{\mu_{i,q}, \tau_{i,q}^{2}}$. Because the outcome model $Y_{i} \sim \mnormal{\mathbb{X}_{i} \bbbeta_{i}, \sigma_{i}^{2}}$ is a linear regression, this forms a conjugate Normal-Normal update. Let $\beta_{i,q}$ denote the regression coefficient corresponding to $X_{i,q}$, and let $\widetilde{\epsilon}_{i, -q} = Y_{i} - \left( \bbbeta_{i,0} + \beta_{i,Z} Z_{i} + \sum_{j \neq q} \beta_{i,j} X_{i,j} \right)$ be the partial residual. The full conditional distribution is:
        \begin{align*}
            X_{i,q}^{(mis)} \mid \dots & \sim \mnormal{ \widetilde{\mu}_{i,q}^{*}, \widetilde{\tau}_{i,q}^{2, *} },
        \end{align*}
        where the updated variance and mean are standard Bayesian linear regression updates:
        \begin{gather*}
            \widetilde{\tau}_{i,q}^{2, *} = \left( \frac{1}{\tau_{i,q}^{2}} + \frac{\beta_{i,q}^{2}}{\sigma_{i}^{2}} \right)^{-1} 
            \quad \& \quad
            \widetilde{\mu}_{i,q}^{*} = \widetilde{\tau}_{i,q}^{2, *} \left( \frac{\mu_{i,q}}{\tau_{i,q}^{2}} + \frac{\beta_{i,q} \widetilde{\epsilon}_{i, -q}}{\sigma_{i}^{2}} \right).
        \end{gather*}

    \end{itemize}
    
    Once the missing covariates $\bbX_{i, mis}$ are imputed, we treat them as fully observed complete data for the remainder of the current MCMC iteration. We then proceed to update the cluster assignments and the local parameters $\left( \bbtheta, \bbomega \right)$ exactly as described in Section~\ref{EDPMqrl.supp.mcmc}, seamlessly propagating the uncertainty of the missing covariates throughout the posterior distribution.

\subsection{Updating Cluster Assignments and Model Parameters} \label{EDPMqrl.supp.mcmc.params}
    The algorithm alternates between sampling the hierarchical cluster assignment $\bbs_{i} = \left( s_{i}^{y}, s_{i}^{\bbx} \right)$ for each subject $i$ via a P\'{o}lya urn scheme, and updating the model parameters conditional on $\bbs$. Let $K$ denote the current number of non-empty \texttt{y}-clusters, and $K_{h \mid k}$ the number of non-empty \texttt{$\bbx$}-subclusters ($z\,\&\,x$-clusters) specific to the $k^{th}$ \texttt{y}-cluster. We denote the parameters associated with these currently occupied clusters as $\bbtheta_{k}^{*}$ and $\bbomega_{h \mid k}^{x}$.
    
    We use the superscript $-i$ to denote quantities computed with subject $i$ excluded. For example, $y^{-i}$ is the outcome vector, $K^{-i}$ represents the number of non-empty clusters, and $n_{k}^{-i}$ and $n_{h \mid k}^{-i}$ are the respective sizes of the clusters omitting subject $i$. During the cluster assignment update, subject $i$ can be allocated to an existing non-empty cluster, to one of $K_{\nnew}$ new \texttt{y}-clusters, or to one of $K_{\nnew}$ new \texttt{$\bbx$}-subclusters within an existing \texttt{y}-cluster. Each potential new cluster is instantiated with auxiliary parameters drawn from the prior. In the simulations and data analysis, we used $K_{\nnew} = 1$.
    \begin{enumerate}[leftmargin=1.5em]
        \item Sample cluster membership for each subject $i$, $\bbs_{i} = \left( s_{i}^{y}, s_{i}^{\bbx} \right)$.
        
        \begin{enumerate}[leftmargin=1.5em]
            \item If the current value of $\bbs_{i}$ is in one of these clusters, then draw values $\bbtheta^{*}$ and $\bbomega^{*}$ from prior distributions $\nG_{0}^{ \bbtheta}$ and $\nG_{0}^{ \bbomega}$ for \texttt{y}-clusters $\left\{ K^{-i}+1,\ldots, K^{-i} + K_{\nnew} \right\}$.
            
            \item If the current value of $s_{i}^{y}$ is in an existing \texttt{y}-cluster (say, \texttt{y}-cluster $k$) but not an existing $\bbx$-subcluster, then set its current cluster to $\bbs_{i} = \left( k, K^{-i}+1 \right)$ and draw $K_{\nnew}-1$ values of $\bbomega^{*}$ from its prior distribution and assign them to $\left\{ K^{-i} + 2, \ldots, K^{-i}+K_{\nnew} \right\}$. In addition, draw $K_{\nnew}$ sets of parameters from the priors for the other clusters and subclusters. 
            
            \item If the current value of $s_{i}^{y}$ does not correspond with any of the $K^{-i}$ \texttt{y}-clusters, then $s_{i}^{y} = K^{-i} + 1$ and then draw $\bbtheta^{*}$ and $\bbomega^{*}$ for $y-$clusters $\left\{ K^{-i} + 2, \ldots, K^{-i}+K_{\nnew} \right\}$ and $\bbx$-subclusters $\left\{ K^{-i} + 2, \ldots, K^{-i} + K_{\nnew} \right\}$ for $k=1,\ldots,K^{-i}$
        
        \end{enumerate}
        At this point, all of the occupied and extra clusters have $\bbtheta^{*}$ and $\bbomega^{*}$ parameters associated with it. We can now draw a new value of $\bbs_{i}$. 
        \begin{align*}
            \mP[ \bbs_{i} = \left( k, h \right) \mid \bbs_{-i}, y, z, \bbx; \bbtheta^{*}, \bbomega^{*} ] 
            \propto
            \zeta_{k,h} f \left( y \mid z, \bbx; \bbtheta_{k}^{*} \right) f \left( z, \bbx; \bbomega_{h \mid k}^{*} \right)
        \end{align*}
        where 
        \begin{align*}
            \zeta_{k,h} = 
            \begin{cases}
                \frac{n_{k}^{-i}n_{h \mid k}^{-i}}{n_{k}^{-i} + \alpha^{\bbomega}} & \text{for $1 \le k \le K^{-i}$ and $1 \le h \le K_{h \mid k}^{-i}$} , \\
                \frac{n_{k}^{-i} \left( \frac{\alpha^{\bbomega}}{K_{\nnew}} \right)}{ n_{k}^{-i} + \alpha^{\bbomega} } & \text{for $1 \le k \le K^{-i}$ and $K_{h \mid k}^{-i} < h \le K_{h \mid k}^{-i} + K_{\nnew}$} ,   \\
                \frac{\alpha^{\bbtheta}}{K_{\nnew}}  & \text{for $K^{-i} < k \le K^{-i} + K_{\nnew}$}.
            \end{cases}
        \end{align*}
        Note that
        \begin{align*}
            \sum_{k=1}^{K^{-i} + K_{\nnew}} \sum_{h=1}^{K_{h \mid k}^{-i} + K_{\nnew}} \mP[ \bbs_{i} = \left( k, h \right) \mid \bbs_{-i}, y, z, \bbx; \bbtheta^{*}, \bbomega^{*} ] = 1.
        \end{align*}

        \item Given $\bbs_{i}$ and data, sample $\bbtheta$ and $\bbomega$ from their conditional distributions
        \begin{enumerate}[leftmargin=1.5em]
            \item For each unique $k$ in $s^{y} = \left\{ s_{1}^{y}, \ldots, s_{N}^{y} \right\}$, update $\bbtheta_{k}^{*}$ from
            \begin{align*}
                \mP[ \bbtheta_{k}^{*} \mid y, z, \bbx ; \bbs, \bbtheta_{-k}^{*}, \bbomega^{*} ] \propto \nG_{0}^{\bbtheta} \left( \bbtheta_{k}^{*} \right) \prod_{i:s_{i}^{y} = k} f \left( y \mid z, \bbx; \bbtheta_{k}^{*} \right).
            \end{align*}
            For the continuous outcome with $\bbtheta^{*} = \left( \bbbeta^{*} , \sigma^{2,*} \right)$, the base measures for $\bbtheta$ are $\nG_{0}^{\bbtheta} \left( \bbbeta \right) = \mnormals{a_{\bbbeta}, \sigma^{2} c_{\bbbeta} \boldsymbol{B}_{\bbbeta} }$, and $\nG_{0}^{\bbtheta} \left( \sigma^{2} \right) = \mscainvchi{ a_{\sigma} , b_{\sigma} }$. Set $a_{\bbbeta}$ and $\boldsymbol{B}_{\bbbeta}$ to the MLE from a parametric AFT model of $\left( T, D \right)$ on $\mathbb{X} = \left( 1, Z, \bbX \right)$ are a design matrix. Also, set $a_{\sigma} = 3$ and $b_{\sigma}=0.1$. Use $c_{\bbbeta} = \frac{N}{5}$. For notation convenience, we omit $c_{\bbbeta}$:
            \begin{align*}
                & \mP[ \bbbeta_{k}^{*}, \sigma_{k}^{2,*} \mid \bbs, y, z, \bbx, \bbtheta_{-k}^{*}, \bbomega^{*} ] \\
                & \propto \nG_{0}^{\bbtheta} \left( \bbbeta_{k}^{*}, \sigma_{k}^{2,*} \right) \prod_{i:s_{i}^{y} = k} f \left( y \mid z, \bbx; \bbbeta_{k}^{*}, \sigma_{k}^{2,*} \right) \\ 
                & \propto \mnormals{ \bbbeta_{k}^{*}; a_{\bbbeta}, \sigma_{k}^{2,*} \boldsymbol{B}_{\bbbeta}} \mscainvchi{ \sigma_{k}^{2,*}; a_{\sigma}, b_{\sigma} } \prod_{i:s_{i}^{y} = k} \mnormals{ y \mid z, \bbx; \bbbeta_{k}^{*}, \sigma_{k}^{2,*}}
            \end{align*}
            We then have the following posteriors: 
            \begin{gather*}
                \sigma_{k}^{2,*} \mid \cdot \sim \mscainvchi{ a_{\sigma}^{*}, b_{\sigma}^{*}} 
                \mbox{  and  }
                \bbbeta_{k}^{*} ; \sigma_{k}^{2,*} \mid \cdot \sim \mnormals{ a_{\bbbeta}^{*}, \sigma_{k}^{2,*} B_{\bbbeta}^{*}}
            \end{gather*}
            where $B_{\bbbeta}^{*} = \left( \boldsymbol{B}_{\bbbeta}^{-1} + \mathbb{X}^{\top} \mathbb{X} \right)^{-1}$, $a_{\bbbeta}^{*} = B_{\bbbeta}^{*} \left( \boldsymbol{B}_{\bbbeta}^{-1} a_{\bbbeta} + \mathbb{X}^{\top} Y \right)$, $a_{\sigma}^{*} = a_{\sigma} + \sum_{i} I \left( s_{i}^{y} = k \right)$, and $a_{\sigma}^{*} b_{\sigma}^{*} = a_{\sigma} b_{\sigma} + \boldsymbol{a}_{\bbbeta}^{\top} \boldsymbol{B}_{\bbbeta}^{-1} \boldsymbol{a}_{\bbbeta} + Y^{\top} Y - a_{\bbbeta}^{*, \top} \boldsymbol{B}_{\bbbeta}^{*,-1} \boldsymbol{a}_{\bbbeta}^{*}$.

            \item For each unique $(k,h)$ in $\bbs =  \left\{ \bbs_{1}, \ldots , \bbs_{N} \right\}$, update $\bbomega_{h \mid k}^{*}$ from
            \begin{align*}
                \mP[ \bbomega_{h \mid k}^{*} \mid y, z, \bbx; \bbs, \bbtheta^{*}, \bbomega_{-h \mid k}^{*} ] 
                \propto
                \nG_{0}^{\bbomega} \left( \bbomega_{h \mid k}^{*} \right) \prod_{i: \bbs_{i} = (k, h)} f \left( z, \bbx; \bbomega_{h \mid k}^{*} \right).
            \end{align*}
            Consider the case where the exposure $Z$ and the first $p_{\bbX,1}$ variables in $\bbX$ are categorical, while the remaining $p_{\bbX,2}$ variables are continuous.
            \begin{enumerate}[leftmargin=1.5em]
                \item For a categorical variable, we assume Dirichlet-Categorical,
                \begin{align*}
                    \mP[ x_{i,q} \mid \bbomega_{i} ] = \mcat{\pi_{i,q}}
                \end{align*}
                with
                \begin{align*}
                    \nG_{0}^{\bbomega} \left( \pi_{i,q} \right) = \mdir{\boldsymbol{a}_{\pi}}, \;\; q= 1, \ldots,1+p_{\bbX,1}.
                \end{align*}
                Thus, for $x_{i,q} \in \left\{ 1, \ldots, L \right\}$, we update 
                \begin{align*}
                    \pi_{h \mid k,q}^{*} \sim \mdir{\left( \boldsymbol{a}_{\pi,1} + n_{1 \mid h \mid k}, \ldots, \boldsymbol{a}_{\pi,L} + n_{L \mid h \mid k} \right)}, \;\; n_{l \mid h \mid k} = \sum_{i:\bbs_{i} = (k,h)} 1 \left( x_{i,q} = l \right).
                \end{align*}
                We use $\boldsymbol{a}_{\pi} = \boldsymbol{1}_{L}$.
    
                When the number of categories is $2$, it is just a Beta-Binomial. For a binary variable,
                \begin{align*}
                    \mP[ x_{i,q} \mid \bbomega_{i} ] = \mbernoulli{\pi_{i,q}} 
                \end{align*}
                with
                \begin{align*}
                    \nG_{0}^{\bbomega} \left( \pi_{i,q} \right) = \mbeta{a_{\pi},b_{\pi}}.
                \end{align*}
                Thus, we update 
                \begin{align*}
                    \pi_{h \mid k,q}^{*} \sim \mbeta{a_{\pi} + \sum_{i:\bbs_{i} = (k,h)} x_{i,q}, b_{\pi} + n_{h \mid k} - \sum_{i:\bbs_{i} = (k,h)} x_{i,q}} .
                \end{align*}
                We use $a_{\pi}=1$ and $b_{\pi}=1$.
                
                \item For a continuous variable, we assume Normal-Sca-Inv-$\chi^{2}$.
                \begin{align*}
                    \mP[ x_{i,q} \mid \bbomega_{i} ] = \mnormals{\mu_{i,q} , \tau_{i,q}^{2}} 
                \end{align*}
                with
                \begin{align*}
                    \nG_{0}^{\bbomega} \left( \tau_{i,q}^{2} \right) = \mscainvchi{a_{\tau}, b_{\tau}} \;\; \& \;\; \nG_{0}^{\bbomega} \left( \mu_{i,q} \mid \tau_{i,q}^{2} \right) = \mnormals{a_{\mu} , \frac{1}{b_{\mu}} \tau_{i,q}^{2}}.
                \end{align*}
                We can then update $\mu_{h \mid k,q}^{*}$ and $\tau_{h \mid k,q}^{2,*}$ from normal and Scale-Inv-$\chi^{2}$ distributions.
                \begin{align*}
                    \tau_{h \mid k,q}^{2,*} \mid \cdot \sim \mscainvchi{a_{\tau} + n_{h \mid k} , \frac{a_{\tau} b_{\tau} + \left( n_{h \mid k} -1 \right) \bar{s}_{h \mid k,q}^{2} + \frac{b_{\mu} n_{h \mid k}}{b_{\mu} + n_{h \mid k}} \left( \Bar{x}_{h \mid k,q} - a_{\mu} \right)^{2} }{a_{\tau} + n_{h \mid k}}}
                \end{align*}
                and
                \begin{align*}
                    \mu_{h \mid k,q}^{*} \mid \cdot \sim \mnormals{\frac{ \frac{b_{\mu}}{\tau_{h \mid k,q}^{2,*} } a_{\mu} + \frac{ n_{h \mid k}}{\tau_{h \mid k,q}^{2,*} } \Bar{x}_{h \mid k,q} }{\frac{ b_{\mu}}{\tau_{h \mid k,q}^{2,*} } + \frac{ n_{h \mid k}}{\tau_{h \mid k,q}^{2,*}} } , \frac{1 }{ \frac{b_{\mu}}{\tau_{h \mid k,q}^{2,*} } + \frac{ n_{h \mid k}}{\tau_{h \mid k,q}^{2,*}} }} 
                \end{align*}
                where $\Bar{x}_{h \mid k,q}$ and $\bar{s}_{h \mid k,q}^{2}$ are the sample mean and sample standard variance, respectively, of the $q^{th}$ confounder among subjects with $s = h \mid k$. Use $a_{\tau} =2$, $b_{\tau} =1$, $a_{\mu} = 0$ and $b_{\mu} = 0.5$.
                
                \end{enumerate}
            \end{enumerate}
    
            \item Update hyperparameters $\alpha^{\bbtheta}$ and $\alpha^{\bbomega} (\bbtheta)$: We specify $\ngamma \left( a_{\bbtheta}, b_{\bbtheta} \right)$. Use $a_{\bbtheta} = b_{\bbtheta} = a_{\bbomega} = b_{\bbomega} = 1$.
            \begin{enumerate}[leftmargin=1.5em]
                \item First draw $\xi \sim \mbeta{\alpha^{\bbtheta} + 1, N}$ and draw $\alpha^{\bbtheta}$ from
                \begin{align*}
                    \alpha^{\bbtheta} \sim \varrho \ngamma \left( a_{\bbtheta} + K , b_{\bbtheta} - \log \left( \xi \right) \right) + \left( 1- \varrho \right) \ngamma \left( a_{\bbtheta} + K - 1, b_{\bbtheta} - \log \left( \xi \right) \right)
                \end{align*}
                where $\varrho = \left( \frac{a_{\bbtheta} + K - 1}{N \left(  b_{\bbtheta}- \log \left( \xi \right) \right)} \right) \Big/ \left( 1 + \frac{a_{\bbtheta} + K - 1}{N \left( b_{\bbtheta} - \log \left( \xi \right) \right)} \right) = \frac{a_{\bbtheta} + K - 1}{ N \left( b_{\bbtheta} - \log \left( \xi \right) \right) + \left( a_{\bbtheta} + K - 1 \right) }$ \citep{escobar1995bayesian}. 
                
                \item To update $\alpha^{\bbomega}$, use Metropolis-Hastings where
                \begin{align*}
                    \mP[ \alpha^{\bbomega} \mid \cdot ] \propto \mP[ \alpha^{\bbomega} ] \left( \alpha^{\bbomega} \right)^{\sum_{k=1}^{K} \left( K - 1\right)} \prod_{k=1}^{K} \left( \alpha^{\bbomega} + n_{k} \right) B \left( \alpha^{\bbomega} + 1, n_{k} \right)
                \end{align*}
                where $B \left( \cdot \right)$ is the beta function \citep{roy2018bayesian}.
            
            \end{enumerate}
    \end{enumerate}
    
\section{Posterior Inference via G-computation} \label{EDPMqrl.supp.poststep}
    After obtaining posterior samples via the MCMC algorithm detailed in Section~\ref{EDPMqrl.supp.mcmc}, our procedure closely follows the post-processing steps described by \citet{roy2018bayesian}. Because this is performed post-MCMC, any functional of the potential outcome distribution can be computed efficiently without rerunning the computationally intensive sampler.

    Based on the identifying assumptions in Section~\ref{EDPMqrl.causal.identification}, we target the conditional survival function of the observed outcome, $\mS[y \mid z, \bbx] = \mP[ Y > y \mid Z=z, \bbX=\bbx ]$. To evaluate this analytically under the EDPMM, we define the marginal base distributions obtained by integrating over the prior measure $\nG_{0}$:
    \begin{gather*}
        f_{0} \left( \bbx \right) = \int f \left( \bbx ; \bbomega^{x} \right) d \nG_{0} \left( \bbomega^{x} \right), \quad
        f_{0} \left( z, \bbx \right) = \int f \left( z, \bbx ; \bbomega \right) d \nG_{0} \left( \bbomega \right), \\
        f_{0} \left( y, z, \bbx \right) = \int f \left( y, z, \bbx ; \bbtheta, \bbomega \right) d \nG_{0} \left( \bbtheta, \bbomega \right), \quad 
        \mS[ y \mid z, \bbx ][0] = \int \mS[ y \mid z, \bbx ; \bbtheta ] d \nG_{0} \left( \bbtheta \right).
    \end{gather*}

    By expanding \eqref{EDPMqrl.eq.edpm.3}, the conditional survival function evaluated at a specific posterior sample $\left\{ \bbtheta^{*}, \bbomega^{*}, \bbs \right\}$ is given by a finite sum over the $K$ active clusters plus a term for the unexplored cluster space:
    \begin{align} \label{EDPqrl.supp.eq.poststep.1}
        \mS[ y \mid z, \bbx ; \bbtheta^{*}, \bbomega^{*}, \bbs ] 
        = \frac{\lambda_{K + 1}^{y} \left( z, \bbx \right) \mS[y \mid z, \bbx][0] + \sum_{k=1}^{K} \lambda_{k}^{y} \left( z, \bbx \right) \mS[ y \mid z, \bbx; \bbtheta_{k}^{*} ]}{\lambda_{K + 1}^{y} \left( z, \bbx \right) + \sum_{k=1}^{K} \lambda_{k}^{y} \left( z, \bbx \right)},
    \end{align}
    where the weights are evaluated as $\lambda_{K + 1}^{y} \left( z, \bbx \right) = \frac{\alpha^{\bbtheta}}{\alpha^{\bbtheta} + N} f_{0} \left( z, \bbx \right)$, and
    \begin{align*}
        \lambda_{k}^{y} \left( z, \bbx \right) = \frac{n_{k}}{\alpha^{\bbtheta} + N} \left\{ \frac{\alpha^{\bbomega}}{\alpha^{\bbomega} + n_{k}} f_{0} \left( z, \bbx \right) + \sum_{h=1}^{K_{k}} \frac{n_{h \mid k}}{\alpha^{\bbomega} + n_{k}} f \left( z, \bbx; \bbomega_{h \mid k}^{*} \right) \right\}.
    \end{align*}
    
    While Equation~\eqref{EDPqrl.supp.eq.poststep.1} provides the exact conditional survival function, marginalizing it analytically over the posterior-predictive covariate distribution is computationally cumbersome. Therefore, for a given posterior sample, we execute this marginalization by generating a synthetic cohort of $M$ individuals, indexed by $m = 1, \ldots, M$. For each synthetic individual, we draw their parameters as follows:
    \begin{enumerate}[leftmargin=1.5em]
        \item Draw a cluster assignment $s_{(m)}^{y} \in \{1, \ldots, K + 1\}$ with probabilities proportional to $\{n_{1}, \ldots, n_{K}, \alpha^{\bbtheta}\}$, where $K$ is the current number of occupied main clusters.

        \item Draw $s_{(m)}^{\bbx} \in \{1, \ldots, K_{k} + 1\}$ with probabilities proportional to $\{n_{1 \mid k}, \ldots, n_{K_{k} \mid k}, \alpha^{\bbomega}\}$ $k \le K$ (an existing main cluster) where $k = s_{(m)}^{y}$. If $k = K + 1$ (a newly generated main cluster), set $s_{(m)}^{\bbx} = 1$.
        
        \item Draw a synthetic covariate vector $\bbx_{(m)}$ from $f \left( \bbx; \bbomega_{s_{(m)}^{\bbx} \mid s_{(m)}^{y}}^{*} \right)$. If a new subcluster was generated in the previous step, its parameters $\bbomega_{s_{(m)}^{\bbx} \mid s_{(m)}^{y}}^{*}$ are first drawn from the prior distribution.
        
    \end{enumerate}

    Using the generated set of $M$ synthetic values $\left\{ \bbx_{(m)}, \bbs_{(m)} \right\}_{m=1}^{M}$ where $\bbs_{(m)} = \left( s_{(m)}^{y}, s_{(m)}^{\bbx} \right)$, the expected marginal joint survival function $\mE[\mS[y_{0}, y_{1} \mid \bbX; \kappa][][0,1]][\bbX]$, can be approximated for any pair of times $y_{0},y_{1} \ge 0$ via Monte Carlo integration:
    \begin{align} \label{EDPqrl.supp.eq.poststep.2}
        \mE[\mS[y_{0}, y_{1} \mid \bbX ; \kappa][][0,1]][\bbX] 
        = \frac{1}{M} \sum_{m=1}^{M} \mS[ y_{0},y_{1} \mid \bbx_{(m)} ; \bbtheta_{s_{( m )}^{y}}^{*} , \bbomega_{s_{( m )}^{\bbx} \mid s_{( m )}^{y}}^{*}, \bbs_{( m )}, \kappa][][0,1].
    \end{align}
    The conditional joint survival function inside the summation is constructed by combining the exposure-specific conditional survival functions in \eqref{EDPqrl.supp.eq.poststep.1} through the Gaussian copula defined in Section~\ref{EDPMqrl.supp.identification}.

    For a specified quantile $\rho \in (0,1)$, we obtain the marginal residual-life quantile $R_{\nu,\rho}^{z}(\kappa)$ under exposure $z$ by solving the empirical analog of \eqref{EDPMqrl.eq.supp.identification.2} with respect to $r$:
    \begin{align*}
        \mS[r;\kappa][\nu][z] =1-\rho.
    \end{align*}
    The expectations appearing in $\mS[r;\kappa][\nu][z]$ are approximated using \eqref{EDPqrl.supp.eq.poststep.2}. Specifically, the numerator is evaluated at $(y_{0},y_{1})=(r+\nu,\nu)$ for $z=0$ and at $(y_{0},y_{1})=(\nu,r+\nu)$ for $z=1$, while the common denominator is evaluated at $(y_{0},y_{1})=(\nu,\nu)$. The corresponding PSQC is computed as
    \begin{align*}
        \Delta(\nu,\rho;\kappa)
        =
        R_{\nu,\rho}^{1}(\kappa)
        -
        R_{\nu,\rho}^{0}(\kappa).
    \end{align*}

    To estimate causal effects conditional on a specific covariate subset $\bbX_{2}=\bbx_{2}$, where $\bbX=(\bbX_{1},\bbX_{2})$, we replace Steps 1--3 of the marginal posterior-predictive sampling procedure with draws from the conditional posterior-predictive distribution of $\bbX_{1}$ given $\bbX_{2}=\bbx_{2}$. Instead of drawing the full covariate vector, we sample the remaining covariates $\bbx_{1}$ from their conditional distribution $\mP[ \bbx_{1} \mid \bbX_{2} = \bbx_{2}; \bbtheta^{*}, \bbomega^{*}, \bbs ]$. This is defined as:
    \begin{align*} 
        \mP[ \bbx_{1} \mid \bbX_{2} = \bbx_{2}; \bbtheta^{*}, \bbomega^{*}, \bbs ]
        & = \frac{\lambda_{K + 1} \left( \bbx_{2} \right) f_{0} \left( \bbx_{1} \mid \bbx_{2} \right) + \sum_{k=1}^{K} \sum_{h=1}^{K_{k}} \lambda_{h \mid k} \left( \bbx_{2} \right) f \left( \bbx_{1} \mid \bbx_{2}; \bbomega_{h \mid k} \right) }{ \lambda_{K + 1} \left( \bbx_{2} \right) + \sum_{k=1}^{K} \sum_{h=1}^{K_{k}} \lambda_{h \mid k} \left( \bbx_{2} \right)} \\
        & = \frac{\lambda_{K + 1} \left( \bbx_{2} \right) f_{0} \left( \bbx_{1} \right) + \sum_{k=1}^{K} \sum_{h=1}^{K_{k}} \lambda_{h \mid k} \left( \bbx_{2} \right) f \left( \bbx_{1} ; \bbomega_{h \mid k} \right) }{ \lambda_{K + 1} \left( \bbx_{2} \right) + \sum_{k=1}^{K} \sum_{h=1}^{K_{k}} \lambda_{h \mid k} \left( \bbx_{2} \right)},
    \end{align*}
    where the simplification in the second equality arises from the assumption that covariates are locally independent within clusters and the factorized base measure for $\bbomega^{x}$. The marginal weights evaluated at the conditioned values $\bbx_{2}$ are given by:

    \begin{gather*}
        \lambda_{K + 1} \left( \bbx_{2} \right) 
        = \left\{ \frac{\alpha^{\bbtheta}}{\alpha^{\bbtheta} + N} + \sum_{k=1}^{K} \frac{n_{k}}{\alpha^{\bbtheta} + N} \frac{\alpha^{\bbomega}}{\alpha^{\bbomega} + n_{k}} \right\} f_{0} \left( \bbx_{2} \right), \\
        \lambda_{h \mid k} \left( \bbx_{2} \right) 
        = \frac{n_{k}}{\alpha^{\bbtheta} + N} \frac{n_{h \mid k}}{\alpha^{\bbomega} + n_{k}} f \left( \bbx_{2}; \bbomega_{h \mid k}^{x*} \right).
    \end{gather*}

    For each synthetic individual $m$, sample either the unexplored component or an occupied subcluster $(k,h)$ with probabilities proportional to $\{\lambda_{\mathrm{new}}(\bbx_{2}),\lambda_{h \mid  k}(\bbx_{2})\}$. Draw $\bbx_{1(m)}$ from $f_{0}(\bbx_{1}  \mid \bbx_{2})$ for the unexplored component and from $f(\bbx_{1} \mid \bbx_{2};\bbomega_{h \mid  k}^{*})$ for an occupied subcluster. Under the factorized base measure and local conditional independence, these distributions reduce to $f_{0}(\bbx_{1})$ and $f(\bbx_{1};\bbomega_{h \mid  k}^{*})$, respectively. We then set $\bbx_{(m)}=(\bbx_{1(m)},\bbx_{2})$.
    
\section{Simulation Details} \label{EDPMqrl.supp.simulation}

\subsection{Simulation Scenarios} \label{EDPMqrl.supp.simulation.scenarios} 
    We generate simulated datasets with primary sample sizes $N \in \{500, 1500\}$ and an additional large-sample setting with $N=5000$. The data-generating mechanism incorporates dependence among baseline covariates, covariate-dependent exposure assignment, heterogeneous survival distributions, and covariate-dependent censoring.
    
    For subject $i=1,\dots,N$, we generate a five-dimensional baseline covariate vector $\bbX_{i}=(X_{i1},X_{i2},X_{i3},X_{i4},X_{i5})^{\top}$, where two covariates are binary and three are continuous:
    \begin{align*}
        X_{i1} &\sim \mbernoulli{0.5}, \\
        X_{i2} \mid X_{i1} &\sim \mbernoulli{0.4 + 0.2 X_{i1}}, \\
        X_{i3} &\sim \mnormals{0, 1}, \\
        X_{i4} \mid X_{i1}, X_{i3} &\sim \mnormals{-0.1 + 0.2 X_{i1} - 0.15 X_{i3}, 1}, \\
        X_{i5} \mid X_{i2}, X_{i4} &\sim \mnormals{0.1 - 0.2 X_{i2} + 0.15 X_{i4}, 0.5^2}.
    \end{align*}
    
    The binary exposure $Z_{i} \in \{0,1\}$ is generated from the probit model
    \begin{align*}
        P(Z_{i}=1 \mid \bbX_{i})
        =
        \Phi(0.2 + 0.1 X_{i1} + 0.2 X_{i3} - 0.1 X_{i5}),
    \end{align*}
    where $\Phi(\cdot)$ denotes the standard normal cumulative distribution function. This setup induces confounding through the shared dependence of the exposure and outcome on the baseline covariates.

    For each exposure level $z \in \{0,1\}$, let $\mathbb{X}_{i}(z)=(1,z,X_{i1},X_{i2},X_{i3},X_{i4},X_{i5})^{\top}$ denote the full design vector. Conditional on $\bbX_{i}$, the potential log-failure times are generated jointly as
    \begin{align*}
        \log Y_{i}(z) \mid \bbX_{i}
        \sim
        0.4 \times \mt{\mu_{i1}(z), 0.3, 10}
        +
        0.6 \times \mnormals{\mu_{i2}(z), 0.4^{2}},
    \end{align*}
    where $\mt{\mu,\sigma^{2},\nu}$ denotes a Student's $t$ distribution with location parameter $\mu$, scale parameter $\sigma$, and $\nu$ degrees of freedom and $\mnormals{\mu,\sigma^{2}}$ denotes a normal distribution with mean $\mu$, and variance $\sigma^{2}$. The component-specific location parameters are $\mu_{ij}(z)=\mathbb{X}_{i}^{\top}(z)\bbeta_{j}$, $j=1,2$, with
    \begin{align*}
        \bbeta_{1} &= (0.3, 0.2, -0.3, -0.5, 0.6, -0.5, -0.3)^{\top}, \\
        \bbeta_{2} &= (2.1, 0.6, -0.5, -0.3, 0.2, -0.3, -0.5)^{\top}.
    \end{align*}
    The mixture indicators and residual errors are generated independently across the two exposure levels. Consequently, the potential log-failure times are conditionally independent given the baseline covariates.
    
    The potential log-censoring time is generated from
    \begin{align*}
        \log C_{i}(z) \mid \bbX_{i}
        \sim
        \mnormals{c^{*}+\mathbb{X}_{i}^{\top}(z)\bbeta_{C}, 2^{2}},
        \qquad
        \bbeta_{C}=(0, 0.2, -0.1, 0.1, -0.2, 0.1, -0.2)^{\top}.
    \end{align*}
    The potential censoring times are generated independently of the potential failure times conditional on the baseline covariates. Thus, censoring is covariate dependent but conditionally independent. Given the observed exposure $Z_{i}$, the realized log-failure and log-censoring times are defined as $\log Y_{i}=\log Y_{i}(Z_{i})$ and $\log C_{i}=\log C_{i}(Z_{i})$, respectively. The observed log-time is $\log T_{i}=\min\{\log Y_{i},\log C_{i}\}$, and the event indicator is $D_{i}=I(\log Y_{i}\leq\log C_{i})$.
    
    The simulation scenarios differ only in the choice of the censoring intercept $c^{*}$, which controls the overall censoring rate. The values of $c^{*}$ are calibrated by Monte Carlo simulation to produce the following approximate censoring rates:
    \begin{itemize}
        \item \textbf{Scenario 1:} $c^{*}=3.20$, yielding approximately 20\% censoring;
        \item \textbf{Scenario 2:} $c^{*}=1.79$, yielding approximately 40\% censoring;
        \item \textbf{Scenario 3:} $c^{*}=0.53$, yielding approximately 60\% censoring;
        \item \textbf{Scenario 4:} $c^{*}=-0.95$, yielding approximately 80\% censoring.
    \end{itemize}
    In each scenario, we evaluate performance at time points $\nu \in \{0,1,2,3\}$. For the lighter-censoring settings of 20\% and 40\%, we consider quantile levels $\rho \in \{0.3,0.6\}$, whereas for the heavier-censoring settings of 60\% and 80\%, we consider $\rho \in \{0.1,0.2\}$.

\subsection{Fitted Model Specifications} \label{EDPMqrl.supp.simulation.specifications}

\subsubsection{EDPMM Specification} \label{EDPMqrl.supp.simulation.specifications.edpmm}
    For notational simplicity, in the model specification below $Y_{i}$ denotes the log-transformed event time. We use the same EDPMM framework introduced in Section~\ref{EDPMqrl.edpm}, but now specify the local kernel distributions used in the simulation study and the ADNI application. In particular, both binary and continuous confounders are accommodated through the following model:
    \begin{align} \label{EDPMqrl.supp.simulation.specifications.edpmm.1}
        \begin{split}
            Y_{i} \mid Z_{i}, \bbX_{i}, \bbtheta_{i} & \sim \mnormal{\mathbb{X}_{i} \bbbeta_{i}, \sigma_{i}^{2}} \\
            Z_{i} \mid \omega_{i}^{z} & \sim \mbernoulli{ \omega_{i}^{z} } \\
            X_{i,q} \mid \omega_{i,q}^{\bbx} & \sim \mbernoulli{ \pi_{i,q} }, \quad q = 1, \ldots, p_{\bbX,1} \\
            X_{i,q} \mid \omega_{i,q}^{\bbx} & \sim \mnormals{\mu_{i,q}, \tau_{i,q}^{2}}, \quad q = 1 + p_{\bbX,1}, \ldots, p_{\bbX,1} + p_{\bbX,2} \\
            \left( \bbtheta_{i}, \bbomega_{i} \right) \mid \nG & \sim \nG \\
            \nG & \sim \nedp \left( \alpha^{\bbtheta}, \alpha^{\bbomega}, \nG_{0} \right).
        \end{split}
    \end{align}
    Equation~\eqref{EDPMqrl.supp.simulation.specifications.edpmm.1} is a specific implementation of the general EDPMM in Equation~\eqref{EDPMqrl.eq.edpm.1}, obtained by specifying a normal regression kernel for the log-event time, a Bernoulli kernel for the binary exposure, and Bernoulli and normal kernels for the binary and continuous covariates, respectively. $\bbtheta_{i}=\left(\bbbeta_{i},\sigma_{i}^{2}\right)$ and $\bbomega_{i}=\left(\omega_{i}^{z},\bbomega_{i}^{\bbx}\right)$ collect the outcome, exposure, and covariate parameters, with $\bbomega_{i}^{\bbx}$ containing the corresponding $\pi_{i,q}$, $\mu_{i,q}$, and $\tau_{i,q}^{2}$ terms. The combined design vector is denoted by $\mathbb{X}_{i}=\left(1,Z_{i},\bbX_{i}\right)$, while $p_{\bbX,1}$ and $p_{\bbX,2}$ represent the number of binary and continuous confounders, respectively.
    
    The notation $\nG \sim \nedp \left( \alpha^{\bbtheta}, \alpha^{\bbomega}, \nG_{0} \right)$ indicates that the marginal main-cluster distribution is $\nG^{\bbtheta} \sim \ndp \left( \alpha^{\bbtheta},\nG_{0}^{\bbtheta} \right)$ and the conditional subcluster distribution is $\nG^{\bbomega \mid \bbtheta} \sim \ndp \left( \alpha^{\bbomega}, \nG_{0}^{\bbomega \mid \bbtheta} \right)$, with the joint base measure factoring as $\nG_{0} = \nG_{0}^{\bbtheta} \times \nG_{0}^{\bbomega \mid \bbtheta}$.
    
\subsubsection{DPMM Specification} \label{EDPMqrl.supp.simulation.specifications.dpmm}
    To establish a direct baseline for our proposed framework, we extend the approach of \citet{park2012bayesian} to a joint Dirichlet Process Mixture (DPM) model. This comparator model utilizes the exact same local parametric distributions defined in \eqref{EDPMqrl.supp.simulation.specifications.edpmm.1} and identical prior base measures $\nG_{0}$, but places a single DP prior over the `complete-data' joint distribution: $\left( \bbtheta_{i}, \bbomega_{i} \right) \mid \nG \sim \nG$ and $\nG \sim \ndp \left( \alpha, \nG_{0} \right)$. 
    \begin{align} \label{EDPMqrl.supp.eq.simulation.specifications.dpmm.1}
        \begin{split}
            Y_{i} \mid Z_{i}, \bbX_{i}, \bbtheta_{i} & \sim \mnormal{\mathbb{X}_{i} \bbbeta_{i}, \sigma_{i}^{2}} \\
            Z_{i} \mid \omega_{i}^{z} & \sim \mbernoulli{ \omega_{i}^{z} } \\
            X_{i,q} \mid \omega_{i,q}^{\bbx} & \sim \mbernoulli{ \pi_{i,q} }, \quad q = 1, \ldots, p_{\bbX,1} \\
            X_{i,q} \mid \omega_{i,q}^{\bbx} & \sim \mnormals{\mu_{i,q}, \tau_{i,q}^{2}}, \quad q = p_{\bbX,1} + 1, \ldots, p_{\bbX,1} + p_{\bbX,2} \\
            \left( \bbtheta_{i}, \bbomega_{i} \right) \mid \nG & \sim \nG \\
            \nG & \sim \ndp \left( \alpha, \nG_{0} \right).
        \end{split}
    \end{align}
    
    The joint DPMM defined in \eqref{EDPMqrl.supp.eq.simulation.specifications.dpmm.1} has a standard stick-breaking representation \citep{sethuraman1994constructive}:
    \begin{gather*}
        \begin{split}
            f \left( y, z, \bbx \mid \nG \right)
            &= \int f \left( y \mid z, \bbx ; \bbtheta \right) 
                f \left( z ; \omega^{z} \right) 
                f \left( \bbx ; \bbomega^{\bbx} \right) 
                d \nG \left( \bbtheta, \omega^{z}, \bbomega^{\bbx} \right) \\
            &= \sum_{k=1}^{\infty} \gamma_{k} 
                f \left( y \mid z, \bbx ; \bbtheta_{k} \right)
                f \left( z ; \omega_{k}^{z} \right) 
                f \left( \bbx ; \bbomega_{k}^{\bbx} \right),
        \end{split}
    \end{gather*}
    where $k$ indexes the clusters. The weights are constructed via $\gamma_{k} = \gamma_{k}' \prod_{ l < k } \left( 1 - \gamma_{l}' \right)$ with priors $\gamma_{k}' \sim \mbeta{1, \alpha}$. 
    
    From this joint distribution, the conditional outcome density is derived as:
    \begin{align*}
        \mP[ y \mid z, \bbx ] 
        = \sum_{k=1}^{\infty} \Lambda_{k} \left( z, \bbx \right) f \left( y \mid z, \bbx ; \bbtheta_{k} \right),
    \end{align*}
    where the conditional mixture weights are dynamically updated by the covariate and exposure distributions:
    \begin{align*}
        \Lambda_{k} \left( z, \bbx \right)
        = \frac{\gamma_{k} f \left( z; \omega_{k}^{z} \right) f \left( \bbx; \bbomega_{k}^{\bbx} \right)}{\sum_{d=1}^{\infty} \gamma_{d} f \left( z; \omega_{d}^{z} \right) f \left( \bbx; \bbomega_{d}^{\bbx} \right)}.
    \end{align*}
    
    Based on this conditional density, the conditional survival function $\mS[y \mid z, \bbx]$ is computed analogously to the conditional survival function in \eqref{EDPMqrl.eq.edpm.3}. Finally, for a given time point $\nu$ and quantile level $\rho$, the conditional QRL and the resulting marginal causal estimands are computed using the procedure detailed in Section~\ref{EDPMqrl.supp.poststep}.

\subsection{Prior Specifications} \label{EDPMqrl.supp.simulation.prior}
    We specified the following base measures for the local outcome parameters $\bbtheta_{i}$ defined in Section~\ref{EDPMqrl.supp.simulation.specifications.edpmm.1}: $\nG_{0} (\bbbeta \mid \sigma^{2} ) = \mnormals{a_{\bbbeta}, c_{\bbbeta} \boldsymbol{B}_{\bbbeta} \sigma^{2}}$ and $\nG_{0} (\sigma^{2}) = \mscainvchi{a_{\sigma}, b_{\sigma}}$. We set the hyperparameter $a_{\bbbeta}$ to the maximum likelihood estimates obtained from a standard parametric log-normal AFT model fit using the \texttt{survreg} function \citep{therneau2023package} with survival response $\left( T, D \right)$ and design matrix $\mathbb{X}$. We set the prior covariance scale matrix $\boldsymbol{B}_{\bbbeta}$ to the corresponding inverse observed information matrix from this initial fit \citep{taddy2008bayesian, roy2018bayesian, daniels2023bayesian} and set the variance scaling factor to $c_{\bbbeta} = N/5$. We fixed the hyperparameters for the residual variance at $a_{\sigma} = 3$ and $b_{\sigma} = 0.1$. For the binary exposure and binary confounders, we assumed conjugate priors $\nG_{0} (\pi_{q}) = \mbeta{a_{\pi}, b_{\pi}}$ with $a_{\pi} = b_{\pi} = 1$ for parameters $q = 1, \ldots , p_{\bbX,1}$. For the continuous confounder parameters, we assumed $\nG_{0} (\mu_{q} \mid \tau_{q}^{2}) = \mnormals{a_{\mu}, \frac{1}{b_{\mu}} \tau_{q}^{2}}$ and $\nG_{0} (\tau_{q}^{2}) = \mscainvchi{a_{\tau}, b_{\tau}}$ for $q = p_{\bbX,1} + 1, \ldots , p_{\bbX,1} + p_{\bbX,2}$. For the concentration parameters $\alpha_{\bbtheta}$ and $\alpha_{\bbomega}$, we specified $\mgamma{1, 1}$ priors. 
    
\subsection{Simulation Implementation and Performance Measures}
\label{EDPMqrl.supp.simulation.implementation}
    Throughout the simulation study and the ADNI application in Section~\ref{EDPMqrl.application}, we use the same prior specification; details are provided in Section~\ref{EDPMqrl.supp.simulation.prior}. Posterior inference uses a 20000-iteration burn-in followed by 20000 additional iterations for the simulation study and 40000 additional iterations for the ADNI application, thinned by 20 to retain 1000 posterior samples for the simulation study and 2000 posterior samples for the ADNI application.

    To estimate the marginal causal effects, we integrate over the covariate distribution using $M=1000$ Monte Carlo samples. For each combination of the two primary sample sizes and four censoring scenarios, we generate 1000 simulated datasets. We report bias, RMSE, and 95\% CP at time points $\nu \in \{0,1,2,3\}$. For the lighter-censoring settings of 20\% and 40\%, we consider quantile levels $\rho \in \{0.3,0.6\}$, whereas for the heavier-censoring settings of 60\% and 80\%, we consider $\rho \in \{0.1,0.2\}$ to focus on regions with adequate observed event-time information. Additional analyses under 60\% and 80\% censoring include 99\% credible intervals (CrIs) for $N=500$ and $N=1500$, as well as a larger-sample evaluation with $N=5000$ using 95\% CrIs.
    
\subsection{Additional Simulation Results} \label{EDPMqrl.supp.simulation.additional}
    To better understand the modest bias and undercoverage observed for the EDPMM under heavy censoring and at later time points, we examine the amount of observed failure-time information available within the TAS principal stratum. We then conduct two additional analyses to assess whether increasing the sample size and using wider CrIs improves estimation and uncertainty quantification.

    \begin{table}[!tbp]
\centering
\caption{\label{tableEDPqrl_SIM_pop_supp}
Observed failure-time information in the TAS principal stratum across censoring settings. Each entry gives the percentage of TAS members with an observed failure, $100 \mP[D=1 \mid Y^{0}>\nu, Y^{1}>\nu]$. The percentage of the full Monte Carlo population that both belongs to the TAS principal stratum and has an observed failure, $100 \mP[Y^{0}>\nu, Y^{1}>\nu, D=1]$, is shown in parentheses. These two quantities are related by $\mP[Y^{0}>\nu, Y^{1}>\nu, D=1] = \mP[D=1 \mid Y^{0}>\nu, Y^{1}>\nu] \mP[Y^{0}>\nu, Y^{1}>\nu]$. The TAS principal-stratum probabilities, $100 \mP[Y^{0}>\nu,Y^{1}>\nu]$, are 100.0\%, 66.3\%, 47.5\%, and 36.3\% at $\nu=0$, $1$, $2$, and $3$, respectively. For example, under 20\% censoring at $\nu=2$, 70.7\% of TAS members have an observed failure, and $0.707\times0.475\approx0.336$, corresponding to 33.6\% of the full Monte Carlo population. The Monte Carlo population size is 1,000,000 for each censoring setting.}
\centering
\fontsize{9}{11}\selectfont
\begin{tabular}[t]{rcccc}
\toprule
Censoring (\%) & $\nu=0$ & $\nu=1$ & $\nu=2$ & $\nu=3$\\
\midrule
20 & 80.1 (80.1) & 74.4 (49.3) & 70.7 (33.6) & 68.0 (24.6)\\
40 & 60.1 (60.1) & 50.9 (33.8) & 45.8 (21.8) & 42.4 (15.4)\\
60 & 39.9 (39.9) & 29.5 (19.5) & 24.6 (11.7) & 21.7 (7.9)\\
80 & 19.9 (19.9) & 11.4 (7.6) & 8.5 (4.0) & 6.9 (2.5)\\
\bottomrule
\end{tabular}
\end{table}
    Table~\ref{tableEDPqrl_SIM_pop_supp} summarizes the observed failure-time information available within the TAS principal stratum across the four censoring settings. The proportion with an observed failure time decreases as both the censoring rate and $\nu$ increase. At $\nu=3$, this proportion is 68.0\% under 20\% censoring and 6.9\% under 80\% censoring, corresponding to 24.6\% and 2.5\% of the full Monte Carlo population, respectively. Thus, although approximately 36\% of the population satisfies $Y^{0}>3$ and $Y^{1}>3$, directly observed failure-time information is available for only a small fraction under the heaviest censoring setting. This substantial reduction in effective information provides a plausible explanation for the greater estimation difficulty and remaining undercoverage observed under heavy censoring and at later time points. Moreover, although the average censoring rate in Scenario 4 is approximately 80\%, some individual simulated datasets exhibit even higher censoring. After conditioning on survival beyond $\nu$, these datasets may contain few observed events and provide limited support for estimating the $\rho=0.2$ residual-life quantile. The resulting reliance on model-based extrapolation may make posterior uncertainty more difficult to quantify accurately.

    \begin{table}[!tbp]
\centering
\caption{\label{tableEDPqrl_sim34_suppN5000}
Additional simulation performance of the EDPMM and DPMM for estimating the marginal PSQC under approximately 60\% and 80\% censoring with $N=5000$. The table reports the true PSQC, empirical bias, RMSE, and CP of nominal 95\% credible intervals at landmark time $\nu$ and quantile level $\rho$.}
\resizebox{\ifdim\width>\linewidth\linewidth\else\width\fi}{!}{
\fontsize{8}{10}\selectfont
\begin{tabular}[t]{lcc>{}c>{}c>{}c>{}c>{}c>{}c>{}c}
\toprule
\multicolumn{4}{c}{ } & \multicolumn{6}{c}{N = 5000} \\
\cmidrule(l{3pt}r{3pt}){5-10}
\multicolumn{4}{c}{ } & \multicolumn{3}{c}{60\%} & \multicolumn{3}{c}{80\%} \\
\cmidrule(l{3pt}r{3pt}){5-7} \cmidrule(l{3pt}r{3pt}){8-10}
Model & $\nu$ & $\rho$ & True & Bias & RMSE & CP & Bias & RMSE & CP\\
\midrule
EDPMM & 0 & 0.1 & \cellcolor{white}{0.10} & \cellcolor[HTML]{f2f2f2}{0.00} & \cellcolor[HTML]{f2f2f2}{0.03} & \cellcolor[HTML]{f2f2f2}{92.8} & \cellcolor[HTML]{d9d9d9}{0.00} & \cellcolor[HTML]{d9d9d9}{0.04} & \cellcolor[HTML]{d9d9d9}{91.5}\\
 &  & 0.2 & \cellcolor{white}{0.22} & \cellcolor[HTML]{f2f2f2}{0.00} & \cellcolor[HTML]{f2f2f2}{0.07} & \cellcolor[HTML]{f2f2f2}{95.4} & \cellcolor[HTML]{d9d9d9}{0.00} & \cellcolor[HTML]{d9d9d9}{0.08} & \cellcolor[HTML]{d9d9d9}{95.1}\\
 & 1 & 0.1 & \cellcolor{white}{0.17} & \cellcolor[HTML]{f2f2f2}{0.01} & \cellcolor[HTML]{f2f2f2}{0.05} & \cellcolor[HTML]{f2f2f2}{95.9} & \cellcolor[HTML]{d9d9d9}{0.01} & \cellcolor[HTML]{d9d9d9}{0.07} & \cellcolor[HTML]{d9d9d9}{92.3}\\
 &  & 0.2 & \cellcolor{white}{0.55} & \cellcolor[HTML]{f2f2f2}{0.04} & \cellcolor[HTML]{f2f2f2}{0.13} & \cellcolor[HTML]{f2f2f2}{96.0} & \cellcolor[HTML]{d9d9d9}{0.04} & \cellcolor[HTML]{d9d9d9}{0.20} & \cellcolor[HTML]{d9d9d9}{92.7}\\
 & 2 & 0.1 & \cellcolor{white}{0.54} & \cellcolor[HTML]{f2f2f2}{0.02} & \cellcolor[HTML]{f2f2f2}{0.08} & \cellcolor[HTML]{f2f2f2}{96.4} & \cellcolor[HTML]{d9d9d9}{0.03} & \cellcolor[HTML]{d9d9d9}{0.13} & \cellcolor[HTML]{d9d9d9}{92.0}\\
 &  & 0.2 & \cellcolor{white}{1.28} & \cellcolor[HTML]{f2f2f2}{0.03} & \cellcolor[HTML]{f2f2f2}{0.16} & \cellcolor[HTML]{f2f2f2}{95.9} & \cellcolor[HTML]{d9d9d9}{0.02} & \cellcolor[HTML]{d9d9d9}{0.26} & \cellcolor[HTML]{d9d9d9}{91.1}\\
 & 3 & 0.1 & \cellcolor{white}{0.87} & \cellcolor[HTML]{f2f2f2}{0.02} & \cellcolor[HTML]{f2f2f2}{0.10} & \cellcolor[HTML]{f2f2f2}{96.3} & \cellcolor[HTML]{d9d9d9}{0.02} & \cellcolor[HTML]{d9d9d9}{0.16} & \cellcolor[HTML]{d9d9d9}{92.0}\\
 &  & 0.2 & \cellcolor{white}{1.77} & \cellcolor[HTML]{f2f2f2}{0.02} & \cellcolor[HTML]{f2f2f2}{0.18} & \cellcolor[HTML]{f2f2f2}{95.6} & \cellcolor[HTML]{d9d9d9}{0.02} & \cellcolor[HTML]{d9d9d9}{0.29} & \cellcolor[HTML]{d9d9d9}{91.6}\\

\cmidrule(l{3pt}r{3pt}){1-10}

DPMM & 0 & 0.1 & \cellcolor{white}{0.10} & \cellcolor[HTML]{f2f2f2}{0.11} & \cellcolor[HTML]{f2f2f2}{0.12} & \cellcolor[HTML]{f2f2f2}{1.9} & \cellcolor[HTML]{d9d9d9}{0.08} & \cellcolor[HTML]{d9d9d9}{0.09} & \cellcolor[HTML]{d9d9d9}{37.1}\\
 &  & 0.2 & \cellcolor{white}{0.22} & \cellcolor[HTML]{f2f2f2}{0.18} & \cellcolor[HTML]{f2f2f2}{0.19} & \cellcolor[HTML]{f2f2f2}{5.2} & \cellcolor[HTML]{d9d9d9}{0.13} & \cellcolor[HTML]{d9d9d9}{0.14} & \cellcolor[HTML]{d9d9d9}{51.8}\\
 & 1 & 0.1 & \cellcolor{white}{0.17} & \cellcolor[HTML]{f2f2f2}{0.02} & \cellcolor[HTML]{f2f2f2}{0.03} & \cellcolor[HTML]{f2f2f2}{89.3} & \cellcolor[HTML]{d9d9d9}{0.00} & \cellcolor[HTML]{d9d9d9}{0.03} & \cellcolor[HTML]{d9d9d9}{95.7}\\
 &  & 0.2 & \cellcolor{white}{0.55} & \cellcolor[HTML]{f2f2f2}{-0.15} & \cellcolor[HTML]{f2f2f2}{0.16} & \cellcolor[HTML]{f2f2f2}{20.1} & \cellcolor[HTML]{d9d9d9}{-0.19} & \cellcolor[HTML]{d9d9d9}{0.21} & \cellcolor[HTML]{d9d9d9}{22.6}\\
 & 2 & 0.1 & \cellcolor{white}{0.54} & \cellcolor[HTML]{f2f2f2}{-0.34} & \cellcolor[HTML]{f2f2f2}{0.34} & \cellcolor[HTML]{f2f2f2}{0.0} & \cellcolor[HTML]{d9d9d9}{-0.36} & \cellcolor[HTML]{d9d9d9}{0.36} & \cellcolor[HTML]{d9d9d9}{0.0}\\
 &  & 0.2 & \cellcolor{white}{1.28} & \cellcolor[HTML]{f2f2f2}{-0.85} & \cellcolor[HTML]{f2f2f2}{0.85} & \cellcolor[HTML]{f2f2f2}{0.0} & \cellcolor[HTML]{d9d9d9}{-0.90} & \cellcolor[HTML]{d9d9d9}{0.90} & \cellcolor[HTML]{d9d9d9}{0.0}\\
 & 3 & 0.1 & \cellcolor{white}{0.87} & \cellcolor[HTML]{f2f2f2}{-0.66} & \cellcolor[HTML]{f2f2f2}{0.66} & \cellcolor[HTML]{f2f2f2}{0.0} & \cellcolor[HTML]{d9d9d9}{-0.68} & \cellcolor[HTML]{d9d9d9}{0.68} & \cellcolor[HTML]{d9d9d9}{0.0}\\
 &  & 0.2 & \cellcolor{white}{1.77} & \cellcolor[HTML]{f2f2f2}{-1.31} & \cellcolor[HTML]{f2f2f2}{1.31} & \cellcolor[HTML]{f2f2f2}{0.0} & \cellcolor[HTML]{d9d9d9}{-1.36} & \cellcolor[HTML]{d9d9d9}{1.36} & \cellcolor[HTML]{d9d9d9}{0.0}\\
\bottomrule
\end{tabular}}
\end{table}
    Table~\ref{tableEDPqrl_sim34_suppN5000} reports additional simulation results for Scenarios 3 and 4, corresponding to approximately 60\% and 80\% censoring, respectively, with sample size $N=5000$. The nominal level is fixed at $\alpha=0.05$, as in the main simulation analysis reported in Table~\ref{tableEDPqrl_sim_main}. Relative to the results for $N=500$ and $N=1500$, increasing the sample size to $N=5000$ generally improves the performance of the EDPMM. The absolute empirical bias is below $0.05$ across all reported $(\nu,\rho)$ combinations, and the RMSE is substantially reduced, particularly at later time points and for $\rho=0.2$.

    Empirical coverage for the EDPMM also generally moves closer to the nominal 95\% level as the sample size increases. Under 60\% censoring, coverage ranges from 92.8\% to 96.4\%. The improvement is most apparent relative to $N=500$, whereas the gains over $N=1500$ are more modest. Under 80\% censoring, coverage ranges from 91.1\% to 95.1\% and is generally higher than that observed at either of the smaller sample sizes. Although modest undercoverage remains in some settings, particularly at later time points, these findings suggest that the bias and undercoverage of the EDPMM under heavy censoring are at least partly attributable to finite-sample limitations and the limited observed failure-time information available after $\nu$.

    In contrast, increasing the sample size does not eliminate the poor performance of the DPMM. Although its performance is acceptable in a small number of settings with smaller values of $\nu$, the DPMM continues to exhibit substantially larger bias and RMSE than the EDPMM. Severe undercoverage also persists, particularly at later time points and for $\rho=0.2$, with empirical coverage equal or close to zero in several settings even when $N=5000$. These results indicate that the poor performance of the DPMM cannot be explained solely by limited sample size.

    \begin{table}[!tbp]
\centering
\caption{\label{tableEDPqrl_sim34_supp}
Simulation performance under approximately 60\% and 80\% censoring using nominal 99\% credible intervals for $N \in \{500, 1500\}$. The table reports the true marginal PSQC, empirical bias, RMSE, and CP at landmark time $\nu$ and quantile level $\rho$.}
\centering
\resizebox{\ifdim\width>\linewidth\linewidth\else\width\fi}{!}{
\fontsize{7}{9}\selectfont
\begin{tabular}[t]{lcc>{}c>{}c>{}c>{}c>{}c>{}c>{}c>{}c>{}c>{}c>{}c>{}c>{}c}
\toprule
\multicolumn{4}{c}{ } & \multicolumn{6}{c}{60\%} & \multicolumn{6}{c}{80\%} \\
\cmidrule(l{3pt}r{3pt}){5-10} \cmidrule(l{3pt}r{3pt}){11-16}
\multicolumn{4}{c}{ } & \multicolumn{3}{c}{N = 500} & \multicolumn{3}{c}{N = 1500} & \multicolumn{3}{c}{N = 500} & \multicolumn{3}{c}{N = 1500} \\
\cmidrule(l{3pt}r{3pt}){5-7} \cmidrule(l{3pt}r{3pt}){8-10} \cmidrule(l{3pt}r{3pt}){11-13} \cmidrule(l{3pt}r{3pt}){14-16}
Model & $\nu$ & $\rho$ & True & Bias & RMSE & CP & Bias & RMSE & CP & Bias & RMSE & CP & Bias & RMSE & CP\\
\midrule
EDPMM & 0 & 0.1 & \cellcolor{white}{0.10} & \cellcolor[HTML]{f2f2f2}{0.00} & \cellcolor[HTML]{f2f2f2}{0.07} & \cellcolor[HTML]{f2f2f2}{98.9} & \cellcolor[HTML]{d9d9d9}{0.00} & \cellcolor[HTML]{d9d9d9}{0.05} & \cellcolor[HTML]{d9d9d9}{99.0} & \cellcolor[HTML]{f2f2f2}{0.00} & \cellcolor[HTML]{f2f2f2}{0.09} & \cellcolor[HTML]{f2f2f2}{97.7} & \cellcolor[HTML]{d9d9d9}{0.00} & \cellcolor[HTML]{d9d9d9}{0.06} & \cellcolor[HTML]{d9d9d9}{98.4}\\
 &  & 0.2 & \cellcolor{white}{0.22} & \cellcolor[HTML]{f2f2f2}{0.00} & \cellcolor[HTML]{f2f2f2}{0.18} & \cellcolor[HTML]{f2f2f2}{98.9} & \cellcolor[HTML]{d9d9d9}{0.00} & \cellcolor[HTML]{d9d9d9}{0.11} & \cellcolor[HTML]{d9d9d9}{99.1} & \cellcolor[HTML]{f2f2f2}{0.02} & \cellcolor[HTML]{f2f2f2}{0.25} & \cellcolor[HTML]{f2f2f2}{97.0} & \cellcolor[HTML]{d9d9d9}{0.00} & \cellcolor[HTML]{d9d9d9}{0.14} & \cellcolor[HTML]{d9d9d9}{98.4}\\
 & 1 & 0.1 & \cellcolor{white}{0.17} & \cellcolor[HTML]{f2f2f2}{0.02} & \cellcolor[HTML]{f2f2f2}{0.13} & \cellcolor[HTML]{f2f2f2}{98.8} & \cellcolor[HTML]{d9d9d9}{0.02} & \cellcolor[HTML]{d9d9d9}{0.08} & \cellcolor[HTML]{d9d9d9}{98.1} & \cellcolor[HTML]{f2f2f2}{0.06} & \cellcolor[HTML]{f2f2f2}{0.21} & \cellcolor[HTML]{f2f2f2}{95.7} & \cellcolor[HTML]{d9d9d9}{0.02} & \cellcolor[HTML]{d9d9d9}{0.12} & \cellcolor[HTML]{d9d9d9}{97.4}\\
 &  & 0.2 & \cellcolor{white}{0.55} & \cellcolor[HTML]{f2f2f2}{0.10} & \cellcolor[HTML]{f2f2f2}{0.35} & \cellcolor[HTML]{f2f2f2}{98.5} & \cellcolor[HTML]{d9d9d9}{0.05} & \cellcolor[HTML]{d9d9d9}{0.21} & \cellcolor[HTML]{d9d9d9}{98.6} & \cellcolor[HTML]{f2f2f2}{0.16} & \cellcolor[HTML]{f2f2f2}{0.55} & \cellcolor[HTML]{f2f2f2}{95.2} & \cellcolor[HTML]{d9d9d9}{0.07} & \cellcolor[HTML]{d9d9d9}{0.32} & \cellcolor[HTML]{d9d9d9}{96.3}\\
 & 2 & 0.1 & \cellcolor{white}{0.54} & \cellcolor[HTML]{f2f2f2}{0.10} & \cellcolor[HTML]{f2f2f2}{0.25} & \cellcolor[HTML]{f2f2f2}{97.8} & \cellcolor[HTML]{d9d9d9}{0.04} & \cellcolor[HTML]{d9d9d9}{0.14} & \cellcolor[HTML]{d9d9d9}{98.3} & \cellcolor[HTML]{f2f2f2}{0.14} & \cellcolor[HTML]{f2f2f2}{0.42} & \cellcolor[HTML]{f2f2f2}{94.3} & \cellcolor[HTML]{d9d9d9}{0.06} & \cellcolor[HTML]{d9d9d9}{0.23} & \cellcolor[HTML]{d9d9d9}{97.0}\\
 &  & 0.2 & \cellcolor{white}{1.28} & \cellcolor[HTML]{f2f2f2}{0.15} & \cellcolor[HTML]{f2f2f2}{0.47} & \cellcolor[HTML]{f2f2f2}{98.0} & \cellcolor[HTML]{d9d9d9}{0.06} & \cellcolor[HTML]{d9d9d9}{0.28} & \cellcolor[HTML]{d9d9d9}{98.1} & \cellcolor[HTML]{f2f2f2}{0.17} & \cellcolor[HTML]{f2f2f2}{0.76} & \cellcolor[HTML]{f2f2f2}{94.1} & \cellcolor[HTML]{d9d9d9}{0.07} & \cellcolor[HTML]{d9d9d9}{0.44} & \cellcolor[HTML]{d9d9d9}{96.5}\\
 & 3 & 0.1 & \cellcolor{white}{0.87} & \cellcolor[HTML]{f2f2f2}{0.14} & \cellcolor[HTML]{f2f2f2}{0.34} & \cellcolor[HTML]{f2f2f2}{97.5} & \cellcolor[HTML]{d9d9d9}{0.05} & \cellcolor[HTML]{d9d9d9}{0.18} & \cellcolor[HTML]{d9d9d9}{98.4} & \cellcolor[HTML]{f2f2f2}{0.19} & \cellcolor[HTML]{f2f2f2}{0.57} & \cellcolor[HTML]{f2f2f2}{94.8} & \cellcolor[HTML]{d9d9d9}{0.08} & \cellcolor[HTML]{d9d9d9}{0.30} & \cellcolor[HTML]{d9d9d9}{97.0}\\
 &  & 0.2 & \cellcolor{white}{1.77} & \cellcolor[HTML]{f2f2f2}{0.17} & \cellcolor[HTML]{f2f2f2}{0.54} & \cellcolor[HTML]{f2f2f2}{97.5} & \cellcolor[HTML]{d9d9d9}{0.06} & \cellcolor[HTML]{d9d9d9}{0.31} & \cellcolor[HTML]{d9d9d9}{98.4} & \cellcolor[HTML]{f2f2f2}{0.21} & \cellcolor[HTML]{f2f2f2}{0.95} & \cellcolor[HTML]{f2f2f2}{94.7} & \cellcolor[HTML]{d9d9d9}{0.09} & \cellcolor[HTML]{d9d9d9}{0.51} & \cellcolor[HTML]{d9d9d9}{97.8}\\

\cmidrule(l{3pt}r{3pt}){1-16}

DPMM & 0 & 0.1 & \cellcolor{white}{0.10} & \cellcolor[HTML]{f2f2f2}{0.11} & \cellcolor[HTML]{f2f2f2}{0.15} & \cellcolor[HTML]{f2f2f2}{88.8} & \cellcolor[HTML]{d9d9d9}{0.11} & \cellcolor[HTML]{d9d9d9}{0.13} & \cellcolor[HTML]{d9d9d9}{64.1} & \cellcolor[HTML]{f2f2f2}{0.08} & \cellcolor[HTML]{f2f2f2}{0.15} & \cellcolor[HTML]{f2f2f2}{95.6} & \cellcolor[HTML]{d9d9d9}{0.08} & \cellcolor[HTML]{d9d9d9}{0.10} & \cellcolor[HTML]{d9d9d9}{89.6}\\
 &  & 0.2 & \cellcolor{white}{0.22} & \cellcolor[HTML]{f2f2f2}{0.18} & \cellcolor[HTML]{f2f2f2}{0.25} & \cellcolor[HTML]{f2f2f2}{91.2} & \cellcolor[HTML]{d9d9d9}{0.18} & \cellcolor[HTML]{d9d9d9}{0.21} & \cellcolor[HTML]{d9d9d9}{73.3} & \cellcolor[HTML]{f2f2f2}{0.12} & \cellcolor[HTML]{f2f2f2}{0.26} & \cellcolor[HTML]{f2f2f2}{96.2} & \cellcolor[HTML]{d9d9d9}{0.12} & \cellcolor[HTML]{d9d9d9}{0.17} & \cellcolor[HTML]{d9d9d9}{92.7}\\
 & 1 & 0.1 & \cellcolor{white}{0.17} & \cellcolor[HTML]{f2f2f2}{0.02} & \cellcolor[HTML]{f2f2f2}{0.09} & \cellcolor[HTML]{f2f2f2}{98.4} & \cellcolor[HTML]{d9d9d9}{0.02} & \cellcolor[HTML]{d9d9d9}{0.05} & \cellcolor[HTML]{d9d9d9}{97.9} & \cellcolor[HTML]{f2f2f2}{0.00} & \cellcolor[HTML]{f2f2f2}{0.12} & \cellcolor[HTML]{f2f2f2}{97.9} & \cellcolor[HTML]{d9d9d9}{0.00} & \cellcolor[HTML]{d9d9d9}{0.06} & \cellcolor[HTML]{d9d9d9}{99.1}\\
 &  & 0.2 & \cellcolor{white}{0.55} & \cellcolor[HTML]{f2f2f2}{-0.14} & \cellcolor[HTML]{f2f2f2}{0.22} & \cellcolor[HTML]{f2f2f2}{95.0} & \cellcolor[HTML]{d9d9d9}{-0.14} & \cellcolor[HTML]{d9d9d9}{0.17} & \cellcolor[HTML]{d9d9d9}{85.2} & \cellcolor[HTML]{f2f2f2}{-0.18} & \cellcolor[HTML]{f2f2f2}{0.31} & \cellcolor[HTML]{f2f2f2}{94.2} & \cellcolor[HTML]{d9d9d9}{-0.19} & \cellcolor[HTML]{d9d9d9}{0.23} & \cellcolor[HTML]{d9d9d9}{83.7}\\
 & 2 & 0.1 & \cellcolor{white}{0.54} & \cellcolor[HTML]{f2f2f2}{-0.33} & \cellcolor[HTML]{f2f2f2}{0.35} & \cellcolor[HTML]{f2f2f2}{16.2} & \cellcolor[HTML]{d9d9d9}{-0.34} & \cellcolor[HTML]{d9d9d9}{0.34} & \cellcolor[HTML]{d9d9d9}{0.0} & \cellcolor[HTML]{f2f2f2}{-0.35} & \cellcolor[HTML]{f2f2f2}{0.37} & \cellcolor[HTML]{f2f2f2}{43.2} & \cellcolor[HTML]{d9d9d9}{-0.36} & \cellcolor[HTML]{d9d9d9}{0.37} & \cellcolor[HTML]{d9d9d9}{0.9}\\
 &  & 0.2 & \cellcolor{white}{1.28} & \cellcolor[HTML]{f2f2f2}{-0.84} & \cellcolor[HTML]{f2f2f2}{0.86} & \cellcolor[HTML]{f2f2f2}{9.0} & \cellcolor[HTML]{d9d9d9}{-0.85} & \cellcolor[HTML]{d9d9d9}{0.85} & \cellcolor[HTML]{d9d9d9}{0.0} & \cellcolor[HTML]{f2f2f2}{-0.87} & \cellcolor[HTML]{f2f2f2}{0.92} & \cellcolor[HTML]{f2f2f2}{34.3} & \cellcolor[HTML]{d9d9d9}{-0.89} & \cellcolor[HTML]{d9d9d9}{0.90} & \cellcolor[HTML]{d9d9d9}{0.0}\\
 & 3 & 0.1 & \cellcolor{white}{0.87} & \cellcolor[HTML]{f2f2f2}{-0.65} & \cellcolor[HTML]{f2f2f2}{0.66} & \cellcolor[HTML]{f2f2f2}{0.1} & \cellcolor[HTML]{d9d9d9}{-0.66} & \cellcolor[HTML]{d9d9d9}{0.66} & \cellcolor[HTML]{d9d9d9}{0.0} & \cellcolor[HTML]{f2f2f2}{-0.66} & \cellcolor[HTML]{f2f2f2}{0.68} & \cellcolor[HTML]{f2f2f2}{7.9} & \cellcolor[HTML]{d9d9d9}{-0.68} & \cellcolor[HTML]{d9d9d9}{0.68} & \cellcolor[HTML]{d9d9d9}{0.0}\\
 &  & 0.2 & \cellcolor{white}{1.77} & \cellcolor[HTML]{f2f2f2}{-1.29} & \cellcolor[HTML]{f2f2f2}{1.31} & \cellcolor[HTML]{f2f2f2}{0.5} & \cellcolor[HTML]{d9d9d9}{-1.31} & \cellcolor[HTML]{d9d9d9}{1.31} & \cellcolor[HTML]{d9d9d9}{0.0} & \cellcolor[HTML]{f2f2f2}{-1.32} & \cellcolor[HTML]{f2f2f2}{1.36} & \cellcolor[HTML]{f2f2f2}{12.4} & \cellcolor[HTML]{d9d9d9}{-1.35} & \cellcolor[HTML]{d9d9d9}{1.36} & \cellcolor[HTML]{d9d9d9}{0.0}\\
\bottomrule
\end{tabular}}
\end{table}
    Table~\ref{tableEDPqrl_sim34_supp} reports additional results for Scenarios 3 and 4 using the more conservative nominal level $\alpha=0.01$, corresponding to 99\% CrIs, for sample sizes $N=500$ and $N=1500$. These results are compared with the corresponding 95\% CrI results in Table~\ref{tableEDPqrl_sim_main}. As expected, the empirical bias and RMSE are similar to those in the main analysis because the posterior point estimates are unchanged. The primary consequence of using wider CrIs is improved empirical coverage.

    Under 60\% censoring, empirical coverage based on the 95\% CrIs ranges from approximately 91\% to 95\% across the two sample sizes. With 99\% CrIs, coverage increases to approximately 98\%--99\% and is generally close to the corresponding nominal level for both $N=500$ and $N=1500$. Under 80\% censoring, coverage increases from approximately 87\%--93\% under the 95\% CrIs to approximately 94\%--98\% under the 99\% CrIs. More importantly, the coverage shortfall relative to the corresponding nominal level is substantially reduced. Some undercoverage remains, particularly for $N=500$ and at larger values of $\nu$, but coverage generally improves further as the sample size increases to $N=1500$.

    The DPMM continues to exhibit substantially larger bias and RMSE than the EDPMM when 99\% CrIs are used. Although the wider intervals improve coverage in some settings with smaller values of $\nu$, severe undercoverage persists at larger values of $\nu$, particularly for $\rho=0.2$. Thus, increasing the credible level improves uncertainty quantification for the EDPMM but does not resolve the underlying estimation problems of the DPMM.

    Overall, the additional analyses indicate that the remaining EDPMM undercoverage in the most challenging settings is associated with limited observed failure-time information and finite sample size. By contrast, the persistent deterioration of the DPMM even at $N=5000$ suggests a structural limitation: single-layer clustering can be dominated by the joint exposure--covariate distribution, obscuring local outcome structure, whereas the EDPMM preserves outcome-oriented clustering while accommodating exposure and covariate dependence. The improved calibration of the 99\% CrIs supports their use in the substantially censored ADNI application without changing the main qualitative comparison between the EDPMM and the DPMM.
    
\section{Supplementary Details and Results for ADNI} \label{EDPMqrl.supp.application}

\subsection{Cohort Construction and Descriptive Characteristics}
\label{EDPMqrl.supp.application.cohort}
    The analytic cohort was constructed by linking a subject-level participant roster with a diagnosis dataset used to determine baseline diagnostic status. We began with $5068$ subjects in the roster. Of these, $1358$ were excluded before merging because no corresponding diagnosis record was available, leaving $3710$ subjects in the merged dataset. After applying the analytic eligibility criteria, an additional $2397$ subjects were excluded, yielding a final analytic sample of $N=1313$ participants with baseline CN or MCI, available baseline amyloid status, and valid follow-up information.
    
    There were no missing values in baseline diagnosis, amyloid status, age, or sex in the final analytic sample. We observed $36$ missing values for APOE $\varepsilon$4 carrier status and $1$ missing value for years of education. Rather than excluding these participants, we handled the missing covariates through the EDPMM data augmentation procedure under the within-subcluster MAR assumption described in Section~\ref{EDPMqrl.supp.mcmc.missing}.
    
    In the final analytic cohort, $533$ participants (40.59\%) had elevated amyloid status at baseline ($Z=1$), and $517$ (39.38\%) were APOE $\varepsilon$4 carriers. The mean age was $70.96$ years (SD $=7.19$), 53.39\% of participants were female, and the mean years of education was $16.45$ (SD $=2.48$). During follow-up, $186$ participants (14.17\%) experienced clinical dementia onset and $1127$ (85.83\%) were right censored.
    
    Event incidence differed markedly across baseline risk strata. By baseline amyloid status, the event rate was 4.62\% ($36/780$) in the non-elevated amyloid group and 28.14\% ($150/533$) in the elevated amyloid group. By baseline diagnosis, the event rate was 2.91\% ($20/688$) among CN participants and 26.56\% ($166/625$) among MCI participants. The contrast was most pronounced in the joint baseline amyloid $\times$ diagnosis strata: 1.23\% ($6/487$) for non-elevated amyloid CN, 10.24\% ($30/293$) for non-elevated amyloid MCI, 6.97\% ($14/201$) for elevated amyloid CN, and 40.96\% ($136/332$) for elevated amyloid MCI.
    
    These patterns demonstrate substantial heterogeneity in event information across subgroups, with lower precision expected in lower-risk strata and at later follow-up times. Heavy right censoring, sparse subgroup events, and conditioning on remaining dementia-free beyond $\nu$ make direct empirical estimation of residual-life contrasts difficult and result in substantial variation in the amount of available event-time information. In this setting, the EDPMM borrows strength across covariate profiles to stabilize conditional survival estimation for Bayesian g-computation.
    
\subsection{Dementia-Free Survival Estimates}
\label{EDPMqrl.supp.application.survival}
    \begin{figure}[!tbp]
    \centering
    \includegraphics[width=\textwidth]{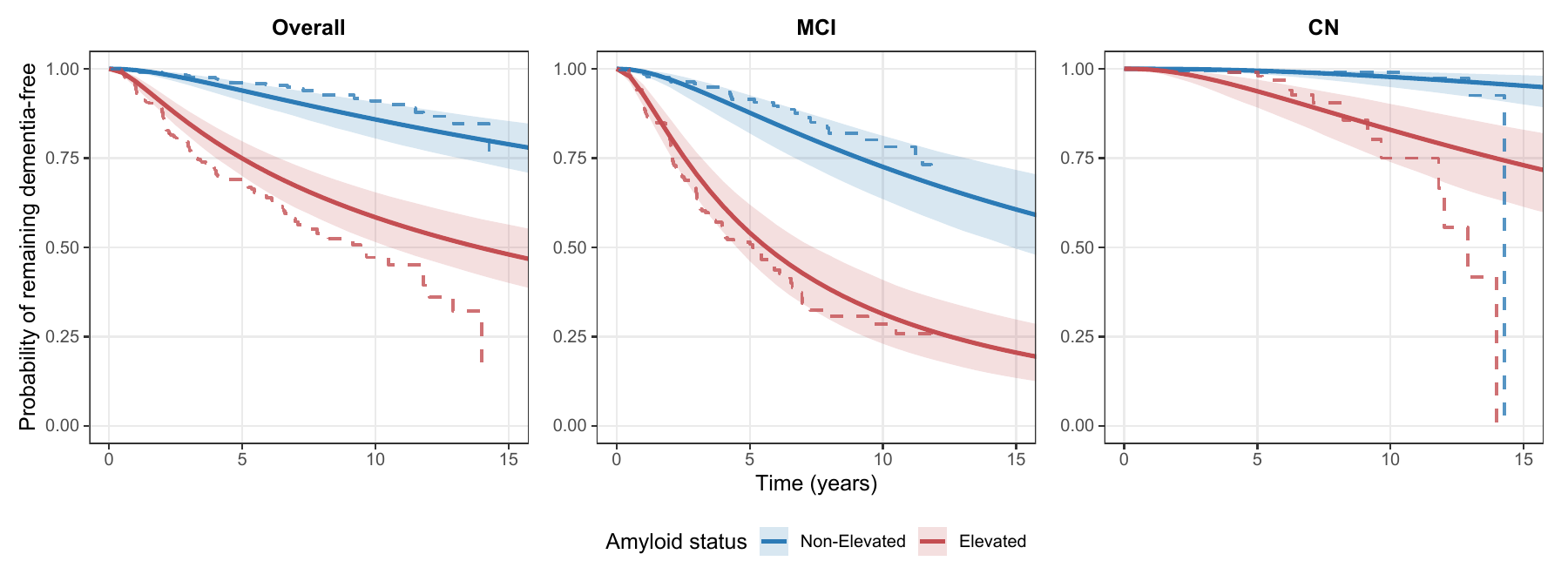}
    \caption{\label{figureEDPqrl_ADNIsurv}
    Dementia-free survival by baseline amyloid status in the overall ADNI cohort and the baseline MCI and CN subgroups. Solid curves and shaded bands denote EDPMM posterior means and 99\% credible intervals; dashed step functions denote unadjusted Kaplan--Meier estimates. Comparisons across panels illustrate heterogeneity in dementia-free survival patterns across the overall, MCI, and CN populations, whereas comparisons between the EDPMM and Kaplan--Meier curves contrast model-based covariate-adjusted estimates with empirical unadjusted estimates.}
\end{figure}
    Figure~\ref{figureEDPqrl_ADNIsurv} displays the estimated dementia-free survival functions under elevated and non-elevated baseline amyloid status for the overall cohort and the baseline MCI and CN subgroups. Differences between the EDPMM and Kaplan--Meier curves are expected because the Kaplan--Meier estimates are unadjusted for baseline confounding. Nevertheless, the EDPMM estimates preserve the broad survival patterns and between-group separation observed in the data. Estimated dementia-free survival is lower under elevated amyloid in all three populations, with the clearest separation in the MCI subgroup. Survival remains high under both amyloid statuses in the CN subgroup, where the small number of dementia events leads to greater posterior uncertainty. The abrupt declines in the tails of the Kaplan--Meier curves reflect small risk sets, while the widening EDPMM CrIs reflect the decreasing event-time information available at later follow-up times.

\subsection{Overall and CN Sensitivity Analyses}
\label{EDPMqrl.supp.application.results}
    \begin{figure}[!tbp]
    \centering
    \includegraphics[width=\textwidth]{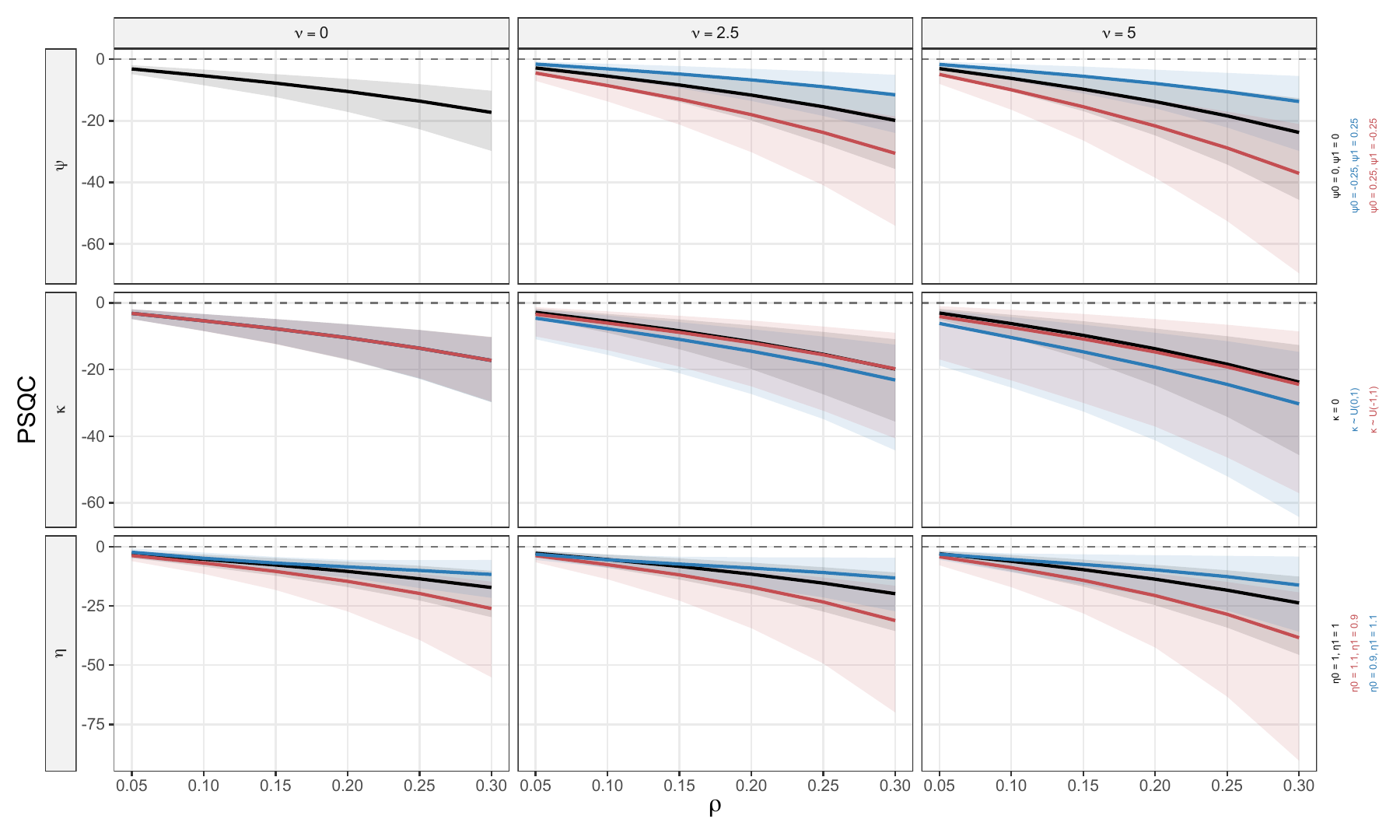}
    \caption{\label{figureEDPqrlPOST_PSQC_OVERALL}
    Primary and sensitivity analyses of the PSQC across quantile levels $\rho$ in the overall ADNI cohort. Columns correspond to landmark times $\nu=0$, $2.5$, and $5$, and rows examine $(\psi_{0},\psi_{1})$, $\kappa$, and $(\eta_{0},\eta_{1})$. Curves and shaded bands denote posterior means and 99\% credible intervals. Black denotes the primary specification in each row, colored curves denote the alternatives shown in the legends, and the horizontal dashed line marks PSQC $=0$.}
\end{figure}
    Figure~\ref{figureEDPqrlPOST_PSQC_OVERALL} presents the primary and sensitivity analyses of the PSQC in the overall ADNI cohort. Under the primary analysis, the posterior mean PSQC remains negative across the displayed time points and quantile levels and generally becomes more negative as $\rho$ increases. The alternative specifications shift the magnitude of the estimated contrast and increase uncertainty in some settings, particularly at higher quantile levels and later time points. Nevertheless, the qualitative conclusion is preserved: within the TAS principal stratum at each time point, quantiles of remaining time to dementia onset are estimated to be shorter under elevated than under non-elevated baseline amyloid status.
    
    \begin{figure}[!tbp]
    \centering
    \includegraphics[width=\textwidth]{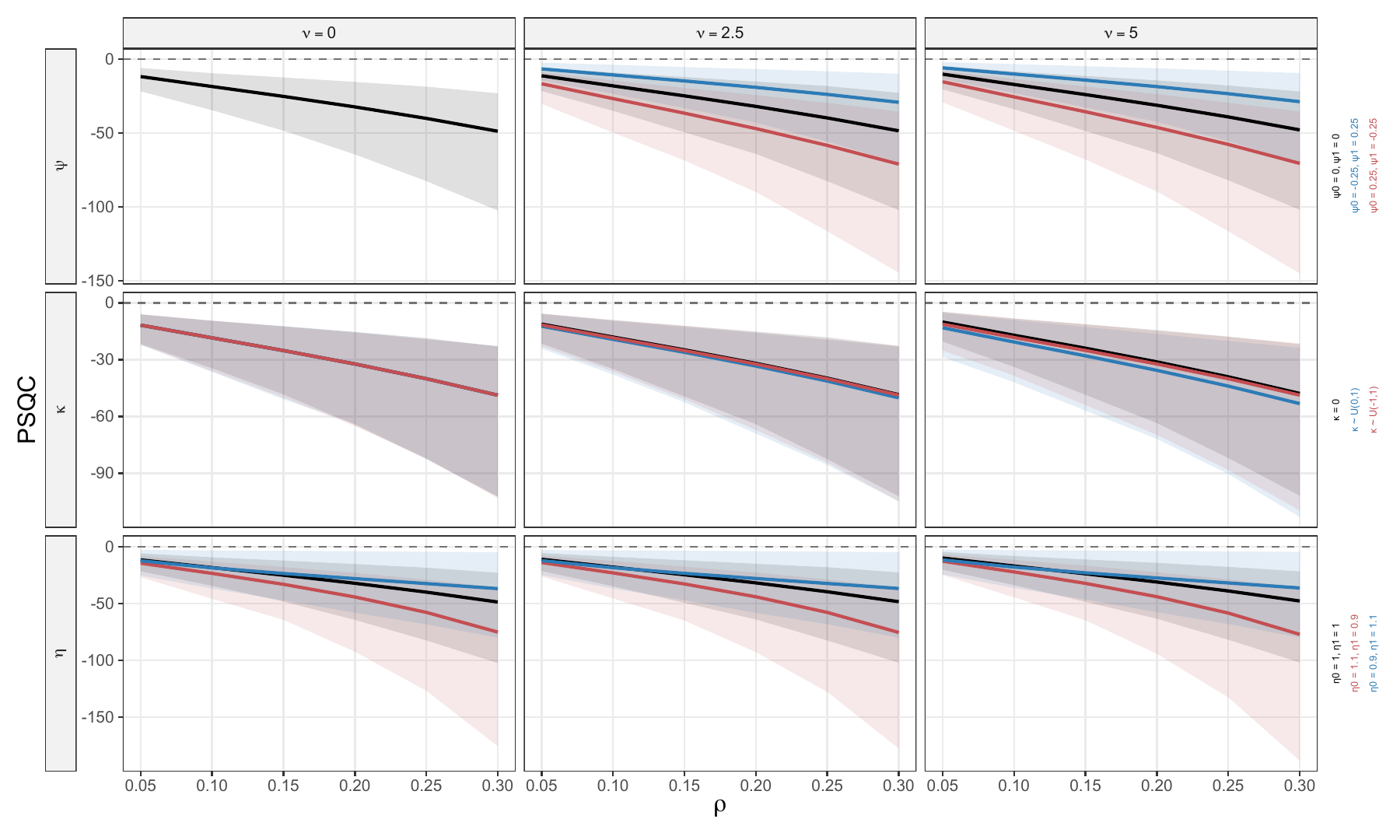}
    \caption{\label{figureEDPqrlPOST_PSQC_CN}
    Primary and sensitivity analyses of the PSQC across quantile levels $\rho$ in the baseline CN subgroup. Columns correspond to landmark times $\nu=0$, $2.5$, and $5$, and rows examine $(\psi_{0},\psi_{1})$, $\kappa$, and $(\eta_{0},\eta_{1})$. Curves and shaded bands denote posterior means and 99\% credible intervals. Black denotes the primary specification in each row, colored curves denote the alternatives shown in the legends, and the horizontal dashed line marks PSQC $=0$.
    }
\end{figure}
    Figure~\ref{figureEDPqrlPOST_PSQC_CN} presents the corresponding primary and sensitivity analyses for the baseline CN subgroup. Under the primary identifying assumptions, the posterior mean PSQC is negative across the displayed time points and quantile levels, with substantially larger magnitudes than those observed in the baseline MCI subgroup. Alternative assumptions regarding residual unmeasured confounding, cross-world dependence, and censoring produce noticeable changes in magnitude and uncertainty, particularly at higher quantile levels, although the negative direction is generally preserved. The CN results should nevertheless be interpreted cautiously because dementia progression is relatively infrequent in this subgroup, resulting in limited event-time information and greater reliance on model-based extrapolation, especially at later time points and in the upper part of the residual-life distribution.


\end{document}